\documentclass[aip, amsmath, amssymb, reprint]{revtex4-2}

\usepackage{graphicx}
\usepackage{dcolumn}
\usepackage{bm}
\usepackage[dvipsnames]{xcolor}
\usepackage[utf8]{inputenc}
\usepackage[T1]{fontenc}
\usepackage{mathptmx}
\usepackage{booktabs}
\usepackage{amsmath}
\usepackage{etoolbox}
\usepackage{orcidlink}
\usepackage{xcolor}
\usepackage{caption}
\usepackage[rightcaption]{sidecap}
\usepackage{ragged2e}
\usepackage[final]{changes}
\usepackage{siunitx}
\usepackage{cancel}

\usepackage{hyperref}
\hypersetup{colorlinks = true, linkcolor = blue, citecolor=blue}

\sidecaptionvpos{figure}{c}
\DeclareCaptionFormat{graybox}{
    \colorbox{gray!13}{
        \parbox{0.95\linewidth}{
          \noindent
          \justifying
          \textbf{#1#2}#3
        }
    }
}

\usepackage{braket}
\usepackage{floatrow}
\usepackage{comment}
\usepackage{makecell}

\DeclareUnicodeCharacter{2009}{\,}

\usepackage{subcaption}
\usepackage[only,llbracket,rrbracket]{stmaryrd}
\newcommand\numberthis{\addtocounter{equation}{1}\tag{\theequation}}

\makeatletter \def\@email#1#2{
    \endgroup
    \patchcmd{\titleblock@produce}
    {\frontmatter@RRAPformat}
    {\frontmatter@RRAPformat{
        \produce@RRAP{*#1\href{mailto:#2}{#2}}
        }
     \frontmatter@RRAPformat}
    {}{}
} \makeatother

\begin{document}

\title{Effects of realistic pulse shapes in two-dimensional spectroscopy}

\author{M. Russo \orcidlink{0009-0006-4596-7380}}
\affiliation{Freie Universität Berlin, Fachbereich Physik, Arnimallee 14, 14195 Berlin, Germany}

\author{R. Gilliot}
\affiliation{Freie Universität Berlin, Fachbereich Physik and Dahlem Center for Complex Quantum Systems, Arnimallee 14, 14195 Berlin, Germany}
\affiliation{Département de Physique, Institut Polytechnique de Paris, France}

\author{A. Blech, \orcidlink{0000-0002-2530-4399}}
\affiliation{Freie Universität Berlin, Fachbereich Physik and Dahlem Center for Complex Quantum Systems, Arnimallee 14, 14195 Berlin, Germany}
 
\author{M. Joffre}
\affiliation{Département de Physique, Institut Polytechnique de Paris, France}
\affiliation{Laboratoire d’Optique et Biosciences, CNRS, Inserm, École Polytechnique, Institut Polytechnique de Paris, 91120 Palaiseau, France.}
  
\author{C. P. Koch\,\orcidlink{0000-0001-6285-5766}}
\affiliation{Freie Universität Berlin, Fachbereich Physik and Dahlem Center for Complex Quantum Systems, Arnimallee 14, 14195 Berlin, Germany}
  
\author{H. Seiler \orcidlink{0000-0003-1521-4418},* }
\affiliation{Freie Universität Berlin, Fachbereich Physik, Arnimallee 14, 14195 Berlin, Germany}
\email{helene.seiler@fu-berlin.de}
  
\date{\today}

\begin{abstract}
    Two-dimensional (2D) spectroscopy is a powerful pump–pump–probe technique for revealing couplings between quantum states and disentangling the different contributions to the optical response of a system. We present an efficient method for 2D spectroscopy simulations in the Markovian limit for the environment,  capable of handling arbitrary pulse shapes and reproducing time-ordering and overlapping pulse effects, while maintaining a computational cost that scales linearly with the number of sampling points. We leverage this framework to investigate how 2D spectra are affected by spectral phase distortions and highly non-Gaussian pulse shapes, such as those produced experimentally by hollow-core fibers or non-collinear optical amplifiers. We show that realistic pulses can induce the appearance of additional spectral features, lineshape distortions and oscillating contributions in the system's dynamics. Notably, even weak temporal pulse tails arising from uncorrected high-order spectral phase terms cause visible changes in the 2D spectra. We also find that homodyne detection schemes employed in experiments can mitigate the presence of such pulse effects. These results emphasize the importance of including realistic pulses in 2D spectroscopy simulations to identify pulse-induced effects and minimize ambiguities in the interpretation of experimental data.
\end{abstract}

\maketitle

\section{Introduction}
\label{sec:Introduction}

Coherent two-dimensional (2D) spectroscopy is a powerful pump–pump–probe technique to investigate couplings between quantum states and disentangle homogeneous and inhomogeneous contributions to lineshapes~\cite{Hamm2011, Biswas2022, Li_2023}. Over the past decades it has been employed to reveal quantum dynamics and dephasing in molecular systems, light-harvesting complexes~\cite{Cao2020, Fuller2014, Schroeter2018}, semiconductor~\cite{Li_2023, Stone2009,Seiler2019, Policht2021} and metallic nanostructures~\cite{Jeffries2020}, and more recently in polaritonic systems~\cite{Li2022-qv, Timmer2026} and quantum materials~\cite{Novelli2020}. 
On the technical side, much effort has gone into shortening pulses and broadening their spectra to improve temporal resolution and address a wider range of transitions. At visible frequencies, broadband continua are typically generated via spectral broadening in non-collinear optical parametric amplifiers (NOPA)~\cite{Wilhelm1997, Cerullo2016} or gas-filled hollow-core fibers~\cite{Stolen1978, Dudley2006, Konorov2004, Seiler2017, Timmer2023}. Pulses derived from these sources often feature highly non-Gaussian spectral shapes~\cite{Seiler2019, Timmer2026, Policht2021}. Furthermore, while second order dispersion can usually be compensated, higher-order spectral phase terms remain challenging to correct. In the time domain, these uncompensated phases introduce long temporal tails that distort the measured dynamics on timescales far exceeding the nominal pulse duration ~\cite{Paleek2019}. Beyond spectral amplitude and phase effects, pulse overlap and time-ordering near the zero delay region can also substantially influence the measured signals, impacting the extraction of physical parameters\cite{Brosseau2023}. A detailed understanding of realistic pulse effects is therefore essential for fully exploiting 2D spectroscopy datasets and minimizing ambiguities in data interpretation.

While the impulsive limit approximation provides an intuitive interpretation of 2D spectra~\cite{Hamm2011, Mukamel1999}, it does not capture pulse shape effects, making simulations indispensable to quantify their impact~\cite{Gelin2009, Gelin2022, Rose2019}. This is due to its inherent assumption of delta function-like pulses that are much shorter than the system dynamics. By contrast, non-perturbative approaches explicitly propagate the density matrix in time such that pulse-shape effects are automatically included~\cite{Seidner1995, Bruschi2022, Leng2017, Manal2006, Seibt2009, Anda2021, Kenneweg2024}. Alternatively, perturbative approaches have also been used to study pulse shape effects~\cite{Hamm2011, GallagherFaeder1999, Abramavicius2010, Do2017, Brosseau2023, Lim2019}, that are incorporated via the convolution of the electric fields with the response function. In general, this requires the numerical evaluation of a computationally demanding triple integral~\cite{GallagherFaeder1999}. Analytical solutions have been derived for Gaussian and Lorentzian pulse shapes~\cite{Smallwood2017, Perlk2017, Schweigert2008}, and the treatment was subsequently extended to arbitrary pulse shapes~\cite{Do2017}, albeit still within the non-overlapping pulse regime. These previous works demonstrate an interest to develop approaches that describe pulse shape effects and their impact on 2D spectra. However, a systematic understanding of how specific characteristics of realistic pulses, in particular non-Gaussian spectral profiles and weak temporal tails, translate into observable features of 2D spectra is still lacking.

Here, we address this gap by establishing a direct connection between these realistic pulse characteristics and the spectral signatures they induce in a 2D spectrum. We present an approach for simulating 2D spectra in which, under Markovian assumption for the environment, the triple convolution integral for the emitted signal can be reshaped into a computationally more convenient form. The method handles arbitrary pulse shapes and accounts for realistic pulse effects, such as pulse overlap and time ordering. We showcase its capabilities on three model systems: an anharmonic oscillator, two coupled excitons, and a dimer, exploring how spectral phase distortions and non-Gaussian spectra influence the 2D spectra. We reveal the appearance of additional spectral features and spectral linewidth distortions in the zero delay region, and we show that spectral chirp can induce oscillating contributions in the early dynamics of the system. Notably, we show that even when the nominal pulse duration is short, the presence of extended temporal tails impacts the 2D spectroscopy signals and gives rise to additional spectral structures. Finally, we show that homodyne detection schemes typically employed in experiments mitigates such pulse-induced features. Overall, our approach efficiently simulates realistic pulse effects in 2D spectroscopy, supporting data interpretation and the exploration of pulse shaping for coherent control \cite{Kuehn2007, Lim2019}.

\section{Theory}
\label{sec:Theory}

This section introduces the theoretical framework used throughout this work. The discussion is structured as follows: in subsection~\ref{sec:Theory - Background}, we start from the non-linear response function formalism and show that, within the Markovian approximation for the environment, the triple integral appearing in the expression for the emitted 2D signal can be recast into a form that is more convenient for the numerical evaluation. Then, in subsection~\ref{sec:Signal Measurement}, we briefly outline the principles of homodyne detection. Lastly, in subsection~\ref{sec:Realistic Effects}, we qualitatively describe realistic pulse effects arising in 2D spectroscopy experiments.

\subsection{Simplification of the 2D signal expression under the Markovian approximation}
\label{sec:Theory - Background}

We adopt the semiclassical perturbative approach to describe the emitted third-order signal \( E_S^{(3)} \), which is a common framework employed in 2D spectroscopy to model the nonlinear response function of a system~\cite{Hamm2011, Mukamel1999}. Within the semiclassical picture, the quantum state of the system is represented by a time-dependent density matrix \( \hat{\rho} \), while the incident laser pulses are described by the classical electric field \( E \), weakly perturbing the system's dynamics. The emitted third-order signal can then be written as a sum of contributions, each arising from a single light-matter interaction pathway (Feynman pathway)~\cite{Hamm2011, Fetherolf2017, Gallego-Valencia2024, Jeske2015}, denoted as \( j \) in the following,
\begin{subequations}
\label{eq:Signal Pathway Decomposition}
\begin{equation}
    E_S^{(3)} ( \tau , \: T , \: t ) = \sum_{j} \, E_{S , \, j}^{(3)} \, ( \tau , \: T , \: t ) \; ,
\end{equation}
with
\begin{widetext}
\begin{equation}
    E_{S , \, j}^{(3)} \, ( \tau , \: T , \: t ) = \int_{-\infty}^{t} du \int_{-\infty}^{u} dv \int_{-\infty}^{v} dw \; E(u) \: E(v) \: E(w) \; R_j^{(3)} \bigl( t - u , \: u - v , \: v - w  \bigr) \; \rho^{(0)} \; .
    \label{eq:Nonlinear Formalism}
\end{equation}
\end{widetext}
\end{subequations}

Here, \( R_j^{(3)} \) is the partial response function corresponding to pathway \( j \), which explicitly depends on the time intervals between the successive light-matter interactions that are taking place at times \( ( u, \, v, \, w ) \). \( \rho^{(0)} \) is the density matrix of the unperturbed initial state of the system. The initial state can be chosen arbitrarily. To reproduce our typical experimental conditions, we will consider that the system is initially at thermal equilibrium in its ground state. This initial condition will be used for all simulations in Sec.~\ref{sec:Results}.

In a 2D spectroscopy experiment, the electric field \( E \) is composed of three pulses, see Fig.~\ref{fig:Pulses and Signal}. The first pump \( E_1 \) is centered on time \( t_1 \), the second pump \( E_2 \) is centered on time \( t_2 \), and the probe \( E_3 \) is centered on time \( t_3 \). Conventionally, \( t_3 \) is chosen as the time-zero. The experimental pulse setup can be equivalently described using the arrival times \( ( t_1 , \: t_2 , \: t_3 ) \) or the delays \( ( \tau , \: T ) \). The coherence time, \( \tau = t_2 - t_1 \), is the delay between the two pump pulses and the waiting-time~\footnote{In the literature of 2D coherent spectroscopy, this time interval is also widely referred to as the \textit{population time}, as it typically tracks the evolution of population states or non-radiative coherences.}, \( T = t_3 - t_2 = - t_2 \), is the delay between the second pump and the probe. These standard spectroscopic delays are also shown in Fig.~\ref{fig:Pulses and Signal}. In either formalisms, \( t - t_3 \) represents the delay between the probe's arrival and the signal emission~\cite{Hamm2011, Mukamel1999}.

The product of the three fields in Eq.~\eqref{eq:Nonlinear Formalism} can be expressed as
\begin{equation*}
    E(u) \, E(v) \, E(w) = \sum_{\sigma} \: E_{\sigma_3}(u - t_{\sigma_3}) \; E_{\sigma_2}(v - t_{\sigma_2}) \; E_{\sigma_1}(w - t_{\sigma_1}) \; ,
\end{equation*}
where \( \sigma = ( \sigma_1 , \: \sigma_2 , \: \sigma_3 ) \in \{ 1 , \, 2 , \, 3 \}^3 \) spans all 27 possible triplets, each corresponding to a certain time-ordering of the pulses. Repetitions are allowed to account for multiple interactions with the same pulse.
Using this decomposition of the electric field, Eq.~\eqref{eq:Signal Pathway Decomposition} can be recast into a form where each contribution associated with a different temporal ordering of the incident pulses is separated,
\begin{subequations}
\label{eq:Decomposed Signal Expression}
\begin{equation}
    E_{S}^{(3)} ( \tau , \: T , \: t ) = \sum_{j} \sum_{\sigma} E_{S , \, j , \, \sigma}^{(3)} \, ( \tau , \: T , \: t ) \; ,
\end{equation}
with
\begin{widetext}    
\begin{equation}
    E_{S , \: j , \: \sigma}^{(3)} \, ( \tau , \: T , \: t ) = \int\displaylimits_{- \infty}^{t} du \int\displaylimits_{- \infty}^{u} dv \int\displaylimits_{- \infty}^{v} dw \; E_{\sigma_3} (u - t_{\sigma_3}) \: E_{\sigma_2} (v - t_{\sigma_2}) \: E_{\sigma_1} (w - t_{\sigma_1}) \; \; R_j^{(3)} ( t - u , \: u - v , \: v - w ) \; \; \rho^{(0)} \; .
    \label{eq:Contribution Expression}
\end{equation}
\end{widetext}
\end{subequations}
This decomposition is key to our approach as it allows to treat independently and efficiently each single pathway \( j \) and time-ordering \( \sigma \). Taking into account every contribution is necessary to observe realistic pulse effects in simulated spectra.
While all pathways \( j \) are always taken into account in the simulations of Sec.~\ref{sec:Results}, only 6 time-orderings \( \sigma \) have to be taken into account to reproduce our typical experimental conditions. These 6 time-orderings correspond to configurations where each pulse interacts with the system exactly once. Experimentally, it is possible to ensure that only these 6 time-orderings contribute to the measured signal by employing phase-matching and phase-cycling techniques.

Solving the triple integral in Eq.~\eqref{eq:Contribution Expression} is notoriously computationally demanding~\cite{GallagherFaeder1999}. Consequently, many different methodologies have been developed over the years to address this problem, each introducing specific trade-offs between physical realism and computational feasibility  \cite{Hamm2011, GallagherFaeder1999, Abramavicius2010, Do2017, Brosseau2023, Lim2019}.
In this work, we introduce a novel approach that overcomes some of the previous limitations, enabling the simulation of finite, time-ordered, overlapping pulses of arbitrary shapes interacting with entirely general discrete level systems. This simplification is achieved under the assumption of Markovian dynamics of the system~\cite{Fetherolf2017, Gallego-Valencia2024}. 
Within this assumption, each partial response function can be expressed as
\begin{widetext}
\begin{equation}
\begin{aligned}
    R^{(3)}_j ( t - u , \: u - v , \: v - w ) = \frac{\mu_{S}^{(j)} \, \mu_{3}^{(j)} \, \mu_{2}^{(j)} \, \mu_{1}^{(j)}}{\hbar^{3}} \: \cdot
    & \: \exp \left( - i \omega_{3}^{(j)} ( t - u ) - i \omega_{2}^{(j)} ( u - v ) - i \omega_{1}^{(j)} ( v - w )  \right) \\
    & \cdot \: \exp \left( - \frac{t - u}{T_{3}^{(j)}} - \frac{u - v}{T_{2}^{(j)}} - \frac{v - w}{T_{1}^{(j)}} \right) \; ,
\end{aligned}
\label{eq:Response Function}
\end{equation}
\end{widetext}
where the transition dipole moments \( ( \mu_{S} , \, \mu_{3} , \, \mu_{2} , \, \mu_{1} ) \), energies \( ( \omega_{3} , \, \omega_{2} , \, \omega_{1} )\), and decay rates \( ( T_{3} , \, T_{2} , \, T_{1} ) \) are system-related parameters, selected according to pathway \( j \).

The partial response function \( R^{(3)}_j \) in Eq.~\eqref{eq:Response Function} depends on the time intervals between the successive interactions that take place at times \( ( u, \, v, \, w ) \) and it can be factorized into three terms, each describing the system's dynamics during a specific time interval, see App.~\ref{app:Derivation Final Expression},
\begin{subequations}
\label{eq:Signal Final Expression}
\begin{equation}
    E_{S, \, j, \, \sigma}^{(3)} \, ( \tau , \: T , \: t ) \; = \; I_{j, \, \sigma}^{(3)} \, ( t , \: T , \: \tau ) \; R_{j, \, \sigma}^{(3)} \, ( t , \: T , \: \tau ) \; \rho^{(0)}
\end{equation}
with
\begin{widetext}
\begin{equation}
    I_{j, \, \sigma}^{(3)} \, ( t , \: T , \: \tau ) \, =  \int\displaylimits_{- \infty}^{t} du \: F^{u}_{j, \, \sigma} \, ( u - t_{\sigma_3} )  \int\displaylimits_{- \infty}^{u} dv \: F^{v}_{j, \, \sigma} \, ( v - t_{\sigma_2} )  \int\displaylimits_{- \infty}^{v} dw \: F^{w}_{j, \, \sigma} \, ( w - t_{\sigma_1} ) \; ,
\end{equation}
\end{widetext}
\end{subequations}
where we have defined
\begin{equation*}
\begin{array}{lllll}
    F^{w}_{j, \, \sigma} \, ( w - t_{\sigma_1} ) & = & R_{j, \, 1} \, ( t_{\sigma_1} - w ) & E_1 \, ( w - t_{\sigma_1} ) & \\
    F^{v}_{j, \, \sigma} \, ( v - t_{\sigma_2} ) & = & R_{j, \, 2} \, ( t_{\sigma_2} - v ) & E_2 \, ( v - t_{\sigma_2} ) & R_{j, \, 1} \, ( v - t_{\sigma_2} ) \\
    F^{u}_{j, \, \sigma} \, ( u - t_{\sigma_3} ) & = & R_{j, \, 3} \, ( t_{\sigma_3} - u ) & E_3 \, ( u - t_{\sigma_3} ) & R_{j, \, 2} \, ( u - t_{\sigma_3} )
\end{array}
\label{eq:FieldeShapes Functions}
\end{equation*}
The expression of \( E_{S, \, j, \, \sigma}^{(3)} \) in Eq.~\eqref{eq:Signal Final Expression} exhibits a clear structure: from right to left, the initial state \( \rho^{(0)} \) is multiplied by the partial response function \( R_{j}^{(3)} \) and by the integral contribution \( I_{j, \, \sigma}^{(3)} \). The first two terms correspond to the expression of the contribution within the impulsive approximation, which is directly proportional to the partial response function~\cite{Hamm2011, Mukamel1999}. \( I_{j, \, \sigma}^{(3)} \) acts as a correcting term to the impulsive approximation and hosts the realistic effects affecting the emitted signal. From a quantum control point of view, \( I_{j, \, \sigma}^{(3)} \) can play the role of an adjustable weighting term in the summation of the partial contributions.
Equation~\eqref{eq:Signal Final Expression} also shows that the computational bottleneck lies in evaluating the integral contribution \( I_{j , \, \sigma}^{(3)} \), as the other terms have been factored out of the integration. Although the integration variables \( ( u, \, v, \, w ) \) still define the integration limits, the three integrals can now be evaluated successively, starting from the innermost integral, while treating the remaining integration variables as fixed parameters. Additionally, the functions to integrate \( \left( F^{u}_{j, \, \sigma} \, , \; F^{v}_{j, \, \sigma} \, , \; F^{w}_{j, \, \sigma} \right) \) rapidly fall to zero on both sides of the arrival time of the pulse. Fulfilling these two conditions is sufficient for the three nested integrals to be computed efficiently. To evaluate the emitted signal, this formulation is thus highly advantageous, reducing the scaling of the numerical complexity from cubic to linear (see Sec.~\ref{sec:Computational Methods}).

\subsection{Signal Measurement: homodyne detection}
\label{sec:Signal Measurement}

As a third-order signal, the isolated field \( E_S^{(3)} \) is too weak to be measured directly and must be mixed with a local oscillator. In the pump-probe geometry, homodyne detection is performed with the probe \( E_3 \), i.e. the last pulse, centered around \( t_3 = 0 \). It can be shown that the experimentally-measured signal \( S \) is given by~\cite{Hamm2011}
\begin{equation*}
    S \, ( \tau , \, T , \, \omega_t ) = 2 \, \mathit{Re} \left( E^*_{3} ( \omega_t ) \, \cdot \, E^{(3)}_S ( \tau , \, T , \, \omega_t ) \right) \; ,
    \label{eq:Measured Signal}
\end{equation*}
where the Fourier transform on the last time variable is performed by the spectrometer. We note that this derivation relies on phase-cycling schemes to cancel other contributions. The experimentally-measured signal \( S \) is typically Fourier transformed along the first delay to obtain the 2D spectrum \( S_{2D} \),
\begin{align*}
    S_{2D} \, ( \omega_\tau , \, T , \, \omega_t ) & = E^*_{3} ( \omega_t ) \, \cdot \, E^{(3)}_S ( \omega_\tau , \, T , \, \omega_t ) \numberthis \label{eq:2D Signal Final} \\
    & \hspace{40pt} + \, E_{3} ( \omega_t ) \, \cdot \, E^{(3) \, *}_S ( - \omega_\tau , \, T , \, \omega_t ) \; .
\end{align*}
In Sec.~\ref{sec:Homodyne Detection}, we will discuss how the homodyne detection scheme mitigates pulse-induced effects in the measured 2D signals by comparing \( S_{2D}( \omega_\tau , \, T , \, \omega_t ) \) to \( E^{(3)}_S ( \omega_\tau , \, T , \, \omega_t ) \, + \, E^{(3) \, *}_S ( - \omega_\tau , \, T , \, \omega_t ) \).

\subsection{Realistic pulse effects}
\label{sec:Realistic Effects}

\begin{figure}[h!]
    \includegraphics[width=.8\linewidth]{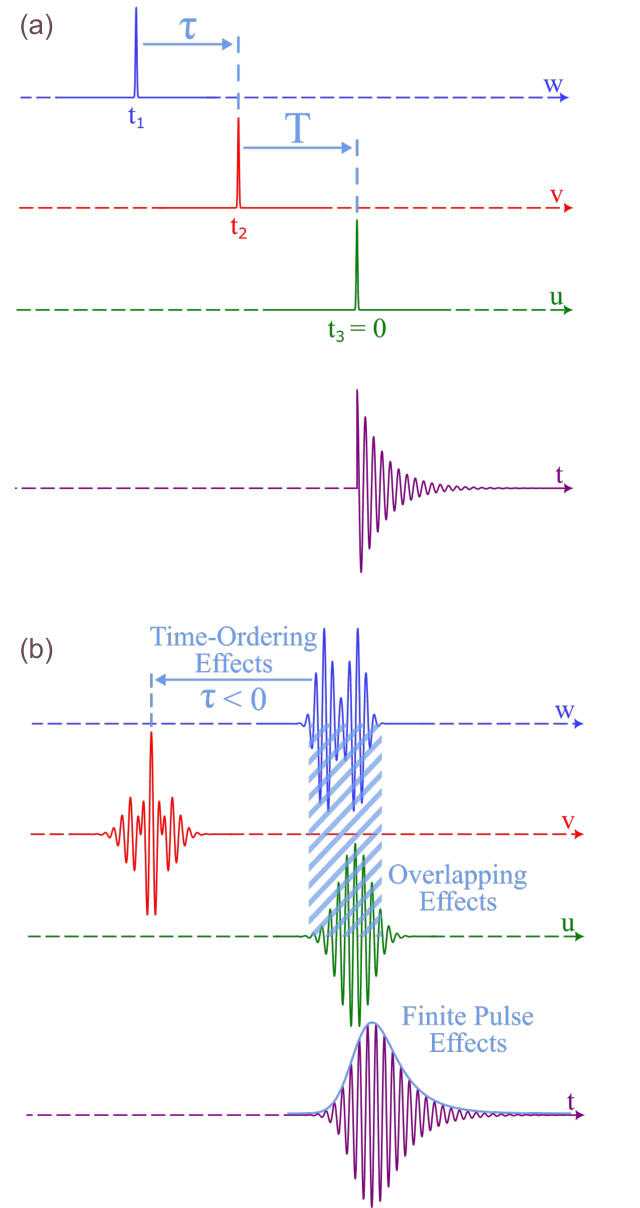}
    \caption{Schematic illustrations of the incident pulses \( ( E_1 , \: E_2 , \: E_3 ) \) (in blue, red, and green) and of the emitted signal \( E_S^{(3)} \) (in purple). Variables \( ( t_{1} , \: t_{2} , \: t_{3} ) \) are the arrival times of each pulse, while \( ( \tau , \: T ) \) are the delays between each pulse. (a) Impulsive picture (b) Realistic picture, where the emitted signal is influenced by time-ordering, overlapping, and finite pulse effects.}
    \label{fig:Pulses and Signal}
\end{figure}

In Sec.~\ref{sec:Theory - Background}, we introduced a convenient approach to simulate pulse shape, overlap, and time-ordering effects in 2D spectra. Here, we provide further details on the distinction between the ideal impulsive limit and the use of realistic pulses. Specifically, we discuss the additional physical effects that arise when finite pulse durations are taken into account.
Fig.~\ref{fig:Pulses and Signal}a displays a sketch of impulsive pulses \( ( E_1 , \: E_2 , \: E_3 ) \) (in blue, red, and green curves) and the resulting emitted signal \( E^{(3)}_S \) (in purple). In this case, the pulses are sufficiently short to be treated as delta functions. Under this assumption, the signal \( E^{(3)}_S \) is directly proportional to the material's response function, represented here as an oscillating exponential decay~\cite{Hamm2011, Mukamel1999}. By contrast, Fig.~\ref{fig:Pulses and Signal}b illustrates additional effects arising from considering realistic pulses. First, pulse shapes employed in experiments can be arbitrarily complex in the spectral and temporal domains. They can be Gaussian, as the green pulse, but also strongly non-Gaussian, as illustrated by the blue and red pulses. Second, realistic pulses have a finite duration, which means that they overlap in the time domain. These overlapping effects are at the origin of multiple additional light-matter interaction pathways due to the various time-orderings of the excitation pulses. This adds complexity to the resonant response measured in experiments, especially around the zero delay region. For example, if the second pulse starts before the tail of the first pulse ends, there is a certain probability that the sample interacts first with the second pulse, giving rise to an unintended light-matter interaction pathway. Importantly, time-ordering effects are not limited to the pulse-overlap region and should therefore be regarded as related but distinct from overlap effects. Even in the absence of direct pulse overlap, such as in the impulsive limit for instance, coherent polarization generated by one pulse can persist beyond the pulse duration. For example, if the probe \( E_3 \) arrives before the pump pulses \( ( E_1 , \: E_2 )\), the probe can generate a long-lived free induction decay. The pump pulses may then interact with this residual coherence, producing signal contributions at negative time delays. These signals are known as perturbed free induction decay (PFID) and have been analyzed several decades ago in transient absorption experiments~\cite{Chemla2001, Leo1990,Hamm1995,Joffre1988, Yan_2011,Mondal2018} as well as in 2D spectroscopy at infrared and THz frequencies~\cite{Kuehn2010, Hamm2000}. More recently, PFID has also been discussed in the context of 2D electronic spectroscopy~\cite{Brosseau2023, Paleek2019, Richter2017, Nguyen2019, Lloyd2021}. 

\section{Computational Methods}
\label{sec:Computational Methods}

Computing the general expression of Eq.~\eqref{eq:Nonlinear Formalism} for the signal \( E^{(3)}_S \) demands the evaluation of a triple integral, which can be a resource-intensive task since it scales cubically as \( O \bigl( N_S^3 \bigr) \), where \( N_S \) is the number of grid points per pulse. Within our approach, however, the triple integral decouples into three successive independent single-variable integrations, see Eq.~\eqref{eq:Signal Final Expression}, lowering the numerical complexity to a linear scaling \( O \bigl( N_S \bigr) \).
Hence our method reaches the same scaling as for simulations within the non-overlapping or impulsive approximations without imposing these approximations.

Eventually, for the simulation of the 2D spectrum \( S_{2D} \) over all three varying time variables \( ( \tau , \: T , \: t ) \), the emitted signal must be evaluated over an array of \( ( N_\tau , \: N_T , \: N_t ) \) points along the \( ( \tau , \: T , \: t ) \) time axes. The computational complexity for evaluating the contribution of a single interaction pathway \( E_{S, \, j, \, \sigma}^{(3)} \) is then \( O \bigl( N_\tau \cdot N_T \cdot N_S \bigr) \). The number of chosen time points \( N_t \) is replaced by the sampling \( N_S\) as each row of \( N_t \) points is interpolated from one series of \( N_S \) points, with \( N_t \ll N_S \) (see also App.~\ref{app:Computational Methods}). Thus the computational complexity for evaluating the complete signal \( E_{S}^{(3)} \) for all values of \( ( \tau , \: T , \: t ) \) is \( O \bigl( J \cdot N_\tau \cdot N_T \cdot N_S \bigr) \), where \( J \) is the number of Feynman pathways. Since an array of \( N_\tau \cdot N_T \cdot N_S \) values is calculated, the scaling of our 2D spectra simulations is minimal.
This is a significant speed-up compared to the exact simulation of 2D spectra using the triple integral of Eq.~\eqref{eq:Nonlinear Formalism}, which scales as \( O \bigl( N_\tau \cdot N_T \cdot N_S^{3} \bigr) \). This is also a crucial improvement in terms of numerical accuracy as the pulse sampling \( N_S \) is the only source of inaccuracy. The linear scaling of the pulse sampling \( N_S \) in our approach allows to use thousands of points and thus to reach excellent accuracy (see App.~\ref{app:Computational Methods}).
Additionally, in our approach, the numerical scaling related to the complexity of the system is the dependency on the number of Feynman pathways \( J \) instead of the size of the density matrix. This opens the opportunity to identify and quantify the relative intensity of each contribution to the emitted signal \( E_{S, \, j, \, \sigma}^{(3)} \), which can prove especially useful for systems with a limited number of relevant pathways.

\section{Results}
\label{sec:Results}
\begin{SCfigure}[0.4][h]
    \centering
    \includegraphics[width=0.7\linewidth, height=4cm]{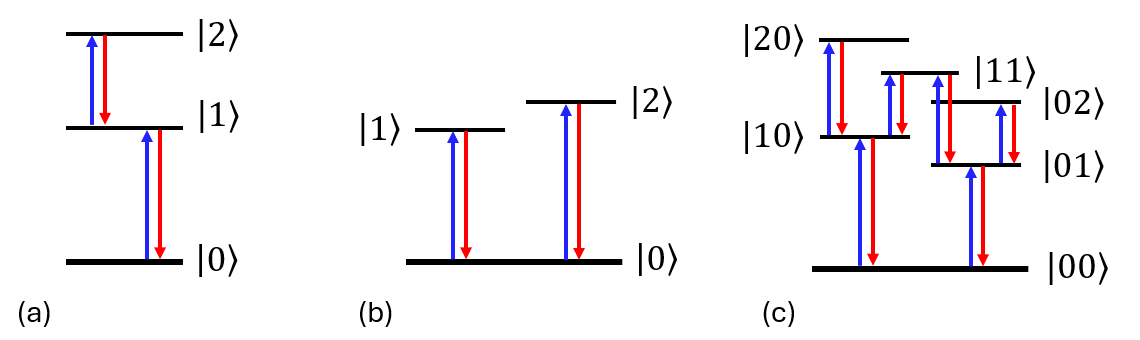}
    \caption{Energy-level diagrams of the model systems investigated in this work: (a) anharmonic oscillator, (b) two-exciton system, and (c) six-level dimer model. These systems represent standard model platforms in 2D spectroscopy.}
    \label{fig:Quantum Systems}
\end{SCfigure}

Here we show how our numerical approach enables the investigation of realistic pulse shape effects on 2D spectra. We consider three model systems that are widely used in 2D spectroscopy: an anharmonic oscillator~\cite{Schneider2017, Slenkamp2014, Pour2017} (Fig.~\ref{fig:Quantum Systems}a), two coupled excitons~\cite{Cassette2015, Guo2018, Rodek2023} (Fig.~\ref{fig:Quantum Systems}b), and a six-level dimer~\cite{Camargo2015, Milota2009, Mandal2018} (Fig.~\ref{fig:Quantum Systems}c). They form a convenient playground to analyze how pulse shape features influence 2D spectra and their dynamics as they involve a small number of transitions, while still capturing the essential physics underlying 2D spectra. The simulation parameters, including the lifetimes of the excited states and coherences, are detailed in App.~\ref{app:model_systems}, \ref{app:pulse_flat}, and \ref{app:param_III}. For all simulations, the system is assumed to start in its ground state. Throughout this work, we consider laser pulses in the visible spectral range. If not stated otherwise explicitly, the pulse spectra are centered around an energy of 1.92 eV (646 nm). In the transform-limited (TL) case, the Gaussian pulses have a full-width at half-maximum (FWHM) of 75 fs, which in the presence of \( + \, 500 \) {fs}$^2$ chirp gets broadened to a FWHM of 84 fs. We note that all these values may vary slightly in specific simulations, as in some cases the pulse parameters were adjusted to optimize the visual clarity of the resulting spectra. Whenever such modifications were introduced, they are explicitly stated in the corresponding figure captions or discussion.

\subsection{Pulse shape and spectral phase effects}

\begin{figure}[H]
    \centering
    \includegraphics[width=6.8in,height=6.12in]{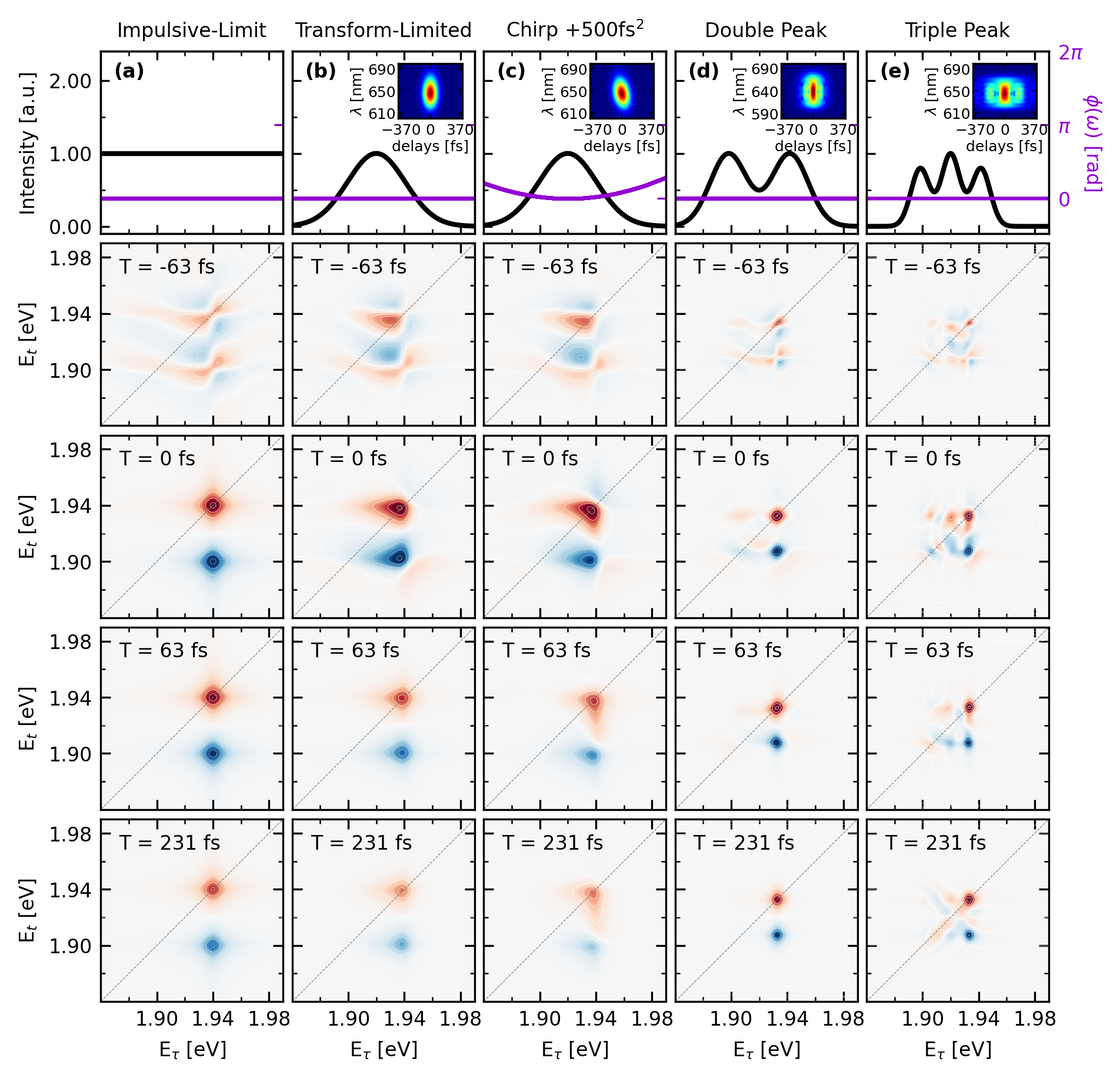}
    \caption{Influence of pulse shape and spectral phase on the 2D spectra for the anharmonic oscillator system. Each column corresponds to a different excitation condition: (a) impulsive limit (IL), (b) Gaussian transform-limited pulse (TL), (c) chirped Gaussian pulse (CH), (d) double-peaked pulse with flat spectral phase, and (e) triple-peaked pulse with flat spectral phase. The first panel of each column shows the spectral amplitude and phase of the corresponding pulse (purple line), together with the associated FROG trace (inset). The panels below display the dynamics of the simulated 2D spectra. The color scale is normalized column-wise for each column. Blue features are negative, while red features are positive.}
    \label{fig:AO}
\end{figure}

In this subsection we present a systematic study to assess how spectral phase and pulse shape affect 2D spectra. The results for the anharmonic oscillator system are summarized in Fig.~\ref{fig:AO}, while additional examples for the coupled two-exciton and the dimer systems are reported in Fig.~\ref{fig:EX} and  Fig.~\ref{fig:DI} in App.~\ref{app:pulse_shape}, respectively. We consider a sequence of scenarios with progressively increasing complexity of pulses: starting from the ideal impulsive-limit case (IL), then moving to transform-limited Gaussian pulses (TL), and finally including Gaussian pulses with a chirped spectral phase (CH). For the anharmonic oscillator system, the dynamics of the 2D spectra corresponding to each one of these situations is shown in columns (a), (b), and (c) of Fig.~\ref{fig:AO}. This progressive approach allows us to disentangle the individual contributions associated with each level of pulse complexity and to identify the mechanisms responsible for the appearance of the different spectral features in the resulting 2D spectra. For completeness, the first row of each column reports the spectral amplitude (black line) and phase (purple line) of the corresponding pulses used in each simulation. An inset showing the simulated frequency-resolved optical gating (FROG) trace of the pulses is also included in these panels. 

Column (a) of Fig.~\ref{fig:AO} shows the dynamics of the 2D spectrum for the anharmonic oscillator system in the IL case. Here, we did not need to employ the numerical approach presented in Sec.~\ref{sec:Theory - Background}, as the expression of the emitted signal can be easily obtained analytically from Eq.~\ref{eq:Decomposed Signal Expression} by considering the excitation pulses as delta-like functions.

\onecolumngrid

\twocolumngrid

In this limit they are assumed to be infinitely short in time compared to the coherence dynamics of the quantum states under investigation, while their spectrum is correspondingly very broad - ideally flat (see first panel of column (a)). As a consequence, the resulting 2D spectra  directly reflects the nonlinear response of the material, free from any distortions induced by finite pulse or spectral filtering effects~\cite{Hamm2011}. In line with theoretical expectations, the 2D spectrum exhibits a redshift of the blue (negative) feature relative to the main diagonal, as well as a non-zero signal prior to the arrival of the probe pulse, i.e. at negative waiting times (see Fig.~\ref{fig:Pulses and Signal}a). By definition, in the impulsive limit such a signal cannot arise from finite pulse overlap effects, but rather reflects the constraints imposed by the causality principle on the time-ordering structure of the third-order nonlinear response function. Consequently, these perturbed free induction decay (PFID) contributions do not introduce any distortion in the 2D spectra at positive waiting times~\cite{Mukamel1999,Chemla2001, Leo1990,Hamm1995,Joffre1988, Yan_2011,Mondal2018, Kuehn2010, Hamm2000,Brosseau2023, Paleek2019, Richter2017, Nguyen2019, Lloyd2021}.

Column (b) of Fig.~\ref{fig:AO} illustrates the corresponding 2D spectra for the anharmonic oscillator when excited with  Gaussian transform-limited pulses (TL). Compared to the IL case, the impact of finite-width pulses is immediately clear, manifesting as visible distortions in the spectral features. This effect persists throughout the dynamics of the system, arising from the fact that the 2D signal is a convolution between the finite-width pulses and the material response~\cite{Mukamel1999}. Moreover, unlike the ideal IL scenario, these pulses cannot excite all frequencies with equal amplitude. This stems from the intrinsic limitations imposed by a finite pulse width and non-uniform spectral coverage, which effectively act as a cut-off on the spectral features~\cite{GallagherFaeder1999}. During the early stages of the system's evolution, we also observe the formation of two additional features adjacent to the main signal, arising from the temporal overlap of finite-width pulses, as it enables additional pulse-ordering pathways~\cite{Anda2021, Leng2017}. Since they are tied to the pulse overlap, these features decay very rapidly during the dynamics. These observations demonstrate that even for ideal transform-limited pulses, a finite width is sufficient to induce additional spectral features in the 2D response.

Column (c) of Fig.~\ref{fig:AO} presents the 2D spectra of the anharmonic oscillator excited by chirped Gaussian (CH) pulses. This analysis extends previous investigations conducted on systems such as a two-level model~\cite{Tekavec2010} and atomic rubidium vapor~\cite{Binz2020}. The spectral chirp affects the pulses in two main ways. First, the initial TL Gaussian pulse with FWHM $\approx 75$ fs is broadened to $\approx 84$ fs due to the positive $500$ {fs}$^2$ chirp. This enhances the temporal overlap between pulses and, subsequently, the pulse-ordering effects during the early stages of the dynamics. As a result, at $T=0$  fs, the features adjacent to the main signal are more pronounced than in TL case. This observation reinforces our interpretation that such additional effects are caused solely by the finite pulse duration: the longer the pulses, the more significantly these features affect the dynamics. Second, the non-flat spectral phase visibly alters the 2D peak profiles compared to the IL case, as consequence of the temporal frequency ordering within the chirped pulses. Unlike the spectral features originating solely from finite pulse effects and rapidly vanishing for $T$ greater than the pulse width, the distortions induced by a non-flat spectral phase persist throughout the time evolution of the system, even beyond the nominal pulse duration. This reinforces the interpretation that this effect is primarily linked to the ordering of frequencies within our pulses.

By comparing the IL, TL, and CH cases, we can effectively disentangle the contributions due to finite pulse width from those arising from nontrivial spectral phase, providing a comprehensive understanding of how different pulse-shape features can influence the resulting 2D spectra.  Going one step further, we explore more exotic pulse shapes in columns (d) and (e). While in the first three columns we investigated the effect of smooth, single-peaked pulses, here we aim to isolate the effect of the pulse shape itself by considering pulses with two and three distinct peaks, respectively. The spectral phase of such pulses has been set to be flat in order to capture the effects due to the pulse shape alone, excluding any other possible source of distortion. In our simulations, we employed one double-peaked pulse with FWHM $\approx 57$ fs and a triple-peaked pulse with FWHM $\approx 80$ fs.

In column (d), the double-peaked pulse spectrum, tailored so that one of its sharp features matches a system transition, shifts the positions of the maxima and localizes the lineshapes, as it acts as a cutoff on the 2D response. This clearly suggests that peak positions and profiles of 2D spectral features can be influenced by the spectral shape of the pulses, paving the way for potential 2D spectroscopy-based quantum control schemes.

Column (e) shows how the use of a more complex multi-peaked pulse results in the appearance of several additional spots in the 2D spectrum. Since these features do not have a counterpart in the IL case, they can be clearly identified as effects produced by the specific pulse shape. This serves as a cautionary reminder that interpreting 2D spectral features requires strict attention to the underlying spectral structure of the pulses in order to prevent misleading conclusions.

Throughout this section we have primarily discussed the anharmonic oscillator, as it represents a simple system which allows for an immediate and transparent interpretation. Nevertheless, we observe the same qualitative behavior in both the coupled two-exciton and the dimer system (Fig.~\ref{fig:EX} and Fig.~\ref{fig:DI} in App.~\ref{app:pulse_shape}).  The consistency of these results across different model systems further strengthens the robustness and general validity of the discussion presented above.

\onecolumngrid

\twocolumngrid

\onecolumngrid

\begin{SCfigure}[0.33][htbp]
    \centering
    \includegraphics[width=5.1in,height=3in]{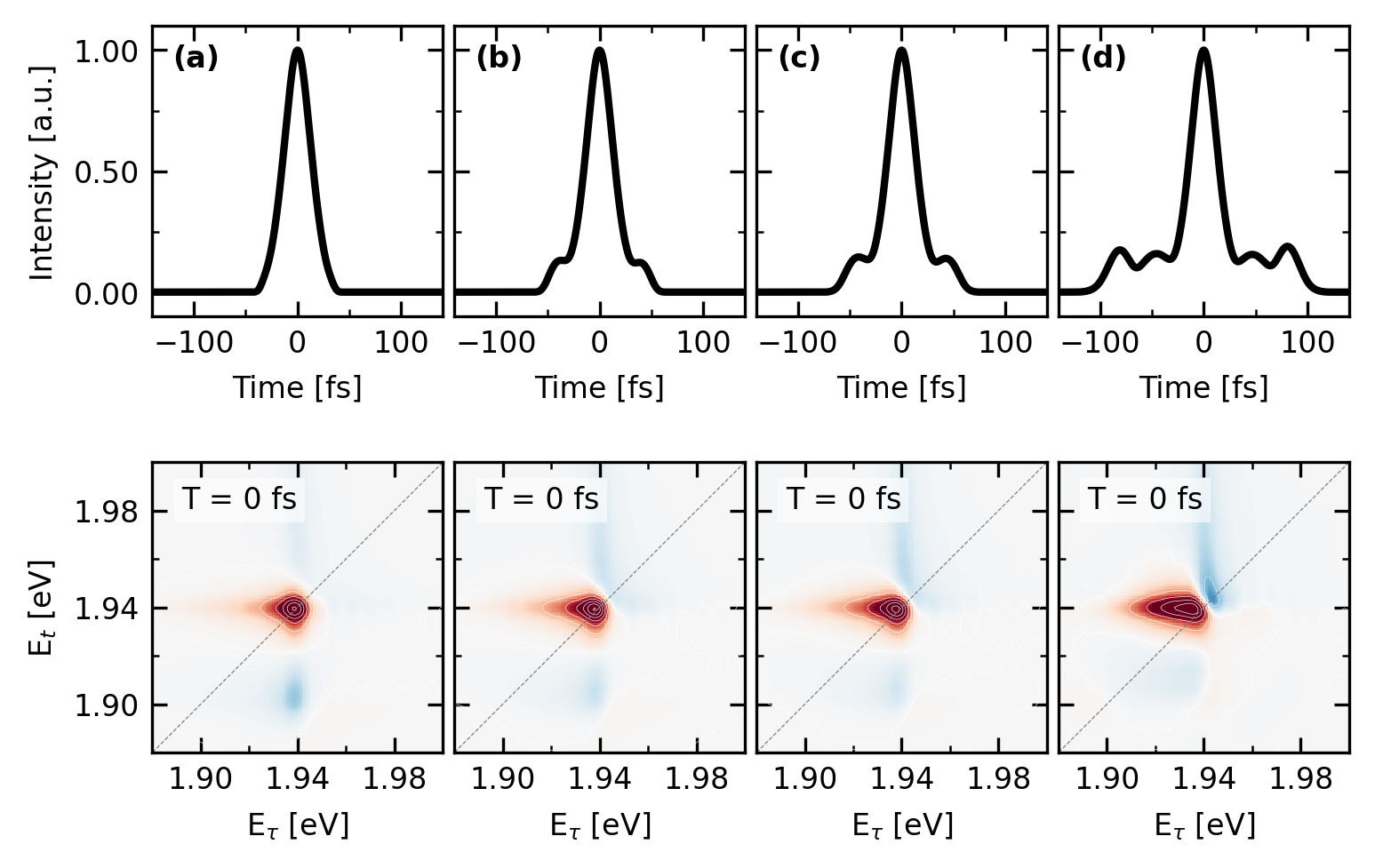}
    \caption{Influence of long temporal tails of pulses on 2D spectra of an anharmonic oscillator system. The first row shows the temporal profiles of simulated fiber pulses, to which a super-Gaussian mask with progressively increasing width (from left to right) has been applied. The panels below display the corresponding 2D spectra calculated for each masked pulse.}
    \label{fig:Tail_Effect}
\end{SCfigure}

\twocolumngrid

\subsection{Realistic pulses and tail effects}
\label{subsec:Realistic pulses and tail effects}

In this subsection, we go beyond ideal pulse shapes to consider more realistic profiles that closely resemble those actually employed in experiments. Specifically, we present 2D spectra computed using simulated pulses~\cite{Palato2020, Bjot2010} emerging from a 2-bar, argon-filled hollow-core fiber, driven by a 100-$\mu$J input femtosecond field centered around 800 nm. These simulated fields have been shown to closely match experimentally measured spectra~\cite{Palato2020, Seiler2017}. We then performed a series of manipulations to mimic standard experimental procedures on fiber-generated broadband pulses. First, the pulse spectrum was low-pass filtered below 800 nm. Subsequently, its spectral phase was flattened, closely replicating experimental dispersion-compensation procedures~\cite{Szipocs1994, Zavelani-Rossi2001, Bor1985, Akturk2006, Treacy1969, Gibson2006, Weiner2000}. The resulting pulses exhibit a characteristic temporal profile consisting of a narrow central peak accompanied by low-amplitude, extended tails. The central idea of this section is to investigate how these tails, routinely observed in experiments, affect the resulting 2D spectra. This is illustrated in Fig.~\ref{fig:Tail_Effect} for the 2D spectrum of the anharmonic oscillator system. In the first row, the temporal profile of a simulated realistic pulse is presented after applying a super-Gaussian mask, selectively suppressing part of its long tails. From left to right, the width of the mask is increased, making the temporal tails progressively more pronounced. In this way, we can systematically examine how the presence of such long temporal tails influences the resulting 2D spectra, which are reported in the second row of Fig.~\ref{fig:Tail_Effect}. A clear trend emerges: as the temporal tails become more prominent, a progressive redistribution of the spectral signal is observed in the 2D spectra of the anharmonic oscillator system. In particular, the blue peak in column (a) gradually decreases in intensity, while an additional peak emerges. Since it has no counterpart in the IL case (see first column in Fig.~\ref{fig:AO}), it represents an additional spectral feature purely arising from pulse tails.

\begin{figure}[htbp]
    \centering
    \includegraphics[width=3.6in,height=5.4in]{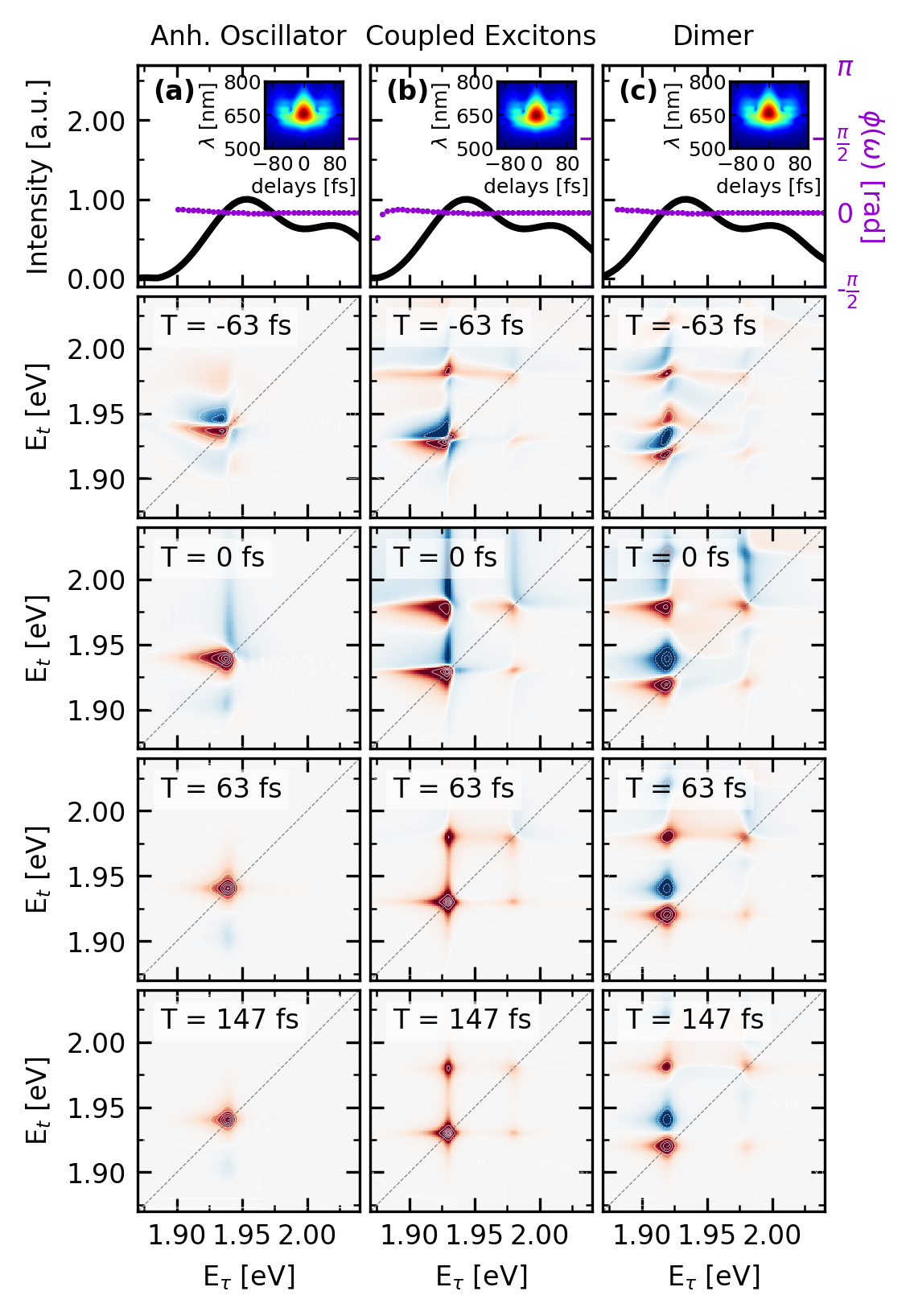}
    \caption{Effect of long temporal tails on the 2D spectra of the three model systems. Each column corresponds to a different system: (a) an anharmonic oscillator, (b) two coupled excitons, and (c) a dimer. The top panel in each column shows the spectral amplitude and phase of the simulated fiber pulse used in the calculation (purple line), together with the corresponding FROG trace (inset). The panels below display the dynamics of the simulated 2D spectra. The color scale is normalized column-wise for each column.}
    \label{fig:Fiber}
\end{figure}

Finally, Fig.~\ref{fig:Fiber} shows the 2D spectra obtained with the temporally and spectrally shaped pulses described above for the three exemplary systems. Even though the pulse remains short in its central peak, the presence of long temporal tails produces significant effects at short waiting times. These effects arise clearly from pulse overlap and time-ordering since they rapidly decay during the dynamics (compare to Fig.~\ref{fig:AO}, Fig.~\ref{fig:EX} and Fig.~\ref{fig:DI}). In Fig.~\ref{fig:Fiber} we employed pulses centered around 1.96 eV  (for anharmonic oscillator and coupled two-excitons systems) and 1.95 eV (for dimer system) with $FWHM \approx 31$ fs . All the other details are reported in App.~\ref{app:pulse_flat}. 
Overall, this study highlights the importance of minimizing tails in pulses employed for 2D spectroscopy in order to suppress pulse effects.

\subsection{Impact of pulse effects on dynamics}
\begin{figure}[H]
    \centering
    \includegraphics[width=3.37in,height=5in]{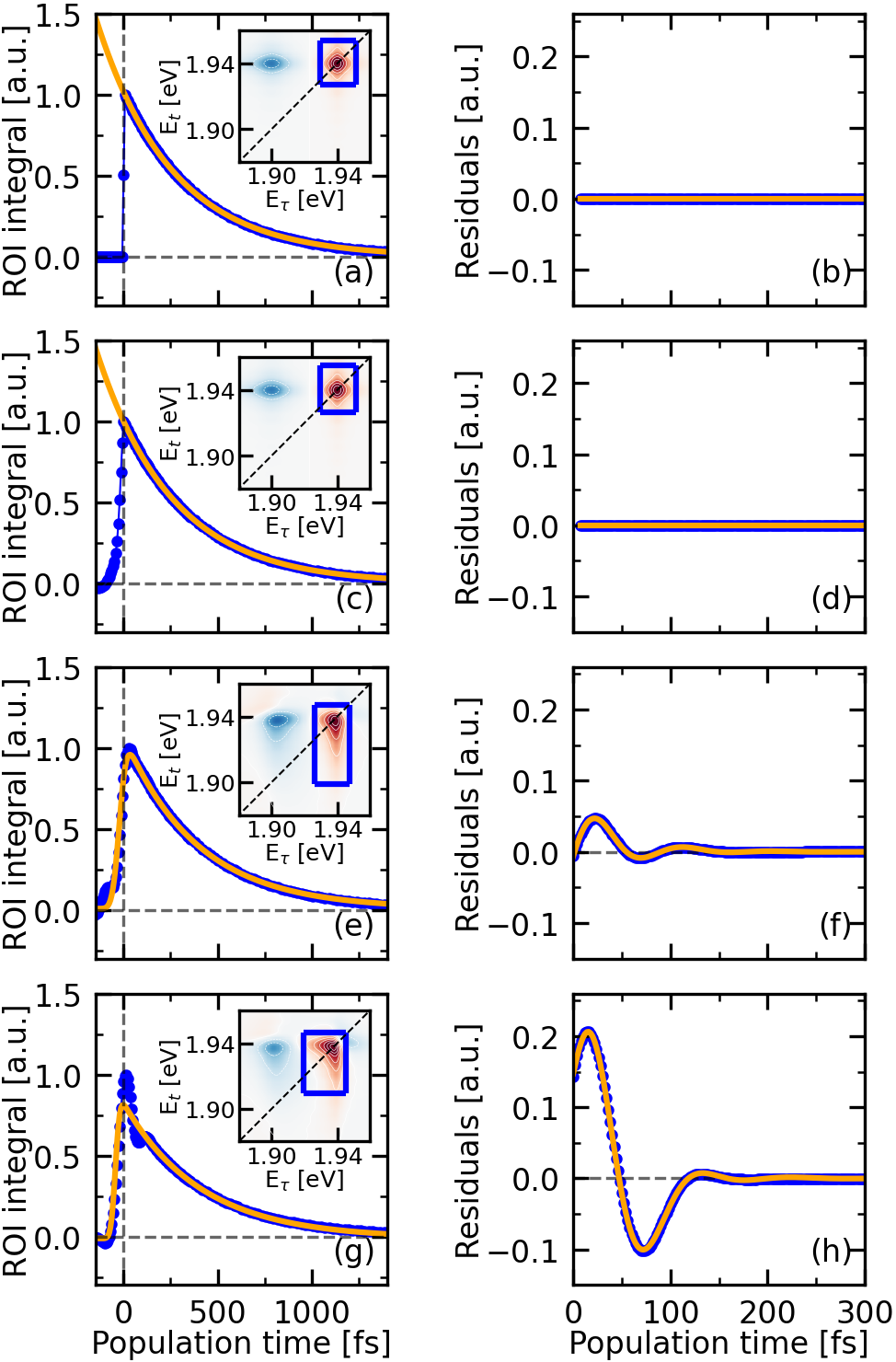}
    \caption{Time evolution of the integrated signal over a selected region of interest (ROI) corresponding to the diagonal (red) peak of the anharmonic oscillator 2D spectrum (blue rectangle in inset figures). The blue dotted line shows the simulated population dynamics, while the superimposed yellow solid line corresponds to the fit. Panels (a, c, e, g) show the dynamics of the ROI signal for increasing levels of model complexity: (a) impulsive limit without time-ordering effects (IL-noTO), (c) impulsive limit with time-ordering (IL-TO), (e) transform-limited pulses with finite duration (TL-TO), (g) and chirped pulses with quadratic spectral phase (CHIRP). Panels (b, d) show the corresponding residuals obtained by subtracting a single-exponential fit from the long-time dynamics, while panels (f, h) report the corresponding residuals obtained after subtraction of an exponentially decaying function convolved with a Gaussian response. The insets show snapshots of the 2D spectra at $T = 0$ fs.}
    \label{fig:ROI_AO}
\end{figure}

To assess how realistic pulse profiles affect the system's dynamics, we study the time evolution of selected regions of interest (ROIs) in the 2D spectra of two previously discussed systems: the anharmonic oscillator (Fig.~\ref{fig:Quantum Systems}(a)) and  two coupled-excitons (Fig.~\ref{fig:Quantum Systems}(b)). Fig.~\ref{fig:ROI_AO} shows the analysis for the anharmonic oscillator. The investigated ROIs are highlighted by the blue rectangles in the inset 2D spectra. These ROIs select the diagonal absorptive peak of the spectrum and therefore directly probe the waiting time dynamics of the excited  population. Panels (a, c, e, g) display the integrated ROI signal as a function of waiting time \( T \) (blue dotted lines) together with superimposed fits (yellow solid lines), obtained from a model consisting of either a single decaying exponential (impulsive-limit cases) or a Gaussian convolved with a decaying exponential (finite-pulse cases). As before, we progressively increase the level of complexity, in order to isolate their individual impact on the dynamics.

Panel (a) reports the simplest case: an anharmonic oscillator in the impulsive limit, neglecting time-ordering effects between the excitation pulses (IL-noTO). In this regime, the measured signal directly reflects the intrinsic response of the system. As expected, the signal exactly vanishes for negative waiting times, where no physical evolution of the system can occur. For positive \( T \), the ROI signal exhibits a purely exponential decay, corresponding to population relaxation dynamics within the Markovian approximation framework. By fitting this decay with a single-exponential function, we extract a characteristic decay constant, which perfectly matches the population lifetime imposed in the simulation (compare App.~\ref{app:model_systems} for the spectrum, App.~\ref{app:pulse_flat} for the lifetimes and all the fitted parameters). The corresponding residuals are shown in panel (b), defined as the difference between the original signal and the exponential decay fit. As expected, the residuals are flat, consistent with the model, where no coupling mechanisms (vibrational or electronic) are possible.

In panel (c), we include time-ordering effects, while still remaining in the impulsive limit (IL-TO). In agreement with response function theory, these additional effects manifest solely at negative waiting times. Indeed, for \( T > 0 \), the signal reflects only the intrinsic dynamics of the system, unaffected by pulse-overlap or time-ordering contributions. As a result, the ROI dynamics for positive waiting times retains its purely exponential character. Consistently, the residuals shown in panel (d) remain flat. 

In panel (e), we consider excitation pulses with a finite temporal duration, while retaining a flat spectral phase. This corresponds to the transform-limited case, now including also time-ordering (TL-TO regime). Because the excitation field now has a finite pulse width, the dynamics as a function of the waiting time is not expected to follow a pure exponential decay, even in the TL case. Instead, the measured signal reflects the convolution of the Gaussian pulse envelope with the exponentially decaying system response. Fitting the curves in panel (e,g) using a Gaussian-convolved exponential model, allows us to recover the underlying exponentially decaying population dynamics, yielding a decay constant in excellent agreement with the expected value (see App.~\ref{app:pulse_flat} and the first column Table~\ref{tab:ROI_AO}). 

The residuals corresponding to this case are reported in panel (f). In contrast to the impulsive-limit scenarios discussed above, the residuals now exhibit oscillatory behavior at short waiting times, which is progressively damped as \( T \) increases. At first glance, such oscillations might be interpreted as signatures of coherent dynamics, typically associated with quantum superpositions between different excited states (see, e.g., the work of Cassette \textit{et al.}\cite{Cassette2015}). However, this interpretation must be carefully reconsidered in the present context. Such quantum superpositions do not exist for the model system considered here, consisting of a simple anharmonic oscillator ~\cite{Salzmann2008}. As a consequence, no coherent superpositions between excited states can be generated during the dynamics. The origin of the observed oscillations must instead be traced back to the finite temporal duration of the excitation pulses. When departing from the impulsive limit, due to the finite pulse duration, the temporal overlap between the pulses becomes non-negligible. As a consequence, time-ordering effects are no longer confined to negative waiting times and they are able to extend into the \( T > 0 \) region. In this regime, different light-matter interaction pathways can interfere, leading to oscillating terms at different transition frequencies. Their mixing generates beat-like modulations at the difference frequency ${\omega}_{01}-{\omega}_{12}$, which appear as rapidly damped oscillations confined to short positive waiting times. 

In our case, to quantitatively characterize this behavior, the residuals were fitted using a model consisting of a sum of damped oscillatory functions. A single damped oscillator was found to be sufficient to accurately reproduce the data in this case. From the Fourier analysis of the short-time oscillatory components we find a frequency of approximately 0.04 eV, which coincides with the anharmonicity of the system, i.e. with the difference between the two optical transition energies ${\omega}_{01}=1.94$ eV and ${\omega}_{12}=1.90$ eV. The extracted damping time $\tau$ is on the order of $\sim 33$ fs, which matches the duration of the excitation pulses (approximately $\sim 32$ fs). This suggests again that the decay of the oscillatory features is governed by the temporal envelope of the driving field, rather than by intrinsic relaxation or dephasing processes of the system.

Finally, we consider the case in which the excitation pulses have both a finite temporal duration and a non-trivial spectral phase. In particular, a quadratic spectral phase corresponding to a chirp of +500 fs$^2$ is introduced. The presence of chirp further modifies the temporal structure of the pulses, effectively stretching them in time and introducing a frequency–time correlation within the pulse envelope. As in the previous cases, we can extract the intrinsic population relaxation time of the dynamics, which is found to be in excellent agreement with the expected value  (see Table~\ref{tab:ROI_AO}). The residuals for this case are shown in panel (h). Similarly to the transform-limited case, clear oscillatory features are observed at short $T$. As before, these oscillations originate purely from time-ordering effects induced by the finite duration and chirped nature of the pulses, and do not reflect genuine coherent dynamics of the system. The introduction of chirp enhances these effects by increasing the temporal overlap, leading to a further extension of time-ordering effects into the region of positive $T$. The corresponding residuals are well reproduced by a model consisting of a sum of two damped harmonic oscillators, which parameters are reported in Table~\ref{tab:ROI_AO}. The extracted frequencies and decay times are comparable to those obtained in the transform-limited case, and therefore share the same interpretation.

In addition, the same analysis was performed for the blue feature in the 2D spectrum of the anharmonic oscillator system (see inset of Fig.~\ref{fig:ROI_AO}a). The study is reported in Fig.~\ref{fig:ROI_AO_C}, while the parameters extracted by the fit are summarized in  App.~\ref{app:ROI} in Table~\ref{tab:ROI_AO_C}. The same qualitative behavior is observed as for the red feature in all investigated regimes. These results highlight an important conceptual point: even in a system that does not support coherent dynamics as a function of $T$, the presence of finite-duration pulses can generate signals that display coherent oscillations. Therefore, particular care must be taken when interpreting oscillatory features in 2D spectroscopy experiments, as they may arise from the interplay between the pulse characteristics and the multi-level structure of the system, rather than from genuine quantum coherences.

\begin{figure}[H]
    \centering
    \includegraphics[width=1\columnwidth]{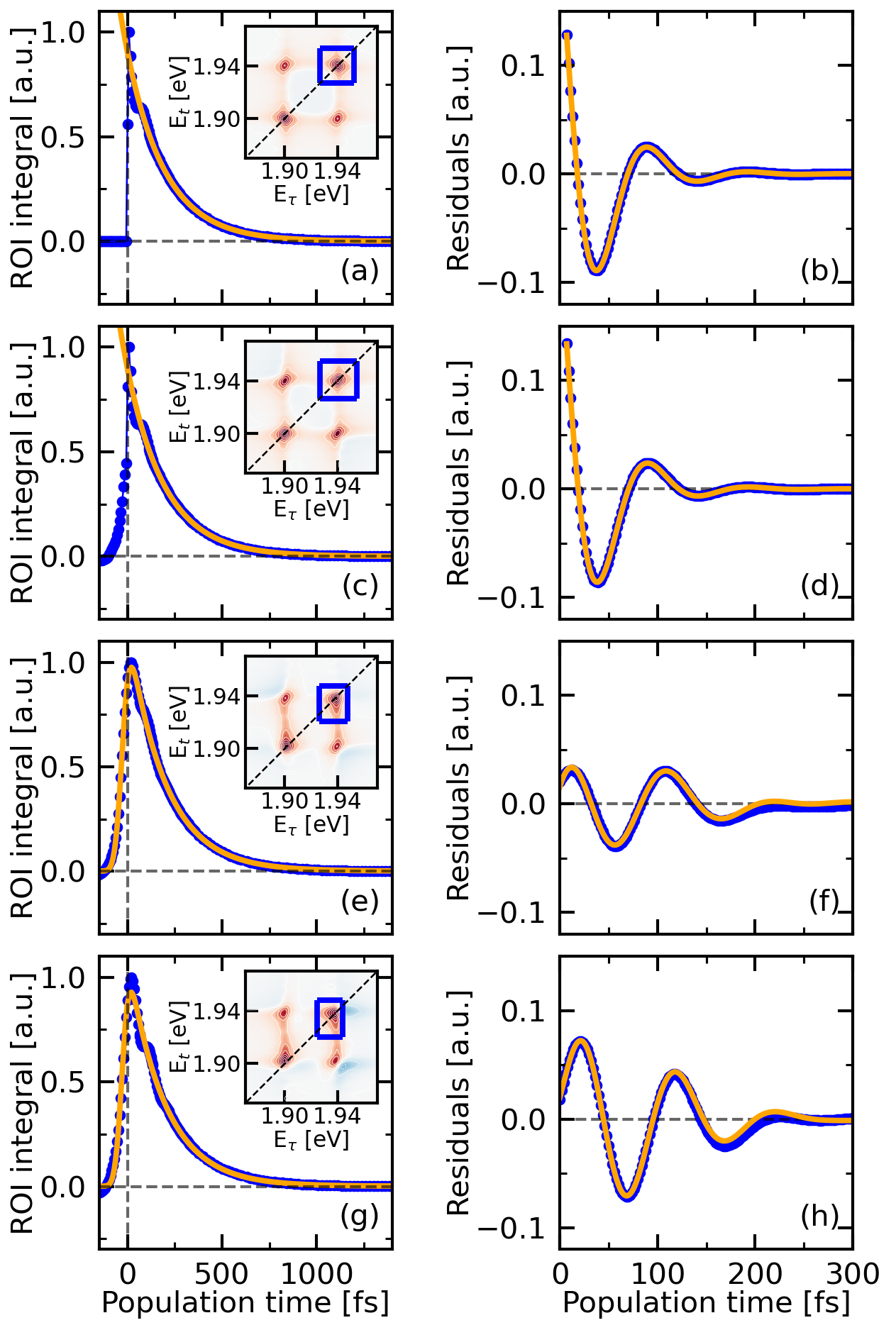}
    \caption{Time evolution of the integrated signal over a selected region of interest (ROI) corresponding to the diagonal (red) peak of a two coupled excitons 2D spectrum (blue rectangle in inset figures). The blue dotted line shows the simulated population dynamics, while the superimposed yellow solid line corresponds to the fit. Panels (a, c, e, g) show the dynamics of the ROI signal for increasing levels of model complexity: (a) impulsive limit without time-ordering effects (IL-noTO), (c) impulsive limit with time-ordering (IL-TO), (e) transform-limited pulses with finite duration (TL-TO), and (g) chirped pulses with quadratic spectral phase (CHIRP). Panels (b, d) show the corresponding residuals obtained by subtracting a single-exponential fit from the long-time dynamics, while panels (f, h) report the corresponding residuals obtained after subtraction of an exponentially decaying function convolved with a Gaussian response. The insets show snapshots of the 2D spectra at $T = 0$ fs.}
    \label{fig:ROI_EX}
\end{figure}

We extended our ROI analysis to a pair of coupled excitons in Fig.~\ref{fig:ROI_EX}. In this case, quantum coherences can be expected, as reported in previous experiments~\cite{Cassette2015}. Fig.~\ref{fig:ROI_EX} (a) corresponds to the impulsive limit in the absence of time-ordering. Even though the signal vanishes for negative waiting times, the dynamics at positive $T$ is no longer pure exponential decay. Instead, clear oscillatory features emerge at early times, suggesting a beating dynamics of the system due to the coupling between the two excitons. This interpretation is confirmed in panel (b), where the residuals are analyzed. In this case, the oscillations are accurately described by a single damped harmonic term, which frequency matches the energy splitting between the two excitonic levels ($\sim$0.040 eV), as reported in Table~\ref{tab:ROI_EX}. This provides clear evidence that the observed oscillations originate from couplings between the two excitons. The same qualitative behavior is observed in the subsequent panels, where additional physical effects are progressively included. 

Even in the presence of time-ordering, finite pulse duration, and chirp, the dominant oscillatory component at the excitonic splitting can be clearly distinguished. In the transform-limited and chirped pulse regimes, the fitting model used to describe the dynamics consist of a Gaussian function convolved with a decaying exponential in order to extract the oscillatory components with the smallest possible residual background contribution. 

The Fourier analysis of the extracted residuals revealed two damped oscillators modes. However, the extra oscillating contribution do not correspond to any intrinsic energy scale of the system and therefore they should be regarded as pulse-overlap and time-ordering effects. 

For completeness, we also analyzed the cross peak of the coupled exciton system of  Fig.~\ref{fig:ROI_EX_C} in App.~\ref{app:ROI}. The resulting behavior is consistent with the scenario discussed for the diagonal peak. In particular, the oscillatory component associated with the excitonic energy splitting is clearly preserved across all configurations, confirming its intrinsic origin. At the same time, additional short-lived oscillatory features appear in the presence of finite pulse duration and time-ordering effects, which can be attributed to pulse-overlap effects. The parameters extracted from the fit are reported in Table~\ref{tab:ROI_EX_C} in App.~\ref{app:ROI}.

\begin{figure}[htbp]
    \centering
    \includegraphics[width=1\columnwidth]{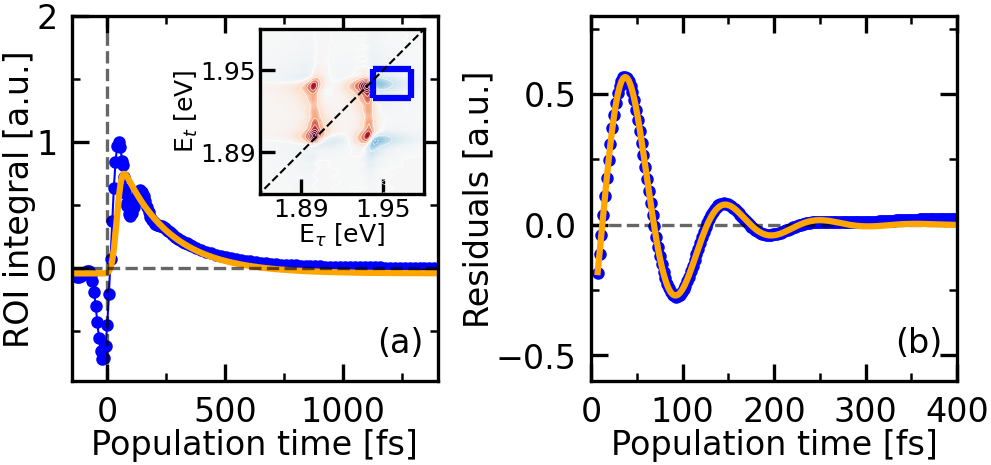}
    \caption{Dynamics of a pulse-induced spectral feature in the two coupled excitons system. The blue dotted line shows the simulated population dynamics, while the superimposed yellow solid line corresponds to the fit. (a) Integrated signal over the ROI indicated in the inset as the blue rectangle. (b) Residuals of the signal in (a), obtained after subtraction of an exponentially decaying function convolved with a Gaussian response. The insets show snapshots of the 2D spectra at $T = 0$ fs.}
    \label{fig:ROI_EX_A}
\end{figure}

Fig.~\ref{fig:ROI_EX_A} extends the ROI analysis to a pulse-induced feature appearing in the chirped case for the anharmonic oscillator system. In panel (a), the extracted decay time (see Table~\ref{tab:ROI_EX_A}) is in excellent agreement with the value reported in App.~\ref{app:model_systems}, showing that even for such a feature the population dynamics has the same decaying behavior that was previously observed for both the diagonal and cross peaks. This is confirmed by the corresponding residual analysis reported in panel (b). In particular, oscillatory components are present with comparable frequencies and decay times. This observation is of crucial importance: it indicates that, based solely on ROI dynamics, it is not always possible to unambiguously discriminate between pulse induced spectral features and those reflecting the intrinsic dynamics of the system. This is further reinforced by the fact that the energy of the extracted oscillation frequency ($\sim$0.040 eV) coincides with characteristic energy scales of the system, which could be interpreted as a signature of coherent dynamics. As discussed in the previous section, such effects can become particularly pronounced when the excitation pulses exhibit a finite temporal duration with long tails, due to significant pulse overlap and enhanced time-ordering effects, that may even dominate the short-time dynamics.

\subsection{Homodyne Detection}
\label{sec:Homodyne Detection}

\begin{figure}[H]
    \centering
    \includegraphics[width=0.62\columnwidth]{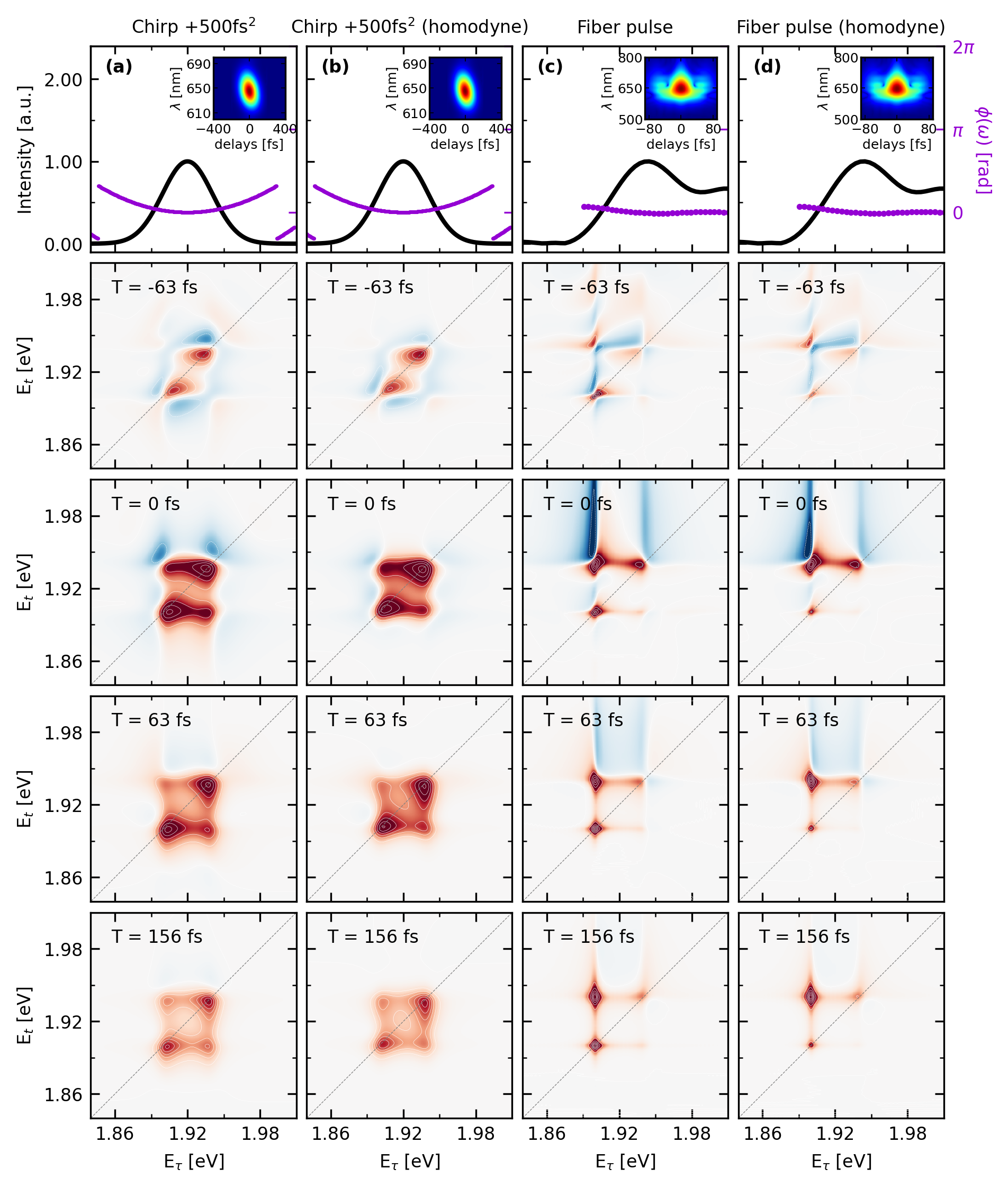}
    \caption{Comparison between the simulated pure 2D signal and the corresponding homodyne-detected response for a system of two coupled excitons. The columns report results obtained using different excitation conditions: a Gaussian pulse with a chirp of +500 fs$^2$ (columns a–b) and a broadband pulse generated from hollow-core fiber simulations (columns c–d). For each case, the left column shows the pure 2D signal, while the right column includes the effect of homodyne detection, where the probe field acts as a local oscillator. The rows correspond to increasing waiting time T, as indicated in each panel. The color scale is normalized column-wise for each column.}
    \label{fig:homodyne}
\end{figure}

Finally, we study the role of the local oscillator on selected pulse induced spectral features. To this end, we consider a pump-probe geometry in which the local oscillator is the probe field itself, corresponding to a homodyne detection scheme~\cite{Hamm2011}.  Thus, the measured signal arises from the interference between the probe field and the emitted 2D signal, see Sec.~\ref{sec:Signal Measurement}. Fig.~\ref{fig:homodyne} presents a comparison highlighting the differences between simulations that include homodyne detection versus only the 2D signal for two coupled excitons. We consider two types of excitation pulses: (i) a Gaussian pulse with a chirp of +500 fs$^2$, and (ii) a broadband pulse obtained from simulations of a hollow-core fiber output (FWHM $\approx  31$ fs). Both cases are investigated in the absence and in the presence of the local oscillator. The results are shown in columns (a) and (b) for the Gaussian pulse, and in columns (c) and (d) for the fiber-generated pulse.

In both cases, homodyne detection partially suppresses finite pulse effects at early times. This can be understood in terms of the interference between the emitted 2D signal and the local oscillator field. In the homodyne scheme, the detected signal is proportional to the cross term between the probe field and the nonlinear signal, which effectively acts as a phase-sensitive filter. It is important to note that the signal obtained through homodyne detection corresponds to a specific phase projection of the emitted 2D response. The apparent reduction of pulse effects in the 2D spectrum arises because those contributions are less phase-correlated with the probe field and  therefore suppressed by interference. As a result, contributions that are not in phase with the local oscillator, such as non-resonant pathways, are strongly reduced. The values of the lifetimes considered in this study are reported in App.~\ref{app:param_III}.

\onecolumngrid

\twocolumngrid

\section{Conclusions}
\label{sec:Conclusions}

We have introduced an approach for 2D spectroscopy simulations of arbitrary pulse shapes, including overlapping pulses and time-ordering effects. We have shown that the computational cost of our approach scales linearly with the number of sampling points in the Markovian limit for the environment. This efficient framework has enabled us to systematically investigate realistic pulse effects on 2D spectra. We have considered a range of scenarios with increasing pulse complexity and their impact on three representative model systems: an anharmonic oscillator, a pair of coupled excitons, and a six-level dimer, revealing qualitative similarities across all three systems behaviors.

We have started with the impulsive-limit (IL) case to benchmark our approach. For positive waiting times \( ( T > 0 ) \), we have retrieved the dynamics of the system response functions. For negative waiting times \( ( T < 0 ) \), signals arising from time-ordering effects were observed, in agreement with response function theory. These perturbed free-induction decay signals were consistent with those reported in previous studies~\cite{Chemla2001, Leo1990,Hamm1995,Joffre1988, Yan_2011,Mondal2018,Kuehn2010, Hamm2000, Brosseau2023, Paleek2019, Richter2017, Nguyen2019, Lloyd2021}. Following the IL case, we have introduced transform-limited pulses of finite duration (TL case). This has led to the appearance of additional spectral features, showing how temporal pulse overlap alone can generate new contributions to the 2D spectra.  Then, we have considered chirped pulses (CH case), i.e. pulses with a non-flat spectral phase, building-up on previous works~\cite{Tekavec2010, Binz2020}. We have identified two main effects. On the one hand, chirp increases the pulse duration, thereby enhancing the formation of additional spectral features at early positive waiting times. On the other hand, the temporal ordering of frequencies within the pulse gives rise to persistent distortions of the spectral lineshapes.

Next, we have investigated the influence of pulse shapes on 2D spectra by designing specific pulse profiles with flat spectral phase, thereby isolating the effect of pulse shape itself. We have observed the emergence of two additional pulse shape effects: unintended shape-induced spectral features and shifts of the 2D spectra peak positions. In particular, we have employed broadband fiber pulses with weak but long temporal tails. We have shown that, in general, 2D spectra obtained using short pulses from a FWHM point of view are not necessarily exempted from pulse shape effects, since long temporal tails can still give rise to significant distortions~\cite{Paleek2019}. Since the IL 2D spectrum is not accessible in experimental conditions, it can be challenging to disentangle intrinsic system features from pulse-induced effects in the measured data. This highlights the importance of carefully characterizing the pulses employed in 2D spectroscopy experiments.

In the second part of the results section, we have focused on the dynamics of selected regions of interest (ROIs) in the 2D spectra of the anharmonic oscillator and the two-excitons systems. Following the same strategy adopted previously, we have considered a sequence of scenarios with increasing pulse complexity: impulsive limit without time-ordering effects, impulsive limit including time ordering, transform-limited case, and finally chirped pulses. We have examined the residuals of the extracted dynamics to identify possible oscillatory components and we have confirmed the expected exponential decay behavior. We have found that oscillations can occur in the dynamics of the spectral features during pulse overlap, even in systems where coherences between excited states are not expected from selection rules. This behavior have been observed in the case of the anharmonic oscillator. In contrast, for the two-exciton system, the analysis of the residuals has revealed oscillations that correspond exactly to the energy difference between the two allowed transitions in the system, as expected from the model.

Our ROI analysis highlights the appearance of pulse-induced oscillations in the dynamics of 2D spectra, in particular when using chirped pulses. In theoretical simulations, the availability of the impulsive, no-time-ordering limit as well as prior knowledge of system parameters provide valuable references that allows one to clearly identify pulse effects. In contrast, in experimental conditions one only has access to finite-duration pulses, which are never perfectly transform-limited due to residual spectral phase distortions. As a consequence, the raw experimental data may contain oscillatory features which origin is not immediately clear. Therefore, careful modeling and systematic comparison with realistic simulations, such as those presented here, are essential to correctly interpret those features. 

Finally, we have simulated the experimentally-measured signal obtained via a homodyne detection scheme in a pump-probe geometry. We have performed this analysis both for a chirped Gaussian pulse and for a simulated broadband fiber spectrum with a flat spectral phase. In both cases, we have observed a reduction of pulse effects in the 2D spectra with respect to the non-homodyning TL case .

Overall, our work emphasizes that a reliable interpretation of 2D spectroscopy data requires a careful choice and characterization of the employed pulses. The simulation framework introduced here can be used for systematic investigations of realistic pulse effects as existing in experiments. More generally, our approach can be exploited to explore pulse shaping not only to probe the system but also to steer its nonlinear response, opening promising perspectives for implementing quantum control strategies within the framework of 2D spectroscopy.

\acknowledgments{M.R and H.S. acknowledge funding by the Deutsche Forschungsgemeinschaft (DFG, German Research Foundation) under project ID: 531215165 (Research Unit “OPTIMAL”, Project P4) and project ID: 555467911 (CRC 1772 on Heterostructures of Molecules and Two-Dimensional Materials, Project A02). The authors acknowledge David G. Y. Koch for providing the simulated electric fields from the hollow-core fiber.}

\section*{Data Availability Statement}

Upon final publication of the study, the data used in this manuscript will be posted in an online repository.

\bibliography{aipsamp}

\begin{thebibliography}{77}%
\makeatletter
\providecommand \@ifxundefined [1]{%
 \@ifx{#1\undefined}
}%
\providecommand \@ifnum [1]{%
 \ifnum #1\expandafter \@firstoftwo
 \else \expandafter \@secondoftwo
 \fi
}%
\providecommand \@ifx [1]{%
 \ifx #1\expandafter \@firstoftwo
 \else \expandafter \@secondoftwo
 \fi
}%
\providecommand \natexlab [1]{#1}%
\providecommand \enquote  [1]{``#1''}%
\providecommand \bibnamefont  [1]{#1}%
\providecommand \bibfnamefont [1]{#1}%
\providecommand \citenamefont [1]{#1}%
\providecommand \href@noop [0]{\@secondoftwo}%
\providecommand \href [0]{\begingroup \@sanitize@url \@href}%
\providecommand \@href[1]{\@@startlink{#1}\@@href}%
\providecommand \@@href[1]{\endgroup#1\@@endlink}%
\providecommand \@sanitize@url [0]{\catcode `\\12\catcode `\$12\catcode
  `\&12\catcode `\#12\catcode `\^12\catcode `\_12\catcode `\%12\relax}%
\providecommand \@@startlink[1]{}%
\providecommand \@@endlink[0]{}%
\providecommand \url  [0]{\begingroup\@sanitize@url \@url }%
\providecommand \@url [1]{\endgroup\@href {#1}{\urlprefix }}%
\providecommand \urlprefix  [0]{URL }%
\providecommand \Eprint [0]{\href }%
\providecommand \doibase [0]{https://doi.org/}%
\providecommand \selectlanguage [0]{\@gobble}%
\providecommand \bibinfo  [0]{\@secondoftwo}%
\providecommand \bibfield  [0]{\@secondoftwo}%
\providecommand \translation [1]{[#1]}%
\providecommand \BibitemOpen [0]{}%
\providecommand \bibitemStop [0]{}%
\providecommand \bibitemNoStop [0]{.\EOS\space}%
\providecommand \EOS [0]{\spacefactor3000\relax}%
\providecommand \BibitemShut  [1]{\csname bibitem#1\endcsname}%
\let\auto@bib@innerbib\@empty
\bibitem [{\citenamefont {Hamm}\ and\ \citenamefont {Zanni}(2011)}]{Hamm2011}%
  \BibitemOpen
  \bibfield  {author} {\bibinfo {author} {\bibfnamefont {P.}~\bibnamefont
  {Hamm}}\ and\ \bibinfo {author} {\bibfnamefont {M.}~\bibnamefont {Zanni}},\
  }\href {https://doi.org/10.1017/cbo9780511675935} {\emph {\bibinfo {title}
  {Concepts and Methods of 2D Infrared Spectroscopy}}}\ (\bibinfo  {publisher}
  {Cambridge University Press},\ \bibinfo {year} {2011})\BibitemShut {NoStop}%
\bibitem [{\citenamefont {Biswas}\ \emph {et~al.}(2022)\citenamefont {Biswas},
  \citenamefont {Kim}, \citenamefont {Zhang},\ and\ \citenamefont
  {Scholes}}]{Biswas2022}%
  \BibitemOpen
  \bibfield  {author} {\bibinfo {author} {\bibfnamefont {S.}~\bibnamefont
  {Biswas}}, \bibinfo {author} {\bibfnamefont {J.}~\bibnamefont {Kim}},
  \bibinfo {author} {\bibfnamefont {X.}~\bibnamefont {Zhang}},\ and\ \bibinfo
  {author} {\bibfnamefont {G.~D.}\ \bibnamefont {Scholes}},\ }\bibfield
  {title} {\enquote {\bibinfo {title} {Coherent two-dimensional and broadband
  electronic spectroscopies},}\ }\href
  {https://doi.org/10.1021/acs.chemrev.1c00623} {\bibfield  {journal} {\bibinfo
   {journal} {Chemical Reviews}\ }\textbf {\bibinfo {volume} {122}},\ \bibinfo
  {pages} {4257–4321} (\bibinfo {year} {2022})}\BibitemShut {NoStop}%
\bibitem [{\citenamefont {Li}\ \emph {et~al.}(2023)\citenamefont {Li},
  \citenamefont {Lomsadze}, \citenamefont {Moody}, \citenamefont {Smallwood},\
  and\ \citenamefont {Cundiff}}]{Li_2023}%
  \BibitemOpen
  \bibfield  {author} {\bibinfo {author} {\bibfnamefont {H.}~\bibnamefont
  {Li}}, \bibinfo {author} {\bibfnamefont {B.}~\bibnamefont {Lomsadze}},
  \bibinfo {author} {\bibfnamefont {G.}~\bibnamefont {Moody}}, \bibinfo
  {author} {\bibfnamefont {C.}~\bibnamefont {Smallwood}},\ and\ \bibinfo
  {author} {\bibfnamefont {S.}~\bibnamefont {Cundiff}},\ }\href
  {https://doi.org/10.1093/oso/9780192843869.001.0001} {\emph {\bibinfo {title}
  {Optical Multidimensional Coherent Spectroscopy}}}\ (\bibinfo  {publisher}
  {Oxford University PressOxford},\ \bibinfo {year} {2023})\BibitemShut
  {NoStop}%
\bibitem [{\citenamefont {Cao}\ \emph {et~al.}(2020)\citenamefont {Cao},
  \citenamefont {Cogdell}, \citenamefont {Coker}, \citenamefont {Duan},
  \citenamefont {Hauer}, \citenamefont {Kleinekath\"{o}fer}, \citenamefont
  {Jansen}, \citenamefont {Mančal}, \citenamefont {Miller}, \citenamefont
  {Ogilvie}, \citenamefont {Prokhorenko}, \citenamefont {Renger}, \citenamefont
  {Tan}, \citenamefont {Tempelaar}, \citenamefont {Thorwart}, \citenamefont
  {Thyrhaug}, \citenamefont {Westenhoff},\ and\ \citenamefont
  {Zigmantas}}]{Cao2020}%
  \BibitemOpen
  \bibfield  {author} {\bibinfo {author} {\bibfnamefont {J.}~\bibnamefont
  {Cao}}, \bibinfo {author} {\bibfnamefont {R.~J.}\ \bibnamefont {Cogdell}},
  \bibinfo {author} {\bibfnamefont {D.~F.}\ \bibnamefont {Coker}}, \bibinfo
  {author} {\bibfnamefont {H.-G.}\ \bibnamefont {Duan}}, \bibinfo {author}
  {\bibfnamefont {J.}~\bibnamefont {Hauer}}, \bibinfo {author} {\bibfnamefont
  {U.}~\bibnamefont {Kleinekath\"{o}fer}}, \bibinfo {author} {\bibfnamefont
  {T.~L.~C.}\ \bibnamefont {Jansen}}, \bibinfo {author} {\bibfnamefont
  {T.}~\bibnamefont {Mančal}}, \bibinfo {author} {\bibfnamefont {R.~J.~D.}\
  \bibnamefont {Miller}}, \bibinfo {author} {\bibfnamefont {J.~P.}\
  \bibnamefont {Ogilvie}}, \bibinfo {author} {\bibfnamefont {V.~I.}\
  \bibnamefont {Prokhorenko}}, \bibinfo {author} {\bibfnamefont
  {T.}~\bibnamefont {Renger}}, \bibinfo {author} {\bibfnamefont {H.-S.}\
  \bibnamefont {Tan}}, \bibinfo {author} {\bibfnamefont {R.}~\bibnamefont
  {Tempelaar}}, \bibinfo {author} {\bibfnamefont {M.}~\bibnamefont {Thorwart}},
  \bibinfo {author} {\bibfnamefont {E.}~\bibnamefont {Thyrhaug}}, \bibinfo
  {author} {\bibfnamefont {S.}~\bibnamefont {Westenhoff}},\ and\ \bibinfo
  {author} {\bibfnamefont {D.}~\bibnamefont {Zigmantas}},\ }\bibfield  {title}
  {\enquote {\bibinfo {title} {Quantum biology revisited},}\ }\href
  {https://doi.org/10.1126/sciadv.aaz4888} {\bibfield  {journal} {\bibinfo
  {journal} {Science Advances}\ }\textbf {\bibinfo {volume} {6}} (\bibinfo
  {year} {2020}),\ 10.1126/sciadv.aaz4888}\BibitemShut {NoStop}%
\bibitem [{\citenamefont {Fuller}\ \emph {et~al.}(2014)\citenamefont {Fuller},
  \citenamefont {Pan}, \citenamefont {Gelzinis}, \citenamefont {Butkus},
  \citenamefont {Senlik}, \citenamefont {Wilcox}, \citenamefont {Yocum},
  \citenamefont {Valkunas}, \citenamefont {Abramavicius},\ and\ \citenamefont
  {Ogilvie}}]{Fuller2014}%
  \BibitemOpen
  \bibfield  {author} {\bibinfo {author} {\bibfnamefont {F.~D.}\ \bibnamefont
  {Fuller}}, \bibinfo {author} {\bibfnamefont {J.}~\bibnamefont {Pan}},
  \bibinfo {author} {\bibfnamefont {A.}~\bibnamefont {Gelzinis}}, \bibinfo
  {author} {\bibfnamefont {V.}~\bibnamefont {Butkus}}, \bibinfo {author}
  {\bibfnamefont {S.~S.}\ \bibnamefont {Senlik}}, \bibinfo {author}
  {\bibfnamefont {D.~E.}\ \bibnamefont {Wilcox}}, \bibinfo {author}
  {\bibfnamefont {C.~F.}\ \bibnamefont {Yocum}}, \bibinfo {author}
  {\bibfnamefont {L.}~\bibnamefont {Valkunas}}, \bibinfo {author}
  {\bibfnamefont {D.}~\bibnamefont {Abramavicius}},\ and\ \bibinfo {author}
  {\bibfnamefont {J.~P.}\ \bibnamefont {Ogilvie}},\ }\bibfield  {title}
  {\enquote {\bibinfo {title} {Vibronic coherence in oxygenic
  photosynthesis},}\ }\href {https://doi.org/10.1038/nchem.2005} {\bibfield
  {journal} {\bibinfo  {journal} {Nature Chemistry}\ }\textbf {\bibinfo
  {volume} {6}},\ \bibinfo {pages} {706–711} (\bibinfo {year}
  {2014})}\BibitemShut {NoStop}%
\bibitem [{\citenamefont {Schröter}, \citenamefont {Pullerits},\ and\
  \citenamefont {Kühn}(2018)}]{Schroeter2018}%
  \BibitemOpen
  \bibfield  {author} {\bibinfo {author} {\bibfnamefont {M.}~\bibnamefont
  {Schröter}}, \bibinfo {author} {\bibfnamefont {T.}~\bibnamefont
  {Pullerits}},\ and\ \bibinfo {author} {\bibfnamefont {O.}~\bibnamefont
  {Kühn}},\ }\bibfield  {title} {\enquote {\bibinfo {title} {Using
  fluorescence detected two-dimensional spectroscopy to investigate initial
  exciton delocalization between coupled chromophores},}\ }\href
  {https://doi.org/10.1063/1.5046645} {\bibfield  {journal} {\bibinfo
  {journal} {The Journal of Chemical Physics}\ }\textbf {\bibinfo {volume}
  {149}} (\bibinfo {year} {2018}),\ 10.1063/1.5046645}\BibitemShut {NoStop}%
\bibitem [{\citenamefont {Stone}\ \emph {et~al.}(2009)\citenamefont {Stone},
  \citenamefont {Gundogdu}, \citenamefont {Turner}, \citenamefont {Li},
  \citenamefont {Cundiff},\ and\ \citenamefont {Nelson}}]{Stone2009}%
  \BibitemOpen
  \bibfield  {author} {\bibinfo {author} {\bibfnamefont {K.~W.}\ \bibnamefont
  {Stone}}, \bibinfo {author} {\bibfnamefont {K.}~\bibnamefont {Gundogdu}},
  \bibinfo {author} {\bibfnamefont {D.~B.}\ \bibnamefont {Turner}}, \bibinfo
  {author} {\bibfnamefont {X.}~\bibnamefont {Li}}, \bibinfo {author}
  {\bibfnamefont {S.~T.}\ \bibnamefont {Cundiff}},\ and\ \bibinfo {author}
  {\bibfnamefont {K.~A.}\ \bibnamefont {Nelson}},\ }\bibfield  {title}
  {\enquote {\bibinfo {title} {Two-quantum 2d ft electronic spectroscopy of
  biexcitons in gaas quantum wells},}\ }\href
  {https://doi.org/10.1126/science.1170274} {\bibfield  {journal} {\bibinfo
  {journal} {Science}\ }\textbf {\bibinfo {volume} {324}},\ \bibinfo {pages}
  {1169–1173} (\bibinfo {year} {2009})}\BibitemShut {NoStop}%
\bibitem [{\citenamefont {Seiler}\ \emph {et~al.}(2019)\citenamefont {Seiler},
  \citenamefont {Palato}, \citenamefont {Sonnichsen}, \citenamefont {Baker},
  \citenamefont {Socie}, \citenamefont {Strandell},\ and\ \citenamefont
  {Kambhampati}}]{Seiler2019}%
  \BibitemOpen
  \bibfield  {author} {\bibinfo {author} {\bibfnamefont {H.}~\bibnamefont
  {Seiler}}, \bibinfo {author} {\bibfnamefont {S.}~\bibnamefont {Palato}},
  \bibinfo {author} {\bibfnamefont {C.}~\bibnamefont {Sonnichsen}}, \bibinfo
  {author} {\bibfnamefont {H.}~\bibnamefont {Baker}}, \bibinfo {author}
  {\bibfnamefont {E.}~\bibnamefont {Socie}}, \bibinfo {author} {\bibfnamefont
  {D.~P.}\ \bibnamefont {Strandell}},\ and\ \bibinfo {author} {\bibfnamefont
  {P.}~\bibnamefont {Kambhampati}},\ }\bibfield  {title} {\enquote {\bibinfo
  {title} {Two-dimensional electronic spectroscopy reveals liquid-like
  lineshape dynamics in cspbi3 perovskite nanocrystals},}\ }\href
  {https://doi.org/10.1038/s41467-019-12830-1} {\bibfield  {journal} {\bibinfo
  {journal} {Nature Communications}\ }\textbf {\bibinfo {volume} {10}}
  (\bibinfo {year} {2019}),\ 10.1038/s41467-019-12830-1}\BibitemShut {NoStop}%
\bibitem [{\citenamefont {Policht}\ \emph {et~al.}(2021)\citenamefont
  {Policht}, \citenamefont {Russo}, \citenamefont {Liu}, \citenamefont
  {Trovatello}, \citenamefont {Maiuri}, \citenamefont {Bai}, \citenamefont
  {Zhu}, \citenamefont {Dal~Conte},\ and\ \citenamefont
  {Cerullo}}]{Policht2021}%
  \BibitemOpen
  \bibfield  {author} {\bibinfo {author} {\bibfnamefont {V.~R.}\ \bibnamefont
  {Policht}}, \bibinfo {author} {\bibfnamefont {M.}~\bibnamefont {Russo}},
  \bibinfo {author} {\bibfnamefont {F.}~\bibnamefont {Liu}}, \bibinfo {author}
  {\bibfnamefont {C.}~\bibnamefont {Trovatello}}, \bibinfo {author}
  {\bibfnamefont {M.}~\bibnamefont {Maiuri}}, \bibinfo {author} {\bibfnamefont
  {Y.}~\bibnamefont {Bai}}, \bibinfo {author} {\bibfnamefont {X.}~\bibnamefont
  {Zhu}}, \bibinfo {author} {\bibfnamefont {S.}~\bibnamefont {Dal~Conte}},\
  and\ \bibinfo {author} {\bibfnamefont {G.}~\bibnamefont {Cerullo}},\
  }\bibfield  {title} {\enquote {\bibinfo {title} {Dissecting interlayer hole
  and electron transfer in transition metal dichalcogenide heterostructures via
  two-dimensional electronic spectroscopy},}\ }\href
  {https://doi.org/10.1021/acs.nanolett.1c01098} {\bibfield  {journal}
  {\bibinfo  {journal} {Nano Letters}\ }\textbf {\bibinfo {volume} {21}},\
  \bibinfo {pages} {4738–4743} (\bibinfo {year} {2021})}\BibitemShut
  {NoStop}%
\bibitem [{\citenamefont {Jeffries}\ \emph {et~al.}(2020)\citenamefont
  {Jeffries}, \citenamefont {Park}, \citenamefont {Vaia},\ and\ \citenamefont
  {Knappenberger}}]{Jeffries2020}%
  \BibitemOpen
  \bibfield  {author} {\bibinfo {author} {\bibfnamefont {W.~R.}\ \bibnamefont
  {Jeffries}}, \bibinfo {author} {\bibfnamefont {K.}~\bibnamefont {Park}},
  \bibinfo {author} {\bibfnamefont {R.~A.}\ \bibnamefont {Vaia}},\ and\
  \bibinfo {author} {\bibfnamefont {K.~L.}\ \bibnamefont {Knappenberger}},\
  }\bibfield  {title} {\enquote {\bibinfo {title} {Resolving
  electron–electron scattering in plasmonic nanorod ensembles using
  two-dimensional electronic spectroscopy},}\ }\href
  {https://doi.org/10.1021/acs.nanolett.0c03272} {\bibfield  {journal}
  {\bibinfo  {journal} {Nano Letters}\ }\textbf {\bibinfo {volume} {20}},\
  \bibinfo {pages} {7722–7727} (\bibinfo {year} {2020})}\BibitemShut
  {NoStop}%
\bibitem [{\citenamefont {Li}\ \emph {et~al.}(2022)\citenamefont {Li},
  \citenamefont {Shan}, \citenamefont {Rupprecht}, \citenamefont {Knopf},
  \citenamefont {Watanabe}, \citenamefont {Taniguchi}, \citenamefont {Qin},
  \citenamefont {Tongay}, \citenamefont {Nuss}, \citenamefont {Schr{\"o}der},
  \citenamefont {Eilenberger}, \citenamefont {H{\"o}fling}, \citenamefont
  {Schneider},\ and\ \citenamefont {Brixner}}]{Li2022-qv}%
  \BibitemOpen
  \bibfield  {author} {\bibinfo {author} {\bibfnamefont {D.}~\bibnamefont
  {Li}}, \bibinfo {author} {\bibfnamefont {H.}~\bibnamefont {Shan}}, \bibinfo
  {author} {\bibfnamefont {C.}~\bibnamefont {Rupprecht}}, \bibinfo {author}
  {\bibfnamefont {H.}~\bibnamefont {Knopf}}, \bibinfo {author} {\bibfnamefont
  {K.}~\bibnamefont {Watanabe}}, \bibinfo {author} {\bibfnamefont
  {T.}~\bibnamefont {Taniguchi}}, \bibinfo {author} {\bibfnamefont
  {Y.}~\bibnamefont {Qin}}, \bibinfo {author} {\bibfnamefont {S.}~\bibnamefont
  {Tongay}}, \bibinfo {author} {\bibfnamefont {M.}~\bibnamefont {Nuss}},
  \bibinfo {author} {\bibfnamefont {S.}~\bibnamefont {Schr{\"o}der}}, \bibinfo
  {author} {\bibfnamefont {F.}~\bibnamefont {Eilenberger}}, \bibinfo {author}
  {\bibfnamefont {S.}~\bibnamefont {H{\"o}fling}}, \bibinfo {author}
  {\bibfnamefont {C.}~\bibnamefont {Schneider}},\ and\ \bibinfo {author}
  {\bibfnamefont {T.}~\bibnamefont {Brixner}},\ }\bibfield  {title} {\enquote
  {\bibinfo {title} {Hybridized exciton-photon-phonon states in a transition
  metal dichalcogenide {v}an der {W}aals heterostructure microcavity},}\ }\href
  {https://doi.org/10.1103/PhysRevLett.128.087401} {\bibfield  {journal}
  {\bibinfo  {journal} {Phys. Rev. Lett.}\ }\textbf {\bibinfo {volume} {128}}
  (\bibinfo {year} {2022}),\ 10.1103/PhysRevLett.128.087401}\BibitemShut
  {NoStop}%
\bibitem [{\citenamefont {Timmer}\ \emph {et~al.}(2026)\citenamefont {Timmer},
  \citenamefont {Gittinger}, \citenamefont {Quenzel}, \citenamefont {Cadore},
  \citenamefont {Rosa}, \citenamefont {Li}, \citenamefont {Soavi},
  \citenamefont {L\"{u}nemann}, \citenamefont {Stephan}, \citenamefont
  {Greten}, \citenamefont {Richter}, \citenamefont {Knorr}, \citenamefont
  {De~Sio}, \citenamefont {Silies}, \citenamefont {Cerullo}, \citenamefont
  {Ferrari},\ and\ \citenamefont {Lienau}}]{Timmer2026}%
  \BibitemOpen
  \bibfield  {author} {\bibinfo {author} {\bibfnamefont {D.}~\bibnamefont
  {Timmer}}, \bibinfo {author} {\bibfnamefont {M.}~\bibnamefont {Gittinger}},
  \bibinfo {author} {\bibfnamefont {T.}~\bibnamefont {Quenzel}}, \bibinfo
  {author} {\bibfnamefont {A.~R.}\ \bibnamefont {Cadore}}, \bibinfo {author}
  {\bibfnamefont {B.~L.~T.}\ \bibnamefont {Rosa}}, \bibinfo {author}
  {\bibfnamefont {W.}~\bibnamefont {Li}}, \bibinfo {author} {\bibfnamefont
  {G.}~\bibnamefont {Soavi}}, \bibinfo {author} {\bibfnamefont {D.~C.}\
  \bibnamefont {L\"{u}nemann}}, \bibinfo {author} {\bibfnamefont
  {S.}~\bibnamefont {Stephan}}, \bibinfo {author} {\bibfnamefont
  {L.}~\bibnamefont {Greten}}, \bibinfo {author} {\bibfnamefont
  {M.}~\bibnamefont {Richter}}, \bibinfo {author} {\bibfnamefont
  {A.}~\bibnamefont {Knorr}}, \bibinfo {author} {\bibfnamefont
  {A.}~\bibnamefont {De~Sio}}, \bibinfo {author} {\bibfnamefont
  {M.}~\bibnamefont {Silies}}, \bibinfo {author} {\bibfnamefont
  {G.}~\bibnamefont {Cerullo}}, \bibinfo {author} {\bibfnamefont {A.~C.}\
  \bibnamefont {Ferrari}},\ and\ \bibinfo {author} {\bibfnamefont
  {C.}~\bibnamefont {Lienau}},\ }\bibfield  {title} {\enquote {\bibinfo {title}
  {Ultrafast transition from coherent to incoherent polariton nonlinearities in
  a hybrid 1l-ws2/plasmon structure},}\ }\href
  {https://doi.org/10.1038/s41565-025-02054-4} {\bibfield  {journal} {\bibinfo
  {journal} {Nature Nanotechnology}\ } (\bibinfo {year} {2026}),\
  10.1038/s41565-025-02054-4}\BibitemShut {NoStop}%
\bibitem [{\citenamefont {Novelli}\ \emph {et~al.}(2020)\citenamefont
  {Novelli}, \citenamefont {Tollerud}, \citenamefont {Prabhakaran},\ and\
  \citenamefont {Davis}}]{Novelli2020}%
  \BibitemOpen
  \bibfield  {author} {\bibinfo {author} {\bibfnamefont {F.}~\bibnamefont
  {Novelli}}, \bibinfo {author} {\bibfnamefont {J.~O.}\ \bibnamefont
  {Tollerud}}, \bibinfo {author} {\bibfnamefont {D.}~\bibnamefont
  {Prabhakaran}},\ and\ \bibinfo {author} {\bibfnamefont {J.~A.}\ \bibnamefont
  {Davis}},\ }\bibfield  {title} {\enquote {\bibinfo {title} {Persistent
  coherence of quantum superpositions in an optimally doped cuprate revealed by
  2d spectroscopy},}\ }\href {https://doi.org/10.1126/sciadv.aaw9932}
  {\bibfield  {journal} {\bibinfo  {journal} {Science Advances}\ }\textbf
  {\bibinfo {volume} {6}} (\bibinfo {year} {2020}),\
  10.1126/sciadv.aaw9932}\BibitemShut {NoStop}%
\bibitem [{\citenamefont {Wilhelm}, \citenamefont {Piel},\ and\ \citenamefont
  {Riedle}(1997)}]{Wilhelm1997}%
  \BibitemOpen
  \bibfield  {author} {\bibinfo {author} {\bibfnamefont {T.}~\bibnamefont
  {Wilhelm}}, \bibinfo {author} {\bibfnamefont {J.}~\bibnamefont {Piel}},\ and\
  \bibinfo {author} {\bibfnamefont {E.}~\bibnamefont {Riedle}},\ }\bibfield
  {title} {\enquote {\bibinfo {title} {Sub-20-fs pulses tunable across the
  visible from a blue-pumped single-pass noncollinear parametric converter},}\
  }\href {https://doi.org/10.1364/OL.22.001494} {\bibfield  {journal} {\bibinfo
   {journal} {Optics Letters}\ }\textbf {\bibinfo {volume} {22}},\ \bibinfo
  {pages} {1494--1496} (\bibinfo {year} {1997})}\BibitemShut {NoStop}%
\bibitem [{\citenamefont {Manzoni}\ and\ \citenamefont
  {Cerullo}(2016)}]{Cerullo2016}%
  \BibitemOpen
  \bibfield  {author} {\bibinfo {author} {\bibfnamefont {C.}~\bibnamefont
  {Manzoni}}\ and\ \bibinfo {author} {\bibfnamefont {G.}~\bibnamefont
  {Cerullo}},\ }\bibfield  {title} {\enquote {\bibinfo {title} {Design criteria
  for ultrafast optical parametric amplifiers},}\ }\href
  {https://doi.org/10.1088/2040-8978/18/10/103501} {\bibfield  {journal}
  {\bibinfo  {journal} {Journal of Optics}\ }\textbf {\bibinfo {volume} {18}},\
  \bibinfo {pages} {103501} (\bibinfo {year} {2016})}\BibitemShut {NoStop}%
\bibitem [{\citenamefont {Stolen}\ and\ \citenamefont
  {Lin}(1978)}]{Stolen1978}%
  \BibitemOpen
  \bibfield  {author} {\bibinfo {author} {\bibfnamefont {R.~H.}\ \bibnamefont
  {Stolen}}\ and\ \bibinfo {author} {\bibfnamefont {C.}~\bibnamefont {Lin}},\
  }\bibfield  {title} {\enquote {\bibinfo {title} {Self-phase-modulation in
  silica optical fibers},}\ }\href {https://doi.org/10.1103/PhysRevA.17.1448}
  {\bibfield  {journal} {\bibinfo  {journal} {Physical Review A}\ }\textbf
  {\bibinfo {volume} {17}},\ \bibinfo {pages} {1448} (\bibinfo {year}
  {1978})}\BibitemShut {NoStop}%
\bibitem [{\citenamefont {Dudley}, \citenamefont {Genty},\ and\ \citenamefont
  {Coen}(2006)}]{Dudley2006}%
  \BibitemOpen
  \bibfield  {author} {\bibinfo {author} {\bibfnamefont {J.~M.}\ \bibnamefont
  {Dudley}}, \bibinfo {author} {\bibfnamefont {G.}~\bibnamefont {Genty}},\ and\
  \bibinfo {author} {\bibfnamefont {S.}~\bibnamefont {Coen}},\ }\bibfield
  {title} {\enquote {\bibinfo {title} {Supercontinuum generation in photonic
  crystal fiber},}\ }\href {https://doi.org/10.1103/RevModPhys.78.1135}
  {\bibfield  {journal} {\bibinfo  {journal} {Review of Modern Physics}\
  }\textbf {\bibinfo {volume} {78}},\ \bibinfo {pages} {1135} (\bibinfo {year}
  {2006})}\BibitemShut {NoStop}%
\bibitem [{\citenamefont {Konorov}\ \emph {et~al.}(2004)\citenamefont
  {Konorov}, \citenamefont {Sidorov-Biryukov}, \citenamefont {Bugar},
  \citenamefont {Chorvat}, \citenamefont {Beloglazov}, \citenamefont {Skibina},
  \citenamefont {Mel'nikov}, \citenamefont {Shcherbakov}, \citenamefont
  {Chorvat},\ and\ \citenamefont {Zheltikov}}]{Konorov2004}%
  \BibitemOpen
  \bibfield  {author} {\bibinfo {author} {\bibfnamefont {S.~O.}\ \bibnamefont
  {Konorov}}, \bibinfo {author} {\bibfnamefont {D.~A.}\ \bibnamefont
  {Sidorov-Biryukov}}, \bibinfo {author} {\bibfnamefont {I.}~\bibnamefont
  {Bugar}}, \bibinfo {author} {\bibfnamefont {D.~J.}\ \bibnamefont {Chorvat}},
  \bibinfo {author} {\bibfnamefont {V.~I.}\ \bibnamefont {Beloglazov}},
  \bibinfo {author} {\bibfnamefont {N.~B.}\ \bibnamefont {Skibina}}, \bibinfo
  {author} {\bibfnamefont {L.~A.}\ \bibnamefont {Mel'nikov}}, \bibinfo {author}
  {\bibfnamefont {A.~V.}\ \bibnamefont {Shcherbakov}}, \bibinfo {author}
  {\bibfnamefont {D.}~\bibnamefont {Chorvat}},\ and\ \bibinfo {author}
  {\bibfnamefont {A.~M.}\ \bibnamefont {Zheltikov}},\ }\bibfield  {title}
  {\enquote {\bibinfo {title} {Self-phase modulation of femtosecond pulses in
  hollow photonic-crystal fibres},}\ }\href
  {https://doi.org/10.1070/QE2004v034n01ABEH002580} {\bibfield  {journal}
  {\bibinfo  {journal} {Quantum Electronics}\ }\textbf {\bibinfo {volume}
  {34}},\ \bibinfo {pages} {56--58} (\bibinfo {year} {2004})}\BibitemShut
  {NoStop}%
\bibitem [{\citenamefont {Seiler}\ \emph {et~al.}(2017)\citenamefont {Seiler},
  \citenamefont {Palato}, \citenamefont {Schmidt},\ and\ \citenamefont
  {Kambhampati}}]{Seiler2017}%
  \BibitemOpen
  \bibfield  {author} {\bibinfo {author} {\bibfnamefont {H.}~\bibnamefont
  {Seiler}}, \bibinfo {author} {\bibfnamefont {S.}~\bibnamefont {Palato}},
  \bibinfo {author} {\bibfnamefont {B.~E.}\ \bibnamefont {Schmidt}},\ and\
  \bibinfo {author} {\bibfnamefont {P.}~\bibnamefont {Kambhampati}},\
  }\bibfield  {title} {\enquote {\bibinfo {title} {Simple fiber-based solution
  for coherent multidimensional spectroscopy in the visible regime},}\ }\href
  {https://doi.org/10.1364/ol.42.000643} {\bibfield  {journal} {\bibinfo
  {journal} {Optics Letters}\ }\textbf {\bibinfo {volume} {42}},\ \bibinfo
  {pages} {643} (\bibinfo {year} {2017})}\BibitemShut {NoStop}%
\bibitem [{\citenamefont {Timmer}\ \emph {et~al.}(2023)\citenamefont {Timmer},
  \citenamefont {L\"{u}nemann}, \citenamefont {Riese}, \citenamefont {Sio},\
  and\ \citenamefont {Lienau}}]{Timmer2023}%
  \BibitemOpen
  \bibfield  {author} {\bibinfo {author} {\bibfnamefont {D.}~\bibnamefont
  {Timmer}}, \bibinfo {author} {\bibfnamefont {D.~C.}\ \bibnamefont
  {L\"{u}nemann}}, \bibinfo {author} {\bibfnamefont {S.}~\bibnamefont {Riese}},
  \bibinfo {author} {\bibfnamefont {A.~D.}\ \bibnamefont {Sio}},\ and\ \bibinfo
  {author} {\bibfnamefont {C.}~\bibnamefont {Lienau}},\ }\bibfield  {title}
  {\enquote {\bibinfo {title} {Full visible range two-dimensional electronic
  spectroscopy with high time resolution},}\ }\href
  {https://doi.org/10.1364/oe.511906} {\bibfield  {journal} {\bibinfo
  {journal} {Optics Express}\ }\textbf {\bibinfo {volume} {32}},\ \bibinfo
  {pages} {835} (\bibinfo {year} {2023})}\BibitemShut {NoStop}%
\bibitem [{\citenamefont {Paleček}\ \emph {et~al.}(2019)\citenamefont
  {Paleček}, \citenamefont {Edlund}, \citenamefont {Gustavsson}, \citenamefont
  {Westenhoff},\ and\ \citenamefont {Zigmantas}}]{Paleek2019}%
  \BibitemOpen
  \bibfield  {author} {\bibinfo {author} {\bibfnamefont {D.}~\bibnamefont
  {Paleček}}, \bibinfo {author} {\bibfnamefont {P.}~\bibnamefont {Edlund}},
  \bibinfo {author} {\bibfnamefont {E.}~\bibnamefont {Gustavsson}}, \bibinfo
  {author} {\bibfnamefont {S.}~\bibnamefont {Westenhoff}},\ and\ \bibinfo
  {author} {\bibfnamefont {D.}~\bibnamefont {Zigmantas}},\ }\bibfield  {title}
  {\enquote {\bibinfo {title} {Potential pitfalls of the early-time dynamics in
  two-dimensional electronic spectroscopy},}\ }\href
  {https://doi.org/10.1063/1.5079817} {\bibfield  {journal} {\bibinfo
  {journal} {The Journal of Chemical Physics}\ }\textbf {\bibinfo {volume}
  {151}} (\bibinfo {year} {2019}),\ 10.1063/1.5079817}\BibitemShut {NoStop}%
\bibitem [{\citenamefont {Brosseau}\ \emph {et~al.}(2023)\citenamefont
  {Brosseau}, \citenamefont {Seiler}, \citenamefont {Palato}, \citenamefont
  {Sonnichsen}, \citenamefont {Baker}, \citenamefont {Socie}, \citenamefont
  {Strandell},\ and\ \citenamefont {Kambhampati}}]{Brosseau2023}%
  \BibitemOpen
  \bibfield  {author} {\bibinfo {author} {\bibfnamefont {P.}~\bibnamefont
  {Brosseau}}, \bibinfo {author} {\bibfnamefont {H.}~\bibnamefont {Seiler}},
  \bibinfo {author} {\bibfnamefont {S.}~\bibnamefont {Palato}}, \bibinfo
  {author} {\bibfnamefont {C.}~\bibnamefont {Sonnichsen}}, \bibinfo {author}
  {\bibfnamefont {H.}~\bibnamefont {Baker}}, \bibinfo {author} {\bibfnamefont
  {E.}~\bibnamefont {Socie}}, \bibinfo {author} {\bibfnamefont
  {D.}~\bibnamefont {Strandell}},\ and\ \bibinfo {author} {\bibfnamefont
  {P.}~\bibnamefont {Kambhampati}},\ }\bibfield  {title} {\enquote {\bibinfo
  {title} {Perturbed free induction decay obscures early time dynamics in
  two-dimensional electronic spectroscopy: The case of semiconductor
  nanocrystals},}\ }\href {https://doi.org/10.1063/5.0138252} {\bibfield
  {journal} {\bibinfo  {journal} {The Journal of Chemical Physics}\ }\textbf
  {\bibinfo {volume} {158}} (\bibinfo {year} {2023}),\
  10.1063/5.0138252}\BibitemShut {NoStop}%
\bibitem [{\citenamefont {Mukamel}(1999)}]{Mukamel1999}%
  \BibitemOpen
  \bibfield  {author} {\bibinfo {author} {\bibfnamefont {S.}~\bibnamefont
  {Mukamel}},\ }\href@noop {} {\emph {\bibinfo {title} {Principles of Nonlinear
  Optical Spectroscopy}}}\ (\bibinfo  {publisher} {Oxford University Press},\
  \bibinfo {year} {1999})\BibitemShut {NoStop}%
\bibitem [{\citenamefont {Gelin}, \citenamefont {Egorova},\ and\ \citenamefont
  {Domcke}(2009)}]{Gelin2009}%
  \BibitemOpen
  \bibfield  {author} {\bibinfo {author} {\bibfnamefont {M.~F.}\ \bibnamefont
  {Gelin}}, \bibinfo {author} {\bibfnamefont {D.}~\bibnamefont {Egorova}},\
  and\ \bibinfo {author} {\bibfnamefont {W.}~\bibnamefont {Domcke}},\
  }\bibfield  {title} {\enquote {\bibinfo {title} {Efficient calculation of
  time- and frequency-resolved four-wave-mixing signals},}\ }\href
  {https://doi.org/10.1021/ar900045d} {\bibfield  {journal} {\bibinfo
  {journal} {Accounts of Chemical Research}\ }\textbf {\bibinfo {volume}
  {42}},\ \bibinfo {pages} {1290–1298} (\bibinfo {year} {2009})}\BibitemShut
  {NoStop}%
\bibitem [{\citenamefont {Gelin}, \citenamefont {Chen},\ and\ \citenamefont
  {Domcke}(2022)}]{Gelin2022}%
  \BibitemOpen
  \bibfield  {author} {\bibinfo {author} {\bibfnamefont {M.~F.}\ \bibnamefont
  {Gelin}}, \bibinfo {author} {\bibfnamefont {L.}~\bibnamefont {Chen}},\ and\
  \bibinfo {author} {\bibfnamefont {W.}~\bibnamefont {Domcke}},\ }\bibfield
  {title} {\enquote {\bibinfo {title} {Equation-of-motion methods for the
  calculation of femtosecond time-resolved 4-wave-mixing and n-wave-mixing
  signals},}\ }\href {https://doi.org/10.1021/acs.chemrev.2c00329} {\bibfield
  {journal} {\bibinfo  {journal} {Chemical Reviews}\ }\textbf {\bibinfo
  {volume} {122}},\ \bibinfo {pages} {17339–17396} (\bibinfo {year}
  {2022})}\BibitemShut {NoStop}%
\bibitem [{\citenamefont {Rose}\ and\ \citenamefont {Krich}(2019)}]{Rose2019}%
  \BibitemOpen
  \bibfield  {author} {\bibinfo {author} {\bibfnamefont {P.~A.}\ \bibnamefont
  {Rose}}\ and\ \bibinfo {author} {\bibfnamefont {J.~J.}\ \bibnamefont
  {Krich}},\ }\bibfield  {title} {\enquote {\bibinfo {title} {Numerical method
  for nonlinear optical spectroscopies: Ultrafast ultrafast spectroscopy},}\
  }\href {https://doi.org/10.1063/1.5094062} {\bibfield  {journal} {\bibinfo
  {journal} {The Journal of Chemical Physics}\ }\textbf {\bibinfo {volume}
  {150}} (\bibinfo {year} {2019}),\ 10.1063/1.5094062}\BibitemShut {NoStop}%
\bibitem [{\citenamefont {Seidner}, \citenamefont {Stock},\ and\ \citenamefont
  {Domcke}(1995)}]{Seidner1995}%
  \BibitemOpen
  \bibfield  {author} {\bibinfo {author} {\bibfnamefont {L.}~\bibnamefont
  {Seidner}}, \bibinfo {author} {\bibfnamefont {G.}~\bibnamefont {Stock}},\
  and\ \bibinfo {author} {\bibfnamefont {W.}~\bibnamefont {Domcke}},\
  }\bibfield  {title} {\enquote {\bibinfo {title} {Nonperturbative approach to
  femtosecond spectroscopy: General theory and application to multidimensional
  nonadiabatic photoisomerization processes},}\ }\href
  {https://doi.org/10.1063/1.469586} {\bibfield  {journal} {\bibinfo  {journal}
  {The Journal of Chemical Physics}\ }\textbf {\bibinfo {volume} {103}},\
  \bibinfo {pages} {3998–4011} (\bibinfo {year} {1995})}\BibitemShut
  {NoStop}%
\bibitem [{\citenamefont {Bruschi}, \citenamefont {Gallina},\ and\
  \citenamefont {Fresch}(2022)}]{Bruschi2022}%
  \BibitemOpen
  \bibfield  {author} {\bibinfo {author} {\bibfnamefont {M.}~\bibnamefont
  {Bruschi}}, \bibinfo {author} {\bibfnamefont {F.}~\bibnamefont {Gallina}},\
  and\ \bibinfo {author} {\bibfnamefont {B.}~\bibnamefont {Fresch}},\
  }\bibfield  {title} {\enquote {\bibinfo {title} {Simulating action-2d
  electronic spectroscopy of quantum dots: insights on the exciton and
  biexciton interplay from detection-mode and time-gating},}\ }\href
  {https://doi.org/10.1039/d2cp04270c} {\bibfield  {journal} {\bibinfo
  {journal} {Physical Chemistry Chemical Physics}\ }\textbf {\bibinfo {volume}
  {24}},\ \bibinfo {pages} {27645–27659} (\bibinfo {year}
  {2022})}\BibitemShut {NoStop}%
\bibitem [{\citenamefont {Leng}\ \emph {et~al.}(2017)\citenamefont {Leng},
  \citenamefont {Yue}, \citenamefont {Weng}, \citenamefont {Song},\ and\
  \citenamefont {Shi}}]{Leng2017}%
  \BibitemOpen
  \bibfield  {author} {\bibinfo {author} {\bibfnamefont {X.}~\bibnamefont
  {Leng}}, \bibinfo {author} {\bibfnamefont {S.}~\bibnamefont {Yue}}, \bibinfo
  {author} {\bibfnamefont {Y.-X.}\ \bibnamefont {Weng}}, \bibinfo {author}
  {\bibfnamefont {K.}~\bibnamefont {Song}},\ and\ \bibinfo {author}
  {\bibfnamefont {Q.}~\bibnamefont {Shi}},\ }\bibfield  {title} {\enquote
  {\bibinfo {title} {Effects of finite laser pulse width on two-dimensional
  electronic spectroscopy},}\ }\href
  {https://doi.org/10.1016/j.cplett.2016.11.030} {\bibfield  {journal}
  {\bibinfo  {journal} {Chemical Physics Letters}\ }\textbf {\bibinfo {volume}
  {667}},\ \bibinfo {pages} {79–86} (\bibinfo {year} {2017})}\BibitemShut
  {NoStop}%
\bibitem [{\citenamefont {Mančal}, \citenamefont {Pisliakov},\ and\
  \citenamefont {Fleming}(2006)}]{Manal2006}%
  \BibitemOpen
  \bibfield  {author} {\bibinfo {author} {\bibfnamefont {T.}~\bibnamefont
  {Mančal}}, \bibinfo {author} {\bibfnamefont {A.~V.}\ \bibnamefont
  {Pisliakov}},\ and\ \bibinfo {author} {\bibfnamefont {G.~R.}\ \bibnamefont
  {Fleming}},\ }\bibfield  {title} {\enquote {\bibinfo {title} {Two-dimensional
  optical three-pulse photon echo spectroscopy. i. nonperturbative approach to
  the calculation of spectra},}\ }\href {https://doi.org/10.1063/1.2200704}
  {\bibfield  {journal} {\bibinfo  {journal} {The Journal of Chemical Physics}\
  }\textbf {\bibinfo {volume} {124}} (\bibinfo {year} {2006}),\
  10.1063/1.2200704}\BibitemShut {NoStop}%
\bibitem [{\citenamefont {Seibt}\ \emph {et~al.}(2009)\citenamefont {Seibt},
  \citenamefont {Renziehausen}, \citenamefont {Voronine},\ and\ \citenamefont
  {Engel}}]{Seibt2009}%
  \BibitemOpen
  \bibfield  {author} {\bibinfo {author} {\bibfnamefont {J.}~\bibnamefont
  {Seibt}}, \bibinfo {author} {\bibfnamefont {K.}~\bibnamefont {Renziehausen}},
  \bibinfo {author} {\bibfnamefont {D.~V.}\ \bibnamefont {Voronine}},\ and\
  \bibinfo {author} {\bibfnamefont {V.}~\bibnamefont {Engel}},\ }\bibfield
  {title} {\enquote {\bibinfo {title} {Probing the geometry dependence of
  molecular dimers with two-dimensional-vibronic spectroscopy},}\ }\href
  {https://doi.org/10.1063/1.3086935} {\bibfield  {journal} {\bibinfo
  {journal} {The Journal of Chemical Physics}\ }\textbf {\bibinfo {volume}
  {130}} (\bibinfo {year} {2009}),\ 10.1063/1.3086935}\BibitemShut {NoStop}%
\bibitem [{\citenamefont {Anda}\ and\ \citenamefont {Cole}(2021)}]{Anda2021}%
  \BibitemOpen
  \bibfield  {author} {\bibinfo {author} {\bibfnamefont {A.}~\bibnamefont
  {Anda}}\ and\ \bibinfo {author} {\bibfnamefont {J.~H.}\ \bibnamefont
  {Cole}},\ }\bibfield  {title} {\enquote {\bibinfo {title} {Two-dimensional
  spectroscopy beyond the perturbative limit: The influence of finite pulses
  and detection modes},}\ }\href {https://doi.org/10.1063/5.0038550} {\bibfield
   {journal} {\bibinfo  {journal} {The Journal of Chemical Physics}\ }\textbf
  {\bibinfo {volume} {154}} (\bibinfo {year} {2021}),\
  10.1063/5.0038550}\BibitemShut {NoStop}%
\bibitem [{\citenamefont {Kenneweg}\ \emph {et~al.}(2024)\citenamefont
  {Kenneweg}, \citenamefont {Mueller}, \citenamefont {Brixner},\ and\
  \citenamefont {Pfeiffer}}]{Kenneweg2024}%
  \BibitemOpen
  \bibfield  {author} {\bibinfo {author} {\bibfnamefont {T.}~\bibnamefont
  {Kenneweg}}, \bibinfo {author} {\bibfnamefont {S.}~\bibnamefont {Mueller}},
  \bibinfo {author} {\bibfnamefont {T.}~\bibnamefont {Brixner}},\ and\ \bibinfo
  {author} {\bibfnamefont {W.}~\bibnamefont {Pfeiffer}},\ }\bibfield  {title}
  {\enquote {\bibinfo {title} {Qdt — a matlab toolbox for the simulation of
  coupled quantum systems and coherent multidimensional spectroscopy},}\ }\href
  {https://doi.org/10.1016/j.cpc.2023.109031} {\bibfield  {journal} {\bibinfo
  {journal} {Computer Physics Communications}\ }\textbf {\bibinfo {volume}
  {296}},\ \bibinfo {pages} {109031} (\bibinfo {year} {2024})}\BibitemShut
  {NoStop}%
\bibitem [{\citenamefont {Gallagher~Faeder}\ and\ \citenamefont
  {Jonas}(1999)}]{GallagherFaeder1999}%
  \BibitemOpen
  \bibfield  {author} {\bibinfo {author} {\bibfnamefont {S.~M.}\ \bibnamefont
  {Gallagher~Faeder}}\ and\ \bibinfo {author} {\bibfnamefont {D.~M.}\
  \bibnamefont {Jonas}},\ }\bibfield  {title} {\enquote {\bibinfo {title}
  {Two-dimensional electronic correlation and relaxation spectra: Theory and
  model calculations},}\ }\href {https://doi.org/10.1021/jp9925738} {\bibfield
  {journal} {\bibinfo  {journal} {The Journal of Physical Chemistry A}\
  }\textbf {\bibinfo {volume} {103}},\ \bibinfo {pages} {10489–10505}
  (\bibinfo {year} {1999})}\BibitemShut {NoStop}%
\bibitem [{\citenamefont {Abramavicius}\ \emph {et~al.}(2010)\citenamefont
  {Abramavicius}, \citenamefont {Butkus}, \citenamefont {Bujokas},\ and\
  \citenamefont {Valkunas}}]{Abramavicius2010}%
  \BibitemOpen
  \bibfield  {author} {\bibinfo {author} {\bibfnamefont {D.}~\bibnamefont
  {Abramavicius}}, \bibinfo {author} {\bibfnamefont {V.}~\bibnamefont
  {Butkus}}, \bibinfo {author} {\bibfnamefont {J.}~\bibnamefont {Bujokas}},\
  and\ \bibinfo {author} {\bibfnamefont {L.}~\bibnamefont {Valkunas}},\
  }\bibfield  {title} {\enquote {\bibinfo {title} {Manipulation of
  two-dimensional spectra of excitonically coupled molecules by
  narrow-bandwidth laser pulses},}\ }\href
  {https://doi.org/10.1016/j.chemphys.2010.04.015} {\bibfield  {journal}
  {\bibinfo  {journal} {Chemical Physics}\ }\textbf {\bibinfo {volume} {372}},\
  \bibinfo {pages} {22–32} (\bibinfo {year} {2010})}\BibitemShut {NoStop}%
\bibitem [{\citenamefont {Do}, \citenamefont {Gelin},\ and\ \citenamefont
  {Tan}(2017)}]{Do2017}%
  \BibitemOpen
  \bibfield  {author} {\bibinfo {author} {\bibfnamefont {T.~N.}\ \bibnamefont
  {Do}}, \bibinfo {author} {\bibfnamefont {M.~F.}\ \bibnamefont {Gelin}},\ and\
  \bibinfo {author} {\bibfnamefont {H.-S.}\ \bibnamefont {Tan}},\ }\bibfield
  {title} {\enquote {\bibinfo {title} {Simplified expressions that incorporate
  finite pulse effects into coherent two-dimensional optical spectra},}\ }\href
  {https://doi.org/10.1063/1.4985888} {\bibfield  {journal} {\bibinfo
  {journal} {The Journal of Chemical Physics}\ }\textbf {\bibinfo {volume}
  {147}} (\bibinfo {year} {2017}),\ 10.1063/1.4985888}\BibitemShut {NoStop}%
\bibitem [{\citenamefont {Lim}\ \emph {et~al.}(2019)\citenamefont {Lim},
  \citenamefont {B\"{o}sen}, \citenamefont {Somoza}, \citenamefont {Koch},
  \citenamefont {Plenio},\ and\ \citenamefont {Huelga}}]{Lim2019}%
  \BibitemOpen
  \bibfield  {author} {\bibinfo {author} {\bibfnamefont {J.}~\bibnamefont
  {Lim}}, \bibinfo {author} {\bibfnamefont {C.~M.}\ \bibnamefont {B\"{o}sen}},
  \bibinfo {author} {\bibfnamefont {A.~D.}\ \bibnamefont {Somoza}}, \bibinfo
  {author} {\bibfnamefont {C.~P.}\ \bibnamefont {Koch}}, \bibinfo {author}
  {\bibfnamefont {M.~B.}\ \bibnamefont {Plenio}},\ and\ \bibinfo {author}
  {\bibfnamefont {S.~F.}\ \bibnamefont {Huelga}},\ }\bibfield  {title}
  {\enquote {\bibinfo {title} {Multicolor quantum control for suppressing
  ground state coherences in two-dimensional electronic spectroscopy},}\ }\href
  {https://doi.org/10.1103/physrevlett.123.233201} {\bibfield  {journal}
  {\bibinfo  {journal} {Physical Review Letters}\ }\textbf {\bibinfo {volume}
  {123}} (\bibinfo {year} {2019}),\ 10.1103/physrevlett.123.233201}\BibitemShut
  {NoStop}%
\bibitem [{\citenamefont {Smallwood}, \citenamefont {Autry},\ and\
  \citenamefont {Cundiff}(2017)}]{Smallwood2017}%
  \BibitemOpen
  \bibfield  {author} {\bibinfo {author} {\bibfnamefont {C.~L.}\ \bibnamefont
  {Smallwood}}, \bibinfo {author} {\bibfnamefont {T.~M.}\ \bibnamefont
  {Autry}},\ and\ \bibinfo {author} {\bibfnamefont {S.~T.}\ \bibnamefont
  {Cundiff}},\ }\bibfield  {title} {\enquote {\bibinfo {title} {Analytical
  solutions to the finite-pulse bloch model for multidimensional coherent
  spectroscopy},}\ }\href {https://doi.org/10.1364/josab.34.000419} {\bibfield
  {journal} {\bibinfo  {journal} {Journal of the Optical Society of America B}\
  }\textbf {\bibinfo {volume} {34}},\ \bibinfo {pages} {419} (\bibinfo {year}
  {2017})}\BibitemShut {NoStop}%
\bibitem [{\citenamefont {Perlík}, \citenamefont {Hauer},\ and\ \citenamefont
  {Šanda}(2017)}]{Perlk2017}%
  \BibitemOpen
  \bibfield  {author} {\bibinfo {author} {\bibfnamefont {V.}~\bibnamefont
  {Perlík}}, \bibinfo {author} {\bibfnamefont {J.}~\bibnamefont {Hauer}},\
  and\ \bibinfo {author} {\bibfnamefont {F.}~\bibnamefont {Šanda}},\
  }\bibfield  {title} {\enquote {\bibinfo {title} {Finite pulse effects in
  single and double quantum spectroscopies},}\ }\href
  {https://doi.org/10.1364/josab.34.000430} {\bibfield  {journal} {\bibinfo
  {journal} {Journal of the Optical Society of America B}\ }\textbf {\bibinfo
  {volume} {34}},\ \bibinfo {pages} {430} (\bibinfo {year} {2017})}\BibitemShut
  {NoStop}%
\bibitem [{\citenamefont {Schweigert}\ and\ \citenamefont
  {Mukamel}(2008)}]{Schweigert2008}%
  \BibitemOpen
  \bibfield  {author} {\bibinfo {author} {\bibfnamefont {I.~V.}\ \bibnamefont
  {Schweigert}}\ and\ \bibinfo {author} {\bibfnamefont {S.}~\bibnamefont
  {Mukamel}},\ }\bibfield  {title} {\enquote {\bibinfo {title} {Simulating
  multidimensional optical wave-mixing signals with finite-pulse envelopes},}\
  }\href {https://doi.org/10.1103/physreva.77.033802} {\bibfield  {journal}
  {\bibinfo  {journal} {Physical Review A}\ }\textbf {\bibinfo {volume} {77}}
  (\bibinfo {year} {2008}),\ 10.1103/physreva.77.033802}\BibitemShut {NoStop}%
\bibitem [{\citenamefont {Kühn}\ and\ \citenamefont
  {Wöste}(2007)}]{Kuehn2007}%
  \BibitemOpen
  \bibfield  {author} {\bibinfo {author} {\bibfnamefont {O.}~\bibnamefont
  {Kühn}}\ and\ \bibinfo {author} {\bibfnamefont {L.}~\bibnamefont {Wöste}},\
  }\href {https://doi.org/10.1007/978-3-540-68038-3} {\emph {\bibinfo {title}
  {Analysis and Control of Ultrafast Photoinduced Reactions}}}\ (\bibinfo
  {publisher} {Springer},\ \bibinfo {year} {2007})\BibitemShut {NoStop}%
\bibitem [{\citenamefont {Fetherolf}\ and\ \citenamefont
  {Berkelbach}(2017)}]{Fetherolf2017}%
  \BibitemOpen
  \bibfield  {author} {\bibinfo {author} {\bibfnamefont {J.~H.}\ \bibnamefont
  {Fetherolf}}\ and\ \bibinfo {author} {\bibfnamefont {T.~C.}\ \bibnamefont
  {Berkelbach}},\ }\bibfield  {title} {\enquote {\bibinfo {title} {Linear and
  nonlinear spectroscopy from quantum master equations},}\ }\href
  {https://doi.org/10.1063/1.5006824} {\bibfield  {journal} {\bibinfo
  {journal} {The Journal of Chemical Physics}\ }\textbf {\bibinfo {volume}
  {147}} (\bibinfo {year} {2017}),\ 10.1063/1.5006824}\BibitemShut {NoStop}%
\bibitem [{\citenamefont {Gallego-Valencia}\ \emph {et~al.}(2024)\citenamefont
  {Gallego-Valencia}, \citenamefont {Mewes}, \citenamefont {Feist},\ and\
  \citenamefont {Sanz-Vicario1}}]{Gallego-Valencia2024}%
  \BibitemOpen
  \bibfield  {author} {\bibinfo {author} {\bibfnamefont {D.}~\bibnamefont
  {Gallego-Valencia}}, \bibinfo {author} {\bibfnamefont {L.}~\bibnamefont
  {Mewes}}, \bibinfo {author} {\bibfnamefont {J.}~\bibnamefont {Feist}},\ and\
  \bibinfo {author} {\bibfnamefont {J.~L.}\ \bibnamefont {Sanz-Vicario1}},\
  }\bibfield  {title} {\enquote {\bibinfo {title} {Coherent multidimensional
  spectroscopy in polariton systems},}\ }\href
  {https://doi.org/10.1103/PhysRevA.109.063704} {\bibfield  {journal} {\bibinfo
   {journal} {Physical Review A}\ }\textbf {\bibinfo {volume} {109}} (\bibinfo
  {year} {2024}),\ 10.1103/PhysRevA.109.063704}\BibitemShut {NoStop}%
\bibitem [{\citenamefont {Jeske}\ \emph {et~al.}(2015)\citenamefont {Jeske},
  \citenamefont {Ing}, \citenamefont {Plenio}, \citenamefont {Huelga},\ and\
  \citenamefont {Cole}}]{Jeske2015}%
  \BibitemOpen
  \bibfield  {author} {\bibinfo {author} {\bibfnamefont {J.}~\bibnamefont
  {Jeske}}, \bibinfo {author} {\bibfnamefont {D.~J.}\ \bibnamefont {Ing}},
  \bibinfo {author} {\bibfnamefont {M.~B.}\ \bibnamefont {Plenio}}, \bibinfo
  {author} {\bibfnamefont {S.~F.}\ \bibnamefont {Huelga}},\ and\ \bibinfo
  {author} {\bibfnamefont {J.~H.}\ \bibnamefont {Cole}},\ }\bibfield  {title}
  {\enquote {\bibinfo {title} {Bloch-redfield equations for modeling
  light-harvesting complexes},}\ }\href {https://doi.org/10.1063/1.4907370}
  {\bibfield  {journal} {\bibinfo  {journal} {The Journal of Chemical Physics}\
  }\textbf {\bibinfo {volume} {142}} (\bibinfo {year} {2015}),\
  10.1063/1.4907370}\BibitemShut {NoStop}%
\bibitem [{Note1()}]{Note1}%
  \BibitemOpen
  \bibinfo {note} {In the literature of 2D coherent spectroscopy, this time
  interval is also widely referred to as the \protect \textit {population
  time}, as it typically tracks the evolution of population states or
  non-radiative coherences.}\BibitemShut {Stop}%
\bibitem [{\citenamefont {Chemla}\ and\ \citenamefont
  {Shah}(2001)}]{Chemla2001}%
  \BibitemOpen
  \bibfield  {author} {\bibinfo {author} {\bibfnamefont {D.~S.}\ \bibnamefont
  {Chemla}}\ and\ \bibinfo {author} {\bibfnamefont {J.}~\bibnamefont {Shah}},\
  }\bibfield  {title} {\enquote {\bibinfo {title} {Many-body and correlation
  effects in semiconductors},}\ }\href {https://doi.org/10.1038/35079000}
  {\bibfield  {journal} {\bibinfo  {journal} {Nature}\ }\textbf {\bibinfo
  {volume} {411}},\ \bibinfo {pages} {549–557} (\bibinfo {year}
  {2001})}\BibitemShut {NoStop}%
\bibitem [{\citenamefont {Leo}\ \emph {et~al.}(1990)\citenamefont {Leo},
  \citenamefont {Wegener}, \citenamefont {Shah}, \citenamefont {Chemla},
  \citenamefont {G\"{o}bel}, \citenamefont {Damen}, \citenamefont
  {Schmitt-Rink},\ and\ \citenamefont {Sch\"{a}fer}}]{Leo1990}%
  \BibitemOpen
  \bibfield  {author} {\bibinfo {author} {\bibfnamefont {K.}~\bibnamefont
  {Leo}}, \bibinfo {author} {\bibfnamefont {M.}~\bibnamefont {Wegener}},
  \bibinfo {author} {\bibfnamefont {J.}~\bibnamefont {Shah}}, \bibinfo {author}
  {\bibfnamefont {D.}~\bibnamefont {Chemla}}, \bibinfo {author} {\bibfnamefont
  {E.}~\bibnamefont {G\"{o}bel}}, \bibinfo {author} {\bibfnamefont
  {T.}~\bibnamefont {Damen}}, \bibinfo {author} {\bibfnamefont
  {S.}~\bibnamefont {Schmitt-Rink}},\ and\ \bibinfo {author} {\bibfnamefont
  {W.}~\bibnamefont {Sch\"{a}fer}},\ }\bibfield  {title} {\enquote {\bibinfo
  {title} {Effects of coherent polarization interactions on time-resolved
  degenerate four-wave mixing},}\ }\href
  {https://doi.org/10.1103/physrevlett.65.1340} {\bibfield  {journal} {\bibinfo
   {journal} {Physical Review Letters}\ }\textbf {\bibinfo {volume} {65}},\
  \bibinfo {pages} {1340–1343} (\bibinfo {year} {1990})}\BibitemShut
  {NoStop}%
\bibitem [{\citenamefont {Hamm}(1995)}]{Hamm1995}%
  \BibitemOpen
  \bibfield  {author} {\bibinfo {author} {\bibfnamefont {P.}~\bibnamefont
  {Hamm}},\ }\bibfield  {title} {\enquote {\bibinfo {title} {Coherent effects
  in femtosecond infrared spectroscopy},}\ }\href
  {https://doi.org/10.1016/0301-0104(95)00262-6} {\bibfield  {journal}
  {\bibinfo  {journal} {Chemical Physics}\ }\textbf {\bibinfo {volume} {200}},\
  \bibinfo {pages} {415–429} (\bibinfo {year} {1995})}\BibitemShut {NoStop}%
\bibitem [{\citenamefont {Joffre}\ \emph {et~al.}(1988)\citenamefont {Joffre},
  \citenamefont {la~Guillaume}, \citenamefont {Peyghambarian}, \citenamefont
  {Lindberg}, \citenamefont {Hulin}, \citenamefont {Migus}, \citenamefont
  {Koch},\ and\ \citenamefont {Antonetti}}]{Joffre1988}%
  \BibitemOpen
  \bibfield  {author} {\bibinfo {author} {\bibfnamefont {M.}~\bibnamefont
  {Joffre}}, \bibinfo {author} {\bibfnamefont {C.~B.}\ \bibnamefont
  {la~Guillaume}}, \bibinfo {author} {\bibfnamefont {N.}~\bibnamefont
  {Peyghambarian}}, \bibinfo {author} {\bibfnamefont {M.}~\bibnamefont
  {Lindberg}}, \bibinfo {author} {\bibfnamefont {D.}~\bibnamefont {Hulin}},
  \bibinfo {author} {\bibfnamefont {A.}~\bibnamefont {Migus}}, \bibinfo
  {author} {\bibfnamefont {S.~W.}\ \bibnamefont {Koch}},\ and\ \bibinfo
  {author} {\bibfnamefont {A.}~\bibnamefont {Antonetti}},\ }\bibfield  {title}
  {\enquote {\bibinfo {title} {Coherent effects in pump–probe spectroscopy of
  excitons},}\ }\href {https://doi.org/10.1364/ol.13.000276} {\bibfield
  {journal} {\bibinfo  {journal} {Optics Letters}\ }\textbf {\bibinfo {volume}
  {13}},\ \bibinfo {pages} {276} (\bibinfo {year} {1988})}\BibitemShut
  {NoStop}%
\bibitem [{\citenamefont {Yan}, \citenamefont {Seidel},\ and\ \citenamefont
  {Tan}(2011)}]{Yan_2011}%
  \BibitemOpen
  \bibfield  {author} {\bibinfo {author} {\bibfnamefont {S.}~\bibnamefont
  {Yan}}, \bibinfo {author} {\bibfnamefont {M.~T.}\ \bibnamefont {Seidel}},\
  and\ \bibinfo {author} {\bibfnamefont {H.-S.}\ \bibnamefont {Tan}},\
  }\bibfield  {title} {\enquote {\bibinfo {title} {Perturbed free induction
  decay in ultrafast mid-ir pump–probe spectroscopy},}\ }\href
  {https://doi.org/10.1016/j.cplett.2011.10.013} {\bibfield  {journal}
  {\bibinfo  {journal} {Chemical Physics Letters}\ }\textbf {\bibinfo {volume}
  {517}},\ \bibinfo {pages} {36–40} (\bibinfo {year} {2011})}\BibitemShut
  {NoStop}%
\bibitem [{\citenamefont {Mondal}\ \emph {et~al.}(2018)\citenamefont {Mondal},
  \citenamefont {Roy}, \citenamefont {Pal},\ and\ \citenamefont
  {Bansal}}]{Mondal2018}%
  \BibitemOpen
  \bibfield  {author} {\bibinfo {author} {\bibfnamefont {R.}~\bibnamefont
  {Mondal}}, \bibinfo {author} {\bibfnamefont {B.}~\bibnamefont {Roy}},
  \bibinfo {author} {\bibfnamefont {B.}~\bibnamefont {Pal}},\ and\ \bibinfo
  {author} {\bibfnamefont {B.}~\bibnamefont {Bansal}},\ }\bibfield  {title}
  {\enquote {\bibinfo {title} {How pump–probe differential reflectivity at
  negative delay yields the perturbed-free-induction-decay: theory of the
  experiment and its verification},}\ }\href
  {https://doi.org/10.1088/1361-648x/aaed79} {\bibfield  {journal} {\bibinfo
  {journal} {Journal of Physics: Condensed Matter}\ }\textbf {\bibinfo {volume}
  {30}},\ \bibinfo {pages} {505902} (\bibinfo {year} {2018})}\BibitemShut
  {NoStop}%
\bibitem [{\citenamefont {Kuehn}\ \emph {et~al.}(2010)\citenamefont {Kuehn},
  \citenamefont {Reimann}, \citenamefont {Woerner}, \citenamefont {Elsaesser},\
  and\ \citenamefont {Hey}}]{Kuehn2010}%
  \BibitemOpen
  \bibfield  {author} {\bibinfo {author} {\bibfnamefont {W.}~\bibnamefont
  {Kuehn}}, \bibinfo {author} {\bibfnamefont {K.}~\bibnamefont {Reimann}},
  \bibinfo {author} {\bibfnamefont {M.}~\bibnamefont {Woerner}}, \bibinfo
  {author} {\bibfnamefont {T.}~\bibnamefont {Elsaesser}},\ and\ \bibinfo
  {author} {\bibfnamefont {R.}~\bibnamefont {Hey}},\ }\bibfield  {title}
  {\enquote {\bibinfo {title} {Two-dimensional terahertz correlation spectra of
  electronic excitations in semiconductor quantum wells},}\ }\href
  {https://doi.org/10.1021/jp1099046} {\bibfield  {journal} {\bibinfo
  {journal} {The Journal of Physical Chemistry B}\ }\textbf {\bibinfo {volume}
  {115}},\ \bibinfo {pages} {5448–5455} (\bibinfo {year} {2010})}\BibitemShut
  {NoStop}%
\bibitem [{\citenamefont {Hamm}\ \emph {et~al.}(2000)\citenamefont {Hamm},
  \citenamefont {Lim}, \citenamefont {DeGrado},\ and\ \citenamefont
  {Hochstrasser}}]{Hamm2000}%
  \BibitemOpen
  \bibfield  {author} {\bibinfo {author} {\bibfnamefont {P.}~\bibnamefont
  {Hamm}}, \bibinfo {author} {\bibfnamefont {M.}~\bibnamefont {Lim}}, \bibinfo
  {author} {\bibfnamefont {W.~F.}\ \bibnamefont {DeGrado}},\ and\ \bibinfo
  {author} {\bibfnamefont {R.~M.}\ \bibnamefont {Hochstrasser}},\ }\bibfield
  {title} {\enquote {\bibinfo {title} {Pump/probe self heterodyned 2d
  spectroscopy of vibrational transitions of a small globular peptide},}\
  }\href {https://doi.org/10.1063/1.480772} {\bibfield  {journal} {\bibinfo
  {journal} {The Journal of Chemical Physics}\ }\textbf {\bibinfo {volume}
  {112}},\ \bibinfo {pages} {1907–1916} (\bibinfo {year} {2000})}\BibitemShut
  {NoStop}%
\bibitem [{\citenamefont {Richter}\ \emph {et~al.}(2017)\citenamefont
  {Richter}, \citenamefont {Branchi}, \citenamefont {Valduga~de
  Almeida~Camargo}, \citenamefont {Zhao}, \citenamefont {Friend}, \citenamefont
  {Cerullo},\ and\ \citenamefont {Deschler}}]{Richter2017}%
  \BibitemOpen
  \bibfield  {author} {\bibinfo {author} {\bibfnamefont {J.~M.}\ \bibnamefont
  {Richter}}, \bibinfo {author} {\bibfnamefont {F.}~\bibnamefont {Branchi}},
  \bibinfo {author} {\bibfnamefont {F.}~\bibnamefont {Valduga~de
  Almeida~Camargo}}, \bibinfo {author} {\bibfnamefont {B.}~\bibnamefont
  {Zhao}}, \bibinfo {author} {\bibfnamefont {R.~H.}\ \bibnamefont {Friend}},
  \bibinfo {author} {\bibfnamefont {G.}~\bibnamefont {Cerullo}},\ and\ \bibinfo
  {author} {\bibfnamefont {F.}~\bibnamefont {Deschler}},\ }\bibfield  {title}
  {\enquote {\bibinfo {title} {Ultrafast carrier thermalization in lead iodide
  perovskite probed with two-dimensional electronic spectroscopy},}\ }\href
  {https://doi.org/10.1038/s41467-017-00546-z} {\bibfield  {journal} {\bibinfo
  {journal} {Nature Communications}\ }\textbf {\bibinfo {volume} {8}} (\bibinfo
  {year} {2017}),\ 10.1038/s41467-017-00546-z}\BibitemShut {NoStop}%
\bibitem [{\citenamefont {Nguyen}\ \emph {et~al.}(2019)\citenamefont {Nguyen},
  \citenamefont {Timmer}, \citenamefont {Rakita}, \citenamefont {Cahen},
  \citenamefont {Steinhoff}, \citenamefont {Jahnke}, \citenamefont {Lienau},\
  and\ \citenamefont {De~Sio}}]{Nguyen2019}%
  \BibitemOpen
  \bibfield  {author} {\bibinfo {author} {\bibfnamefont {X.~T.}\ \bibnamefont
  {Nguyen}}, \bibinfo {author} {\bibfnamefont {D.}~\bibnamefont {Timmer}},
  \bibinfo {author} {\bibfnamefont {Y.}~\bibnamefont {Rakita}}, \bibinfo
  {author} {\bibfnamefont {D.}~\bibnamefont {Cahen}}, \bibinfo {author}
  {\bibfnamefont {A.}~\bibnamefont {Steinhoff}}, \bibinfo {author}
  {\bibfnamefont {F.}~\bibnamefont {Jahnke}}, \bibinfo {author} {\bibfnamefont
  {C.}~\bibnamefont {Lienau}},\ and\ \bibinfo {author} {\bibfnamefont
  {A.}~\bibnamefont {De~Sio}},\ }\bibfield  {title} {\enquote {\bibinfo {title}
  {Ultrafast charge carrier relaxation in inorganic halide perovskite single
  crystals probed by two-dimensional electronic spectroscopy},}\ }\href
  {https://doi.org/10.1021/acs.jpclett.9b01936} {\bibfield  {journal} {\bibinfo
   {journal} {The Journal of Physical Chemistry Letters}\ }\textbf {\bibinfo
  {volume} {10}},\ \bibinfo {pages} {5414–5421} (\bibinfo {year}
  {2019})}\BibitemShut {NoStop}%
\bibitem [{\citenamefont {Lloyd}\ \emph {et~al.}(2021)\citenamefont {Lloyd},
  \citenamefont {Wood}, \citenamefont {Mujid}, \citenamefont {Sohoni},
  \citenamefont {Ji}, \citenamefont {Ting}, \citenamefont {Higgins},
  \citenamefont {Park},\ and\ \citenamefont {Engel}}]{Lloyd2021}%
  \BibitemOpen
  \bibfield  {author} {\bibinfo {author} {\bibfnamefont {L.~T.}\ \bibnamefont
  {Lloyd}}, \bibinfo {author} {\bibfnamefont {R.~E.}\ \bibnamefont {Wood}},
  \bibinfo {author} {\bibfnamefont {F.}~\bibnamefont {Mujid}}, \bibinfo
  {author} {\bibfnamefont {S.}~\bibnamefont {Sohoni}}, \bibinfo {author}
  {\bibfnamefont {K.~L.}\ \bibnamefont {Ji}}, \bibinfo {author} {\bibfnamefont
  {P.-C.}\ \bibnamefont {Ting}}, \bibinfo {author} {\bibfnamefont {J.~S.}\
  \bibnamefont {Higgins}}, \bibinfo {author} {\bibfnamefont {J.}~\bibnamefont
  {Park}},\ and\ \bibinfo {author} {\bibfnamefont {G.~S.}\ \bibnamefont
  {Engel}},\ }\bibfield  {title} {\enquote {\bibinfo {title} {Sub-10 fs
  intervalley exciton coupling in monolayer mos$_2$ revealed by
  helicity-resolved two-dimensional electronic spectroscopy},}\ }\href
  {https://doi.org/10.1021/acsnano.1c02381} {\bibfield  {journal} {\bibinfo
  {journal} {ACS Nano}\ }\textbf {\bibinfo {volume} {15}},\ \bibinfo {pages}
  {10253–10263} (\bibinfo {year} {2021})}\BibitemShut {NoStop}%
\bibitem [{\citenamefont {Schneider}\ \emph {et~al.}(2017)\citenamefont
  {Schneider}, \citenamefont {Kratochvil}, \citenamefont {Zanni},\ and\
  \citenamefont {Boxer}}]{Schneider2017}%
  \BibitemOpen
  \bibfield  {author} {\bibinfo {author} {\bibfnamefont {S.~H.}\ \bibnamefont
  {Schneider}}, \bibinfo {author} {\bibfnamefont {H.~T.}\ \bibnamefont
  {Kratochvil}}, \bibinfo {author} {\bibfnamefont {M.~T.}\ \bibnamefont
  {Zanni}},\ and\ \bibinfo {author} {\bibfnamefont {S.~G.}\ \bibnamefont
  {Boxer}},\ }\bibfield  {title} {\enquote {\bibinfo {title}
  {Solvent-independent anharmonicity for carbonyl oscillators},}\ }\href
  {https://doi.org/10.1021/acs.jpcb.7b00537} {\bibfield  {journal} {\bibinfo
  {journal} {The Journal of Physical Chemistry B}\ }\textbf {\bibinfo {volume}
  {121}},\ \bibinfo {pages} {2331–2338} (\bibinfo {year} {2017})}\BibitemShut
  {NoStop}%
\bibitem [{\citenamefont {Slenkamp}\ \emph {et~al.}(2014)\citenamefont
  {Slenkamp}, \citenamefont {Lynch}, \citenamefont {Van~Kuiken}, \citenamefont
  {Brookes}, \citenamefont {Bannan}, \citenamefont {Daifuku},\ and\
  \citenamefont {Khalil}}]{Slenkamp2014}%
  \BibitemOpen
  \bibfield  {author} {\bibinfo {author} {\bibfnamefont {K.~M.}\ \bibnamefont
  {Slenkamp}}, \bibinfo {author} {\bibfnamefont {M.~S.}\ \bibnamefont {Lynch}},
  \bibinfo {author} {\bibfnamefont {B.~E.}\ \bibnamefont {Van~Kuiken}},
  \bibinfo {author} {\bibfnamefont {J.~F.}\ \bibnamefont {Brookes}}, \bibinfo
  {author} {\bibfnamefont {C.~C.}\ \bibnamefont {Bannan}}, \bibinfo {author}
  {\bibfnamefont {S.~L.}\ \bibnamefont {Daifuku}},\ and\ \bibinfo {author}
  {\bibfnamefont {M.}~\bibnamefont {Khalil}},\ }\bibfield  {title} {\enquote
  {\bibinfo {title} {Investigating vibrational anharmonic couplings in
  cyanide-bridged transition metal mixed valence complexes using
  two-dimensional infrared spectroscopy},}\ }\href
  {https://doi.org/10.1063/1.4866294} {\bibfield  {journal} {\bibinfo
  {journal} {The Journal of Physical Chemistry}\ }\textbf {\bibinfo {volume}
  {140}},\ \bibinfo {pages} {084505} (\bibinfo {year} {2014})}\BibitemShut
  {NoStop}%
\bibitem [{\citenamefont {Galestian~Pour}\ \emph {et~al.}(2017)\citenamefont
  {Galestian~Pour}, \citenamefont {Lincoln}, \citenamefont {Perlík},
  \citenamefont {Šanda},\ and\ \citenamefont {Hauer}}]{Pour2017}%
  \BibitemOpen
  \bibfield  {author} {\bibinfo {author} {\bibfnamefont {A.}~\bibnamefont
  {Galestian~Pour}}, \bibinfo {author} {\bibfnamefont {C.~N.}\ \bibnamefont
  {Lincoln}}, \bibinfo {author} {\bibfnamefont {V.}~\bibnamefont {Perlík}},
  \bibinfo {author} {\bibfnamefont {F.}~\bibnamefont {Šanda}},\ and\ \bibinfo
  {author} {\bibfnamefont {J.}~\bibnamefont {Hauer}},\ }\bibfield  {title}
  {\enquote {\bibinfo {title} {Anharmonic vibrational effects in linear and
  two-dimensional electronic spectra},}\ }\href
  {https://doi.org/10.1039/C7CP05189A} {\bibfield  {journal} {\bibinfo
  {journal} {Physical Chemistry Chemical Physics}\ }\textbf {\bibinfo {volume}
  {19}},\ \bibinfo {pages} {24752--24760} (\bibinfo {year} {2017})}\BibitemShut
  {NoStop}%
\bibitem [{\citenamefont {Cassette}\ \emph {et~al.}(2015)\citenamefont
  {Cassette}, \citenamefont {Pensack}, \citenamefont {Mahler},\ and\
  \citenamefont {Scholes}}]{Cassette2015}%
  \BibitemOpen
  \bibfield  {author} {\bibinfo {author} {\bibfnamefont {E.}~\bibnamefont
  {Cassette}}, \bibinfo {author} {\bibfnamefont {R.~D.}\ \bibnamefont
  {Pensack}}, \bibinfo {author} {\bibfnamefont {B.}~\bibnamefont {Mahler}},\
  and\ \bibinfo {author} {\bibfnamefont {G.~D.}\ \bibnamefont {Scholes}},\
  }\bibfield  {title} {\enquote {\bibinfo {title} {Room-temperature exciton
  coherence and dephasing in two-dimensional nanostructures},}\ }\href
  {https://doi.org/10.1038/ncomms7086} {\bibfield  {journal} {\bibinfo
  {journal} {Nature Communication}\ }\textbf {\bibinfo {volume} {6}},\ \bibinfo
  {pages} {6086} (\bibinfo {year} {2015})}\BibitemShut {NoStop}%
\bibitem [{\citenamefont {Guo}\ \emph {et~al.}(2018)\citenamefont {Guo},
  \citenamefont {Wu}, \citenamefont {Cao}, \citenamefont {Monahan},
  \citenamefont {Lee}, \citenamefont {Louie},\ and\ \citenamefont
  {Fleming}}]{Guo2018}%
  \BibitemOpen
  \bibfield  {author} {\bibinfo {author} {\bibfnamefont {L.}~\bibnamefont
  {Guo}}, \bibinfo {author} {\bibfnamefont {M.}~\bibnamefont {Wu}}, \bibinfo
  {author} {\bibfnamefont {T.}~\bibnamefont {Cao}}, \bibinfo {author}
  {\bibfnamefont {D.~M.}\ \bibnamefont {Monahan}}, \bibinfo {author}
  {\bibfnamefont {Y.-H.}\ \bibnamefont {Lee}}, \bibinfo {author} {\bibfnamefont
  {S.~G.}\ \bibnamefont {Louie}},\ and\ \bibinfo {author} {\bibfnamefont
  {G.~R.}\ \bibnamefont {Fleming}},\ }\bibfield  {title} {\enquote {\bibinfo
  {title} {Exchange-driven intravalley mixing of excitons in monolayer
  transition metal dichalcogenides},}\ }\href
  {https://doi.org/10.1038/s41567-018-0362-y} {\bibfield  {journal} {\bibinfo
  {journal} {Nature Physics}\ }\textbf {\bibinfo {volume} {15}},\ \bibinfo
  {pages} {228–232} (\bibinfo {year} {2018})}\BibitemShut {NoStop}%
\bibitem [{\citenamefont {Rodek}\ \emph {et~al.}(2023)\citenamefont {Rodek},
  \citenamefont {Hahn}, \citenamefont {Howarth}, \citenamefont {Taniguchi},
  \citenamefont {Watanabe}, \citenamefont {Potemski}, \citenamefont {Kossacki},
  \citenamefont {Wigger},\ and\ \citenamefont {Kasprzak}}]{Rodek2023}%
  \BibitemOpen
  \bibfield  {author} {\bibinfo {author} {\bibfnamefont {A.}~\bibnamefont
  {Rodek}}, \bibinfo {author} {\bibfnamefont {T.}~\bibnamefont {Hahn}},
  \bibinfo {author} {\bibfnamefont {J.}~\bibnamefont {Howarth}}, \bibinfo
  {author} {\bibfnamefont {T.}~\bibnamefont {Taniguchi}}, \bibinfo {author}
  {\bibfnamefont {K.}~\bibnamefont {Watanabe}}, \bibinfo {author}
  {\bibfnamefont {M.}~\bibnamefont {Potemski}}, \bibinfo {author}
  {\bibfnamefont {P.}~\bibnamefont {Kossacki}}, \bibinfo {author}
  {\bibfnamefont {D.}~\bibnamefont {Wigger}},\ and\ \bibinfo {author}
  {\bibfnamefont {J.}~\bibnamefont {Kasprzak}},\ }\bibfield  {title} {\enquote
  {\bibinfo {title} {Controlled coherent-coupling and dynamics of exciton
  complexes in a mose$_2$ monolayer},}\ }\href
  {https://doi.org/10.1088/2053-1583/acc59a} {\bibfield  {journal} {\bibinfo
  {journal} {2D Materials}\ }\textbf {\bibinfo {volume} {10}},\ \bibinfo
  {pages} {025027} (\bibinfo {year} {2023})}\BibitemShut {NoStop}%
\bibitem [{\citenamefont {Camargo}\ \emph {et~al.}(2015)\citenamefont
  {Camargo}, \citenamefont {Anderson}, \citenamefont {Meech},\ and\
  \citenamefont {Heisler}}]{Camargo2015}%
  \BibitemOpen
  \bibfield  {author} {\bibinfo {author} {\bibfnamefont {F.~V.~A.}\
  \bibnamefont {Camargo}}, \bibinfo {author} {\bibfnamefont {H.~L.}\
  \bibnamefont {Anderson}}, \bibinfo {author} {\bibfnamefont {S.~R.}\
  \bibnamefont {Meech}},\ and\ \bibinfo {author} {\bibfnamefont {I.~A.}\
  \bibnamefont {Heisler}},\ }\bibfield  {title} {\enquote {\bibinfo {title}
  {Time-resolved twisting dynamics in a porphyrin dimer characterized by
  two-dimensional electronic spectroscopy},}\ }\href
  {https://doi.org/10.1021/acs.jpcb.5b09964} {\bibfield  {journal} {\bibinfo
  {journal} {The Journal of Physical Chemistry B}\ }\textbf {\bibinfo {volume}
  {119}},\ \bibinfo {pages} {14660--14667} (\bibinfo {year}
  {2015})}\BibitemShut {NoStop}%
\bibitem [{\citenamefont {Milota}\ \emph {et~al.}(2009)\citenamefont {Milota},
  \citenamefont {Sperling}, \citenamefont {Nemeth}, \citenamefont {Mančal},\
  and\ \citenamefont {Kauffmann}}]{Milota2009}%
  \BibitemOpen
  \bibfield  {author} {\bibinfo {author} {\bibfnamefont {F.}~\bibnamefont
  {Milota}}, \bibinfo {author} {\bibfnamefont {J.}~\bibnamefont {Sperling}},
  \bibinfo {author} {\bibfnamefont {A.}~\bibnamefont {Nemeth}}, \bibinfo
  {author} {\bibfnamefont {T.}~\bibnamefont {Mančal}},\ and\ \bibinfo {author}
  {\bibfnamefont {H.~F.}\ \bibnamefont {Kauffmann}},\ }\bibfield  {title}
  {\enquote {\bibinfo {title} {Two-dimensional electronic spectroscopy of
  molecular excitons},}\ }\href {https://doi.org/10.1021/ar800282e} {\bibfield
  {journal} {\bibinfo  {journal} {Accounts of Chemical Research}\ }\textbf
  {\bibinfo {volume} {42}},\ \bibinfo {pages} {1364–1374} (\bibinfo {year}
  {2009})}\BibitemShut {NoStop}%
\bibitem [{\citenamefont {Mandal}\ \emph {et~al.}(2018)\citenamefont {Mandal},
  \citenamefont {Chen}, \citenamefont {Foszcz}, \citenamefont {Schultz},
  \citenamefont {Kearns}, \citenamefont {Young}, \citenamefont {Zanni},\ and\
  \citenamefont {Wasielewski}}]{Mandal2018}%
  \BibitemOpen
  \bibfield  {author} {\bibinfo {author} {\bibfnamefont {A.}~\bibnamefont
  {Mandal}}, \bibinfo {author} {\bibfnamefont {M.}~\bibnamefont {Chen}},
  \bibinfo {author} {\bibfnamefont {E.}~\bibnamefont {Foszcz}}, \bibinfo
  {author} {\bibfnamefont {J.~D.}\ \bibnamefont {Schultz}}, \bibinfo {author}
  {\bibfnamefont {N.~M.}\ \bibnamefont {Kearns}}, \bibinfo {author}
  {\bibfnamefont {R.~M.}\ \bibnamefont {Young}}, \bibinfo {author}
  {\bibfnamefont {M.~T.}\ \bibnamefont {Zanni}},\ and\ \bibinfo {author}
  {\bibfnamefont {M.~R.}\ \bibnamefont {Wasielewski}},\ }\bibfield  {title}
  {\enquote {\bibinfo {title} {Two-dimensional electronic spectroscopy reveals
  excitation energy-dependent state mixing during singlet fission in a
  terrylenediimide dimer},}\ }\href {https://doi.org/10.1021/jacs.8b08627}
  {\bibfield  {journal} {\bibinfo  {journal} {Journal of the American Chemical
  Society}\ }\textbf {\bibinfo {volume} {140}},\ \bibinfo {pages}
  {17907–17914} (\bibinfo {year} {2018})}\BibitemShut {NoStop}%
\bibitem [{\citenamefont {Tekavec}\ \emph {et~al.}(2010)\citenamefont
  {Tekavec}, \citenamefont {Myers}, \citenamefont {Lewis}, \citenamefont
  {Fuller},\ and\ \citenamefont {Ogilvie}}]{Tekavec2010}%
  \BibitemOpen
  \bibfield  {author} {\bibinfo {author} {\bibfnamefont {P.~F.}\ \bibnamefont
  {Tekavec}}, \bibinfo {author} {\bibfnamefont {J.~A.}\ \bibnamefont {Myers}},
  \bibinfo {author} {\bibfnamefont {K.~L.~M.}\ \bibnamefont {Lewis}}, \bibinfo
  {author} {\bibfnamefont {F.~D.}\ \bibnamefont {Fuller}},\ and\ \bibinfo
  {author} {\bibfnamefont {J.~P.}\ \bibnamefont {Ogilvie}},\ }\bibfield
  {title} {\enquote {\bibinfo {title} {Effects of chirp on two-dimensional
  fourier transform electronic spectra},}\ }\href
  {https://doi.org/10.1364/oe.18.011015} {\bibfield  {journal} {\bibinfo
  {journal} {Optics Express}\ }\textbf {\bibinfo {volume} {18}},\ \bibinfo
  {pages} {11015} (\bibinfo {year} {2010})}\BibitemShut {NoStop}%
\bibitem [{\citenamefont {Binz}\ \emph {et~al.}(2020)\citenamefont {Binz},
  \citenamefont {Bruder}, \citenamefont {Chen}, \citenamefont {Gelin},
  \citenamefont {Domcke},\ and\ \citenamefont {Stienkemeier}}]{Binz2020}%
  \BibitemOpen
  \bibfield  {author} {\bibinfo {author} {\bibfnamefont {M.}~\bibnamefont
  {Binz}}, \bibinfo {author} {\bibfnamefont {L.}~\bibnamefont {Bruder}},
  \bibinfo {author} {\bibfnamefont {L.}~\bibnamefont {Chen}}, \bibinfo {author}
  {\bibfnamefont {M.~F.}\ \bibnamefont {Gelin}}, \bibinfo {author}
  {\bibfnamefont {W.}~\bibnamefont {Domcke}},\ and\ \bibinfo {author}
  {\bibfnamefont {F.}~\bibnamefont {Stienkemeier}},\ }\bibfield  {title}
  {\enquote {\bibinfo {title} {Effects of high pulse intensity and chirp in
  two-dimensional electronic spectroscopy of an atomic vapor},}\ }\href
  {https://doi.org/10.1364/OE.396108} {\bibfield  {journal} {\bibinfo
  {journal} {Optics Express}\ }\textbf {\bibinfo {volume} {28}},\ \bibinfo
  {pages} {25806--25829} (\bibinfo {year} {2020})}\BibitemShut {NoStop}%
\bibitem [{\citenamefont {Palato}\ \emph {et~al.}(2020)\citenamefont {Palato},
  \citenamefont {Seiler}, \citenamefont {Baker}, \citenamefont {Sonnichsen},
  \citenamefont {Zifkin}, \citenamefont {McGowan},\ and\ \citenamefont
  {Kambhampati}}]{Palato2020}%
  \BibitemOpen
  \bibfield  {author} {\bibinfo {author} {\bibfnamefont {S.}~\bibnamefont
  {Palato}}, \bibinfo {author} {\bibfnamefont {H.}~\bibnamefont {Seiler}},
  \bibinfo {author} {\bibfnamefont {H.}~\bibnamefont {Baker}}, \bibinfo
  {author} {\bibfnamefont {C.}~\bibnamefont {Sonnichsen}}, \bibinfo {author}
  {\bibfnamefont {R.}~\bibnamefont {Zifkin}}, \bibinfo {author} {\bibfnamefont
  {J.}~\bibnamefont {McGowan}},\ and\ \bibinfo {author} {\bibfnamefont
  {P.}~\bibnamefont {Kambhampati}},\ }\bibfield  {title} {\enquote {\bibinfo
  {title} {An analysis of hollow-core fiber for applications in coherent
  femtosecond spectroscopies},}\ }\href {https://doi.org/10.1063/1.5113691}
  {\bibfield  {journal} {\bibinfo  {journal} {Journal of Applied Physics}\
  }\textbf {\bibinfo {volume} {128}},\ \bibinfo {pages} {103107} (\bibinfo
  {year} {2020})}\BibitemShut {NoStop}%
\bibitem [{\citenamefont {Béjot}\ \emph {et~al.}(2010)\citenamefont {Béjot},
  \citenamefont {Schmidt}, \citenamefont {Kasparian}, \citenamefont {Wolf},\
  and\ \citenamefont {Legaré}}]{Bjot2010}%
  \BibitemOpen
  \bibfield  {author} {\bibinfo {author} {\bibfnamefont {P.}~\bibnamefont
  {Béjot}}, \bibinfo {author} {\bibfnamefont {B.~E.}\ \bibnamefont {Schmidt}},
  \bibinfo {author} {\bibfnamefont {J.}~\bibnamefont {Kasparian}}, \bibinfo
  {author} {\bibfnamefont {J.-P.}\ \bibnamefont {Wolf}},\ and\ \bibinfo
  {author} {\bibfnamefont {F.}~\bibnamefont {Legaré}},\ }\bibfield  {title}
  {\enquote {\bibinfo {title} {Mechanism of hollow-core-fiber
  infrared-supercontinuum compression with bulk material},}\ }\href
  {https://doi.org/10.1103/physreva.81.063828} {\bibfield  {journal} {\bibinfo
  {journal} {Physical Review A}\ }\textbf {\bibinfo {volume} {81}} (\bibinfo
  {year} {2010}),\ 10.1103/physreva.81.063828}\BibitemShut {NoStop}%
\bibitem [{\citenamefont {Szipöcs}\ \emph {et~al.}(1994)\citenamefont
  {Szipöcs}, \citenamefont {Ferencz}, \citenamefont {Spielmann},\ and\
  \citenamefont {Krausz}}]{Szipocs1994}%
  \BibitemOpen
  \bibfield  {author} {\bibinfo {author} {\bibfnamefont {R.}~\bibnamefont
  {Szipöcs}}, \bibinfo {author} {\bibfnamefont {K.}~\bibnamefont {Ferencz}},
  \bibinfo {author} {\bibfnamefont {C.}~\bibnamefont {Spielmann}},\ and\
  \bibinfo {author} {\bibfnamefont {F.}~\bibnamefont {Krausz}},\ }\bibfield
  {title} {\enquote {\bibinfo {title} {Chirped multilayer coatings for
  broadband dispersion control in femtosecond lasers},}\ }\href
  {https://doi.org/10.1364/OL.19.000201} {\bibfield  {journal} {\bibinfo
  {journal} {Optics Letters}\ }\textbf {\bibinfo {volume} {19}},\ \bibinfo
  {pages} {201--203} (\bibinfo {year} {1994})}\BibitemShut {NoStop}%
\bibitem [{\citenamefont {Zavelani-Rossi}\ \emph {et~al.}(2001)\citenamefont
  {Zavelani-Rossi}, \citenamefont {Cerullo}, \citenamefont {De~Silvestri},
  \citenamefont {Gallmann}, \citenamefont {Matuschek}, \citenamefont
  {Steinmeyer}, \citenamefont {Keller}, \citenamefont {Angelow}, \citenamefont
  {Scheuer},\ and\ \citenamefont {Tschudi}}]{Zavelani-Rossi2001}%
  \BibitemOpen
  \bibfield  {author} {\bibinfo {author} {\bibfnamefont {M.}~\bibnamefont
  {Zavelani-Rossi}}, \bibinfo {author} {\bibfnamefont {G.}~\bibnamefont
  {Cerullo}}, \bibinfo {author} {\bibfnamefont {S.}~\bibnamefont
  {De~Silvestri}}, \bibinfo {author} {\bibfnamefont {L.}~\bibnamefont
  {Gallmann}}, \bibinfo {author} {\bibfnamefont {N.}~\bibnamefont {Matuschek}},
  \bibinfo {author} {\bibfnamefont {G.}~\bibnamefont {Steinmeyer}}, \bibinfo
  {author} {\bibfnamefont {U.}~\bibnamefont {Keller}}, \bibinfo {author}
  {\bibfnamefont {G.}~\bibnamefont {Angelow}}, \bibinfo {author} {\bibfnamefont
  {V.}~\bibnamefont {Scheuer}},\ and\ \bibinfo {author} {\bibfnamefont
  {T.}~\bibnamefont {Tschudi}},\ }\bibfield  {title} {\enquote {\bibinfo
  {title} {Pulse compression over a 170-thz bandwidth in the visible by use of
  only chirped mirrors},}\ }\href {https://doi.org/10.1364/OL.26.001155}
  {\bibfield  {journal} {\bibinfo  {journal} {Optics Letters}\ }\textbf
  {\bibinfo {volume} {26}},\ \bibinfo {pages} {1155--1157} (\bibinfo {year}
  {2001})}\BibitemShut {NoStop}%
\bibitem [{\citenamefont {Bor}\ and\ \citenamefont {Rácz}(1985)}]{Bor1985}%
  \BibitemOpen
  \bibfield  {author} {\bibinfo {author} {\bibfnamefont {Z.}~\bibnamefont
  {Bor}}\ and\ \bibinfo {author} {\bibfnamefont {B.}~\bibnamefont {Rácz}},\
  }\bibfield  {title} {\enquote {\bibinfo {title} {Extremely simple
  single-prism ultrashort-pulse compressor},}\ }\href
  {https://doi.org/10.1016/0030-4018(85)90284-6} {\bibfield  {journal}
  {\bibinfo  {journal} {Optics Communications}\ }\textbf {\bibinfo {volume}
  {54}},\ \bibinfo {pages} {165--170} (\bibinfo {year} {1985})}\BibitemShut
  {NoStop}%
\bibitem [{\citenamefont {Akturk}\ \emph {et~al.}(2006)\citenamefont {Akturk},
  \citenamefont {Gu}, \citenamefont {Kimmel},\ and\ \citenamefont
  {Trebino}}]{Akturk2006}%
  \BibitemOpen
  \bibfield  {author} {\bibinfo {author} {\bibfnamefont {S.}~\bibnamefont
  {Akturk}}, \bibinfo {author} {\bibfnamefont {X.}~\bibnamefont {Gu}}, \bibinfo
  {author} {\bibfnamefont {M.}~\bibnamefont {Kimmel}},\ and\ \bibinfo {author}
  {\bibfnamefont {R.}~\bibnamefont {Trebino}},\ }\bibfield  {title} {\enquote
  {\bibinfo {title} {Extremely simple single-prism ultrashort-pulse
  compressor},}\ }\href {https://doi.org/10.1364/OE.14.010101} {\bibfield
  {journal} {\bibinfo  {journal} {Optics Express}\ }\textbf {\bibinfo {volume}
  {14}},\ \bibinfo {pages} {10101--10108} (\bibinfo {year} {2006})}\BibitemShut
  {NoStop}%
\bibitem [{\citenamefont {Treacy}(1969)}]{Treacy1969}%
  \BibitemOpen
  \bibfield  {author} {\bibinfo {author} {\bibfnamefont {E.}~\bibnamefont
  {Treacy}},\ }\bibfield  {title} {\enquote {\bibinfo {title} {Optical pulse
  compression with diffraction gratings},}\ }\href
  {https://doi.org/10.1109/JQE.1969.1076303} {\bibfield  {journal} {\bibinfo
  {journal} {IEEE Journal of Quantum Electronics}\ }\textbf {\bibinfo {volume}
  {5}},\ \bibinfo {pages} {454 -- 458} (\bibinfo {year} {1969})}\BibitemShut
  {NoStop}%
\bibitem [{\citenamefont {Gibson}\ \emph {et~al.}(2006)\citenamefont {Gibson},
  \citenamefont {Gaudiosi}, \citenamefont {Kapteyn}, \citenamefont {Jimenez},
  \citenamefont {Kane}, \citenamefont {Huff}, \citenamefont {Durfee},\ and\
  \citenamefont {Squier}}]{Gibson2006}%
  \BibitemOpen
  \bibfield  {author} {\bibinfo {author} {\bibfnamefont {E.~A.}\ \bibnamefont
  {Gibson}}, \bibinfo {author} {\bibfnamefont {D.~M.}\ \bibnamefont
  {Gaudiosi}}, \bibinfo {author} {\bibfnamefont {H.~C.}\ \bibnamefont
  {Kapteyn}}, \bibinfo {author} {\bibfnamefont {R.}~\bibnamefont {Jimenez}},
  \bibinfo {author} {\bibfnamefont {S.}~\bibnamefont {Kane}}, \bibinfo {author}
  {\bibfnamefont {R.}~\bibnamefont {Huff}}, \bibinfo {author} {\bibfnamefont
  {C.}~\bibnamefont {Durfee}},\ and\ \bibinfo {author} {\bibfnamefont
  {J.}~\bibnamefont {Squier}},\ }\bibfield  {title} {\enquote {\bibinfo {title}
  {Efficient reflection grisms for pulse compression and dispersion
  compensation of femtosecond pulses},}\ }\href
  {https://doi.org/10.1364/OL.31.003363} {\bibfield  {journal} {\bibinfo
  {journal} {Optics Letters}\ }\textbf {\bibinfo {volume} {31}},\ \bibinfo
  {pages} {3363--3365} (\bibinfo {year} {2006})}\BibitemShut {NoStop}%
\bibitem [{\citenamefont {Weiner}(2000)}]{Weiner2000}%
  \BibitemOpen
  \bibfield  {author} {\bibinfo {author} {\bibfnamefont {A.~M.}\ \bibnamefont
  {Weiner}},\ }\bibfield  {title} {\enquote {\bibinfo {title} {Femtosecond
  pulse shaping using spatial light modulators},}\ }\href
  {https://doi.org/10.1063/1.1150614} {\bibfield  {journal} {\bibinfo
  {journal} {Review of Scientific Instruments}\ }\textbf {\bibinfo {volume}
  {71}},\ \bibinfo {pages} {1929–1960} (\bibinfo {year} {2000})}\BibitemShut
  {NoStop}%
\bibitem [{\citenamefont {Salzmann}\ \emph {et~al.}(2008)\citenamefont
  {Salzmann}, \citenamefont {Mullins}, \citenamefont {Eng}, \citenamefont
  {Albert}, \citenamefont {Wester}, \citenamefont {Weidemüller}, \citenamefont
  {Merli}, \citenamefont {Weber}, \citenamefont {Sauer}, \citenamefont
  {Plewicki}, \citenamefont {Weise}, \citenamefont {Wöste},\ and\
  \citenamefont {Lindinger†}}]{Salzmann2008}%
  \BibitemOpen
  \bibfield  {author} {\bibinfo {author} {\bibfnamefont {W.}~\bibnamefont
  {Salzmann}}, \bibinfo {author} {\bibfnamefont {T.}~\bibnamefont {Mullins}},
  \bibinfo {author} {\bibfnamefont {J.}~\bibnamefont {Eng}}, \bibinfo {author}
  {\bibfnamefont {M.}~\bibnamefont {Albert}}, \bibinfo {author} {\bibfnamefont
  {R.}~\bibnamefont {Wester}}, \bibinfo {author} {\bibfnamefont
  {M.}~\bibnamefont {Weidemüller}}, \bibinfo {author} {\bibfnamefont
  {A.}~\bibnamefont {Merli}}, \bibinfo {author} {\bibfnamefont {S.~M.}\
  \bibnamefont {Weber}}, \bibinfo {author} {\bibfnamefont {F.}~\bibnamefont
  {Sauer}}, \bibinfo {author} {\bibfnamefont {M.}~\bibnamefont {Plewicki}},
  \bibinfo {author} {\bibfnamefont {F.}~\bibnamefont {Weise}}, \bibinfo
  {author} {\bibfnamefont {L.}~\bibnamefont {Wöste}},\ and\ \bibinfo {author}
  {\bibfnamefont {A.}~\bibnamefont {Lindinger†}},\ }\bibfield  {title}
  {\enquote {\bibinfo {title} {Coherent transients in the femtosecond
  photoassociation of ultracold molecules},}\ }\href
  {https://doi.org/10.1103/PhysRevLett.100.233003} {\bibfield  {journal}
  {\bibinfo  {journal} {Physical Review Letters}\ }\textbf {\bibinfo {volume}
  {100}} (\bibinfo {year} {2008}),\ 10.1103/PhysRevLett.100.233003}\BibitemShut
  {NoStop}%
\end{thebibliography}%

\appendix
\section{Derivation of the 2D Signal Final Expression}
\label{app:Derivation Final Expression}

We derive here the expression for the emitted 2D signal, cf. Eq.~\eqref{eq:Signal Final Expression}, starting from Eq.~\eqref{eq:Response Function}.
Defining
\begin{eqnarray}
    R_{j, \, k} \, ( t ) &=& \exp \left( - i \omega_k^{(j)} t - \frac{t}{T_{k}^{(j)}} \right) \; , \quad \forall k \in \llbracket 1, \, 3 \rrbracket \, , \nonumber \\
    |R_j| &=& \frac{1}{\hbar^{3}} \: \mu_{S}^{(j)} \, \mu_{3}^{(j)} \, \mu_{2}^{(j)} \, \mu_{1}^{(j)} \; ,
    \label{eq:Response Terms (App.)}
\end{eqnarray}
Eq.~\eqref{eq:Response Function} can be rephrased as 
\begin{widetext}
\begin{eqnarray}
    \label{eq:Partial Response Function (App.)}
    R^{(3)}_j ( t - u , \: u - v , \: v - w )
    &=& |R_j| \; R_{j, \, 3} \, ( t - u ) \; R_{j, \, 2} \, ( u - v ) \, R_{j, \, 1} \, ( v - w ) \\
    &=& R_{j}^{(3)} ( t - t_{\sigma_3} , \: t_{\sigma_3} - t_{\sigma_2} , \: t_{\sigma_2} - t_{\sigma_1}  ) \; R_{j , \, 3} \, ( t_{\sigma_3} - u ) \; R_{j, \, 2} \, ( u - t_{\sigma_3} ) \; R_{j, \, 2} \, ( t_{\sigma_2} - v ) \; R_{j, \, 1} \, ( v - t_{\sigma_2} ) \; R_{j, \, 1} \, ( t_{\sigma_1} - w ) \; , \nonumber
\end{eqnarray}
\end{widetext}
where we have used that all the \( R_{j, \, k} \) terms are linear exponentials obeying \( R_{j, \, k} \, (a + b) = R_{j, \, k} \, (a) \; R_{j, \: k} \, (b) \).
This allows for writing the partial response function \( R^{(3)}_j \) as a product of terms, each one depending on a single integration variable \( ( u, \, v, \, w ) \), and one leading term independent of all three.
The first term in Eq.~\eqref{eq:Partial Response Function (App.)}, \( R_{j}^{(3)} ( t - t_{\sigma_3} , \: t_{\sigma_3} - t_{\sigma_2} , \: t_{\sigma_2} - t_{\sigma_1} ) \),  corresponds to the impulsive limit of the partial response function, only dependent on the arrival times of the pulses.
The other terms can be seen as independent corrections to the impulsive limit that modify the weight of the pulses taking into account the simultaneous system dynamics (coherence and decay). These terms can thus be collected with their associated pulses into one function each, obtaining the functions \( F_{j, \, \sigma} \) in Eq.~\eqref{eq:Signal Final Expression}.
Substituting Eq.~\eqref{eq:Partial Response Function (App.)} into Eq.~\eqref{eq:Contribution Expression} and using the functions \( F_{j, \, \sigma} \), we finally obtain Eq.~\eqref{eq:Signal Final Expression}. In this expression of the emitted signal, the complicated dependencies of the partial response function \( R_{j}^{(3)} \) and of the integral contribution \( I_{j, \, \sigma}^{(3)} \) on the time-ordered arrival times of the pulses \( ( t_{\sigma_3} , \: t_{\sigma_2} , \: t_{\sigma_1} ) \) have been simplified using the subscript \( \sigma \) and the delays \( ( t , \: T , \: \tau ) \).

Additionally, using Eq.~\eqref{eq:Response Terms (App.)}, the functions \( F_{j, \, \sigma} \) can be rewritten as
\begin{widetext}
\begin{equation}
    F^{x_k}_{j, \, \sigma} \, ( x_k - t_{\sigma_k} ) = E_{\sigma_k}^{\mathit{shape}} ( x_k - t_{\sigma_k} ) \; \exp \left( - i \left( \omega_{\sigma_k}^{E} - \Delta \omega^{(j)}_{k} \right) ( x_k - t_{\sigma_k} ) \right) \; \exp \left( - \frac{x_k - t_{\sigma_k}}{\Delta T^{(j)}_{k}} \right) \; , \quad \forall k \in \llbracket 1, \, 3 \rrbracket ,
    \label{eq:Elegant FieldeShape (App.)}
\end{equation}    
\end{widetext}
with
\begin{eqnarray*}
x_k &\in& \{w , \, v , \, u\} \; , \\
\Delta \omega^{(j)}_{k} &=& \omega^{(j)}_{k} - \omega^{(j)}_{k - 1} \; , \\
\frac{1}{\Delta T^{(j)}_{k}} &=& \frac{1}{T^{(j)}_{k}} - \frac{1}{T^{(j)}_{k - 1}} \; , \\
E_{\sigma_k} \, (x_k - t_{\sigma_k}) &=& E_{\sigma_k}^{\mathit{shape}} (x - t_{\sigma_k}) \; \exp \left( - i \omega_{\sigma_k}^{E} ( x - t_{\sigma_k} ) \right) \; .   
\end{eqnarray*}
In Eq.~\eqref{eq:Elegant FieldeShape (App.)}, three contributions can be identified. The first one is the contribution of the pulse shape, which is generally described in terms of spectral shape and spectral phase in the Fourier space. This term is playing a major role in the integration process. The second term is the contribution of the mismatch between the pulse center energy and the intended transition \( \Delta \omega \), which is rapidly averaging to zero for non-resonant interactions. The third term is the contribution of the system's dynamics, also called the lineshape contribution, which is modeled according to the Markovian approximation for the environment. Note that to obtain this last expression, we used the rotating wave approximation (RWA) to discard the far off-resonance terms.

 \section{Computational Methods - Details}
\label{app:Computational Methods}

To compute the signal \( E_{S, \, j, \, \sigma} \) in Eq.~\eqref{eq:Signal Final Expression}, the partial response function \( R_j \) and integral contributions \( I_{j, \, \sigma} \) are evaluated independently.
For the partial response functions, we have developed an algorithm to obtain all light-matter interaction pathways for a given system. The selected pathways are then computed according to Eq.~\eqref{eq:Response Function}.
For the integral contributions, the first step is to evaluate the fields \( \left( E^{shape}_{\sigma_1} , \, E^{shape}_{\sigma_2} , \, E^{shape}_{\sigma_3} \right) \), as per Eq.~\ref{eq:Elegant FieldeShape (App.)}. The sampling of the fields is performed by the user over \( N_S \) points with typically similar samplings being employed for all pulses. The sampling is straightforward as the shape function of a pulse rapidly fall to zero on both sides of the arrival time of the pulse. Following that, we obtain the functions \( \left( F^{w}_{j, \, \sigma} \, , \, F^{v}_{j, \, \sigma} \, , \, F^{u}_{j, \, \sigma} \right) \) according to Eq.~\eqref{eq:Elegant FieldeShape (App.)}. For a given pathway \( j \) and time-ordering \( \sigma \), we proceed to integrate the functions \( F_{j, \, \sigma} \) according to Eq.~\eqref{eq:Signal Final Expression}. We therefore obtain \( I_{j, \, \sigma} \) over a fine grid, which is typically composed of thousands of points to ensure accurate field sampling. \( I_{j, \, \sigma} \) is then resampled over a smaller number of user-defined time points \( N_t \), typically composed of hundreds of points. Multiplying \( I_{j, \, \sigma} \) by the partial response function \( R_{\sigma , \, j} \) and the initial state \( \rho_j^{(0)} \), we obtain one signal contribution \( E_{S, \, j, \, \sigma}^{(3)} \) according to Eq.~\eqref{eq:Elegant FieldeShape (App.)}.
Summing these contributions, the complete signal \( E_{S}^{(3)} \) is obtained over time axis \( t \) for fixed chosen delays \( ( \tau , \: T ) \). We repeat this procedure for a grid of delays \( \tau \) and \( T \), which can be chosen accordingly to the experimentally-used delays for instance. The user-defined range and spacing of the three axes \( ( \tau , \: T , \: t ) \) should ensure proper sampling of the dephasing and population dynamics. A Fourier transform along both axes \( \tau \) and \( t \) is performed to obtain a 2D spectrum \( S_{2D} \) at a given waiting time \( T \). Often, the 2D spectrum \( S_{2D} \) is shown as a function of energy coordinates \( ( E_{\tau} , \: T , \: E_{t} ) \) instead of angular momenta \( ( \omega_{\tau} , \: T , \: \omega_{t} ) \), where \( ( E_{\tau} = \hbar \omega_{\tau} ) \) and \( ( E_{t} = \hbar \omega_{t} ) \).
To arrive at the full 2D signal, the process described above can be visually represented. First, the pulses are computed to obtain the emitted signal \( E_{S}^{(3)} \), over time axis \( t \) at fixed delays \( ( \tau , \: T ) \). To obtain the full plot of the emitted signal over times axes \( ( \tau , \: t ) \), the step described previously are repeated over varying delays \( \tau \). This first computed signal over time axis \( t \) then corresponds to one of the columns of this plot, still at fixed delay \( T \). This computed signal over times axes \( ( \tau , \: t ) \) is divided into non-rephasing and rephasing parts, usually represented in two side-by-side plots. These two plots are summed and Fourier-transformed along both time axis \( \tau \) and \( t \) according to Eq.~\eqref{eq:2D Signal Final} to obtain one 2D spectrum, \( S_{2D} \), over angular frequencies \( ( \omega_{\tau} , \: \omega_{t} ) \), still at fixed delay \( T \). To follow the temporal evolution of the 2D spectrum, all the previous steps are repeated over time axis \( T \) to obtain the full 2D signal \( S_{2D} ( \omega_{\tau} , \: T , \: \omega_{t} ) \).

We describe here explicitly why the nested integrals of Eq.~\eqref{eq:Signal Final Expression} can be evaluated in linear time. The first step is to evaluate the first function \( F^{w}_{j, \, \sigma} \) on a set of time points \( w \), chosen around arrival time \( t_{\sigma_1} \) where the field \( E_{\sigma_1} \) is non-zero. This function is then cumulatively integrated over \( w \). The second step is to evaluate the second function \( F^{v}_{j, \, \sigma} \) on a set of time points \( v \), chosen around arrival time \( t_{\sigma_2} \) where the field \( E_{\sigma_2} \) is non-zero. Then, the values obtained are multiplied with the values of the first cumulative integration, and then cumulatively integrated over \( v \). The same process is repeated for the third and last step. The third function \( F^{u}_{j, \, \sigma} \) is evaluated, multiplied and cumulatively integrated over a set of time points \( u \). After each cumulative integration, the results are interpolated to fit the time points of the subsequent step. The interpolation steps are straightforward because the functions \( F_{j, \, \sigma} \) rapidly fall to zero far from the arrival times of the pulses, in the same manner as the pulse shapes \( E^{shape}_{\sigma_k} \). Evaluating the integral contribution \( I_{j , \, \sigma}^{(3)} \) with linear complexity is a key step that makes our approach efficient. This process is rigorous because, in Eq.~\eqref{eq:Signal Final Expression}, the integration variables \( ( u, \, v, \, w ) \) appear in a separated manner inside the nested integrals, and because the corresponding functions \( \left( F^{u}_{j, \, \sigma} \, , \; F^{v}_{j, \, \sigma} \, , \; F^{w}_{j, \, \sigma} \right) \) rapidly fall to zero on both sides of the arrival time of the pulse.

As discussed in Sec.~\ref{sec:Computational Methods}, though the numerical integration is of linear complexity, evaluating the 2D signal \( S_{2D} \) over all three varying time variables \( ( \tau , \: T , \: t ) \) scales as \( O \bigl( J \cdot N_\tau \cdot N_T \cdot N_S \bigr) \). As a reference, for the most computationally-intensive results presented in Sec.~\ref{sec:Results}, we used \( J = 40 \) (six-level dimer), with our six different time-orderings \( \sigma \) for each of these pathways, \( N_\tau = N_t = 256 \) and \( N_S = 1024 \). Such a simulation can run in a few minutes on a standard laptop, which, in comparison, would require up to days of runtime to precisely evaluate triple integral expressions.

\begin{figure}[H]
    \centering
    \includegraphics[width=1\linewidth]{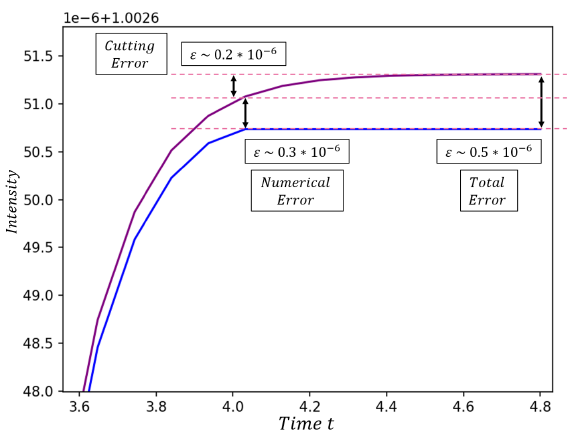}
    \caption{Zoom of the plotting of the integration process with chirped gaussian pulse shapes. The numerical evaluation (in blue) is compared to the analytical evaluation (in purple). The two sources of numerical error are represented: from top to bottom, the first one is the cutting error as the pulse shapes are integrated over a finite interval around their arrival time. The second one is the integration error, arising from the numerical discretization of the interval of integration. The total relative error is \( 0.5 \cdot 10^{-6} \).}
    \label{fig:Integration Inaccuracy}
\end{figure}

As discussed in Sec.~\ref{sec:Computational Methods}, the sampling of the pulses and the subsequent evaluation of the nested integrals of Eq.~\eqref{eq:Signal Final Expression} is the only source of numerical inaccuracy. This inaccuracy is strongly bounded when using fine samplings \( N_S \) of the pulses with thousands of points. In addition, our implementation is flexible, allowing us to use any numerical integration routine. As a reference, when simulating gaussian and chirped gaussian pulse shapes such as those studied in Sec.~\ref{sec:Results}, the relative inaccuracy of the integration process was below 10$^{-6}$, as illustrated in Fig.~\ref{fig:Integration Inaccuracy}.

The computational complexity discussed above refers only to the evaluation of the 2D time-domain signal \( S(\tau,T,t) \). The complete computation of the 2D spectra additionally requires the evaluation of two Fourier transforms along the \( \tau \) and \( t \) axes, introducing an additional cost of \( O(N_{\tau} \cdot log_2 (N_{\tau} \cdot N_T \cdot N_s)) + O(N_{\tau} \cdot N_T \cdot N_s \cdot log_2 (N_s)) \). For the typical parameters  \( N_\tau = N_t = 256 \) and \( N_S = 1024 \), the runtime of the Fourier-transforms is of the same order as that of the signal evaluation itself, indicating that the proposed algorithm reduces the cost of the numerical integration to the point where the overall runtime is largely determined by the unavoidable Fourier transforms.

\onecolumngrid

\section{System parameters I}
\label{app:model_systems}

In this section we report the values of the parameters used in our simulations in Sec.~\ref{sec:Results} to model the three quantum systems described above: the anharmonic oscillator, the two-excitons and the dimer systems. Their corresponding spectra are shown below in terms of energy levels.

\[
\begin{array}{| w{c}{4cm} | w{c}{4cm} | w{c}{3cm}  w{c}{3cm} |}
\hline
\textbf{Anharmonic oscillator} &
\textbf{Two-excitons} &
\multicolumn{2}{c|}{\textbf{Dimer}} \\
\hline
E_0=\SI{0}{\electronvolt} &
E_0=\SI{0}{\electronvolt} &
E_0=\SI{0}{\electronvolt} &
E_3=\SI{3.68}{\electronvolt} \\[0.3em]

E_1=\SI{1.94}{\electronvolt} &
E_1=\SI{1.90}{\electronvolt} &
E_1=\SI{1.86}{\electronvolt} &
E_4=\SI{3.83}{\electronvolt} \\[0.3em]

E_2=\SI{3.84}{\electronvolt} &
E_2=\SI{1.94}{\electronvolt} &
E_2=\SI{1.98}{\electronvolt} &
E_5=\SI{3.92}{\electronvolt} \\
\hline
\end{array}
\]

Below we report the dipole operators $\hat{\mu}$ that describe the light–matter interactions for each one of the three systems.

\[
\begin{array}{c@{\hspace{2cm}}c@{\hspace{2cm}}c}
\textbf{Anharmonic oscillator} & \textbf{Two-exciton} & \textbf{Dimer} \\[6pt]
{\hat{\mu}}_{AO} =
\begin{pmatrix}
0 &     1    &      0    \\
1 &     0    &  \sqrt{2} \\
0 & \sqrt{2} &      0
\end{pmatrix}
&
{\hat{\mu}}_{2X} =
\begin{pmatrix}
0 &     1    &      1    \\
1 &     0    &      0    \\
1 &     0    &      0
\end{pmatrix}
&
{\hat{\mu}}_{DI} =
\begin{pmatrix}
0 &     1    &      1     &      0     &      0     &      0     &\\
1 &     0    &      0     &  \sqrt{2}  &      1     &      0     &\\
1 &     0    &      0     &      0     &      1     &   \sqrt{2} &\\
0 & \sqrt{2} &      0     &      0     &      0     &      0     &\\
0 &     1    &      1     &      0     &      0     &      0     &\\
0 &     0    &   \sqrt{2} &      0     &      0     &      0     &
\end{pmatrix}

\end{array}
\]

Finally, for convenience, we report the lifetimes of the excited-state populations and of the coherences generated by the interaction between the excitation pulse and the material in a compact form as an operator $\hat{T}$.

\[
\begin{array}{c@{\hspace{0.9cm}}c@{\hspace{0.9cm}}c}
\textbf{Anharmonic oscillator} & \textbf{Two-exciton system} & \textbf{Dimer} \\[6pt]
{\hat{T}}_{AO} =
\begin{pmatrix}
500 &     100    &      80    \\
100 &     500    &      100   \\
80  &     100    &      500
\end{pmatrix} \text{fs}
&
{\hat{T}}_{2X} =
\begin{pmatrix}
200 &     80    &      80    \\
80  &    200    &      40    \\
80  &     40    &      200
\end{pmatrix} \text{fs}
&
{\hat{T}}_{DI} =
\begin{pmatrix}
150.00 &    75.00&      75.00 &      25.00  &      25.00  &      25.00 &\\
75.00  &   150.00&      37.50 &      50.00  &      50.00  &      18.75 &\\
75.00  &    37.50&      150.00&      18.75  &      50.00  &      50.00 &\\
25.00  &    50.00&      18.75 &      150.00 &      21.43  &      12.50 &\\
25.00  &    50.00&      50.00 &      21.43  &      150.00 &      21.43 &\\
25.00  &    18.75&      50.00 &      12.50  &      21.43  &     150.00 &
\end{pmatrix} \text{fs}

\end{array}
\]

\section{System parameters II}
\label{app:pulse_flat} 

Then we report the lifetimes of the excited-state populations and of the coherences used to model the three systems in Sec.~\ref{sec:Results} C.

\[
\begin{array}{c@{\hspace{0.9cm}}c@{\hspace{0.9cm}}c}
\textbf{Anharmonic oscillator} & \textbf{Two-exciton system} & \textbf{Dimer} \\[6pt]
{\hat{T}}_{AO} =
\begin{pmatrix}
400 &     140    &      70    \\
140 &     400    &      70   \\
70  &     70    &      400
\end{pmatrix} \text{fs}
&
{\hat{T}}_{2X} =
\begin{pmatrix}
200 &     80    &      80    \\
80  &    200    &      40    \\
80  &     40    &      200
\end{pmatrix} \text{fs}
&
{\hat{T}}_{DI} =
\begin{pmatrix}
300.00 &    150.00&      150.00 &      50.00  &      50.00  &      50.00 &\\
150.00  &   300.00&      37.50 &      100.00  &      100.00  &      37.50 &\\
150.00  &    37.50&      300.00&      37.50  &      100.00  &      100.00 &\\
50.00  &    100.00&      37.50 &      300.00 &      42.83  &      25 &\\
50.00  &    100.00&      100.00 &       42.83  &      300.00 &       42.83 &\\
50.00  &    37.50&      100.00 &      25  &      42.83  &     300.00 &
\end{pmatrix} \text{fs}

\end{array}
\]

\section{System parameters III}
\label{app:param_III}

Below we report the lifetimes of the excited-state populations and of the coherences used to model the two excitons system in Sec.~\ref{sec:Results} D.

\begin{equation*}
\begin{aligned}
\hat{T}_{2X} &=
\begin{pmatrix}
500 & 250 & 140 \\
250 & 500 & 70 \\
140 & 70 & 500
\end{pmatrix} \text{fs}
\end{aligned}
\end{equation*}

\section{Pulse shape and spectral phase effects}
\label{app:pulse_shape}

\begin{figure}[H]
    \centering
    \includegraphics[width=6.8in,height=6.12in]{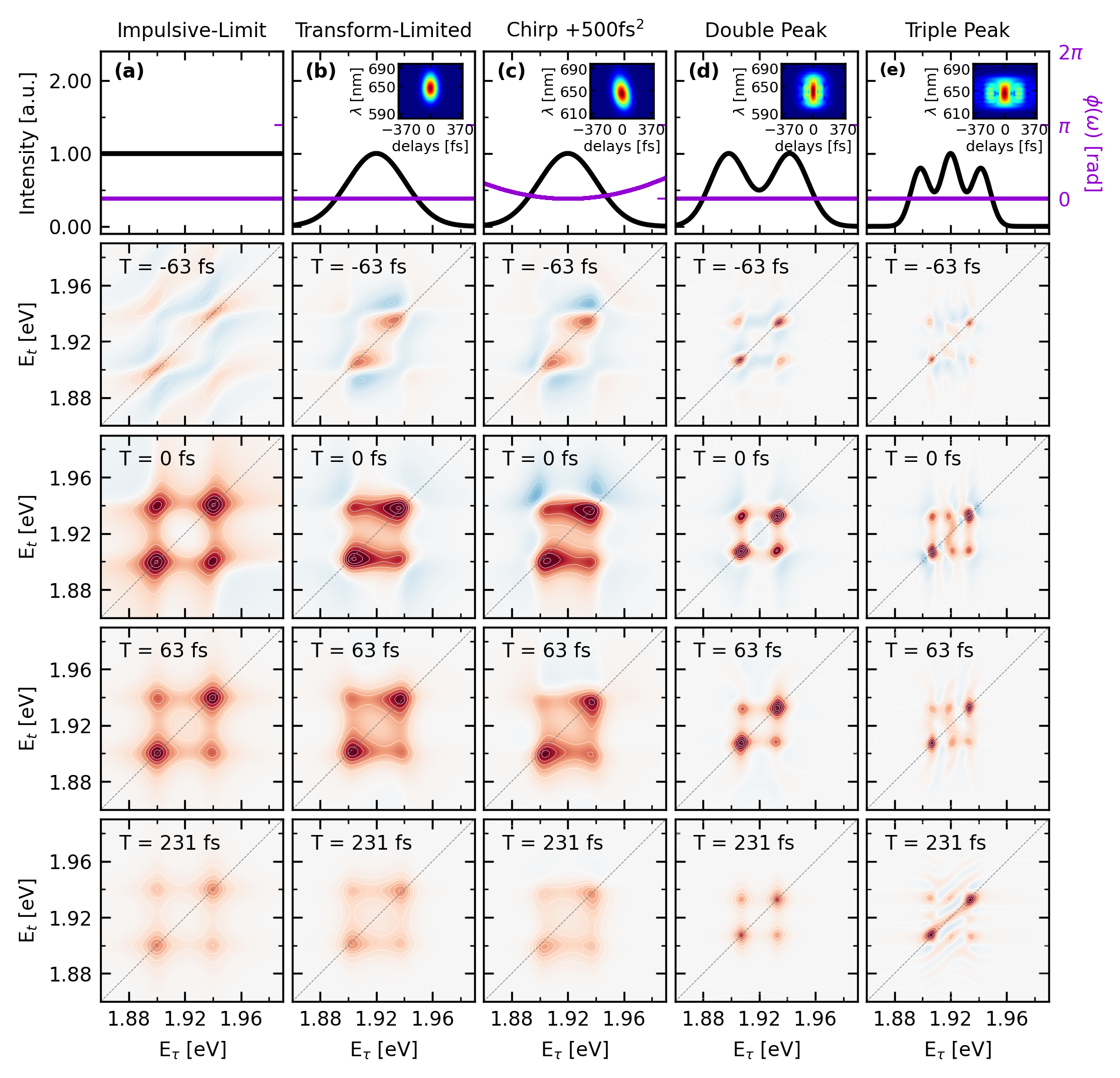}
    \vspace{-10pt}
    \caption{Influence of the pulse shape on the 2D spectra of the two-exciton system. Each column corresponds to a different excitation condition: (a) impulsive limit (IL), (b) Gaussian transform-limited pulse (TL), (c) chirped Gaussian pulse (CH), (d) double-peaked pulse with flat spectral phase, and (e) triple-peaked pulse with flat spectral phase. The first panel of each column shows the spectral amplitude and phase of the corresponding pulse (purple line), together with the associated FROG trace (inset). The panels below display the temporal evolution of the calculated 2D spectra. The same pump and probe pulse shapes are used in each simulation. The color scale is normalized column-wise for each column. The double-peak has $FWHM \approx 57 fs$ while the triple-peak has $FWHM \approx 80 fs$}
    \label{fig:EX}
\end{figure}

\begin{figure}[H]
    \centering
    \includegraphics[width=6.8in,height=6.12in]{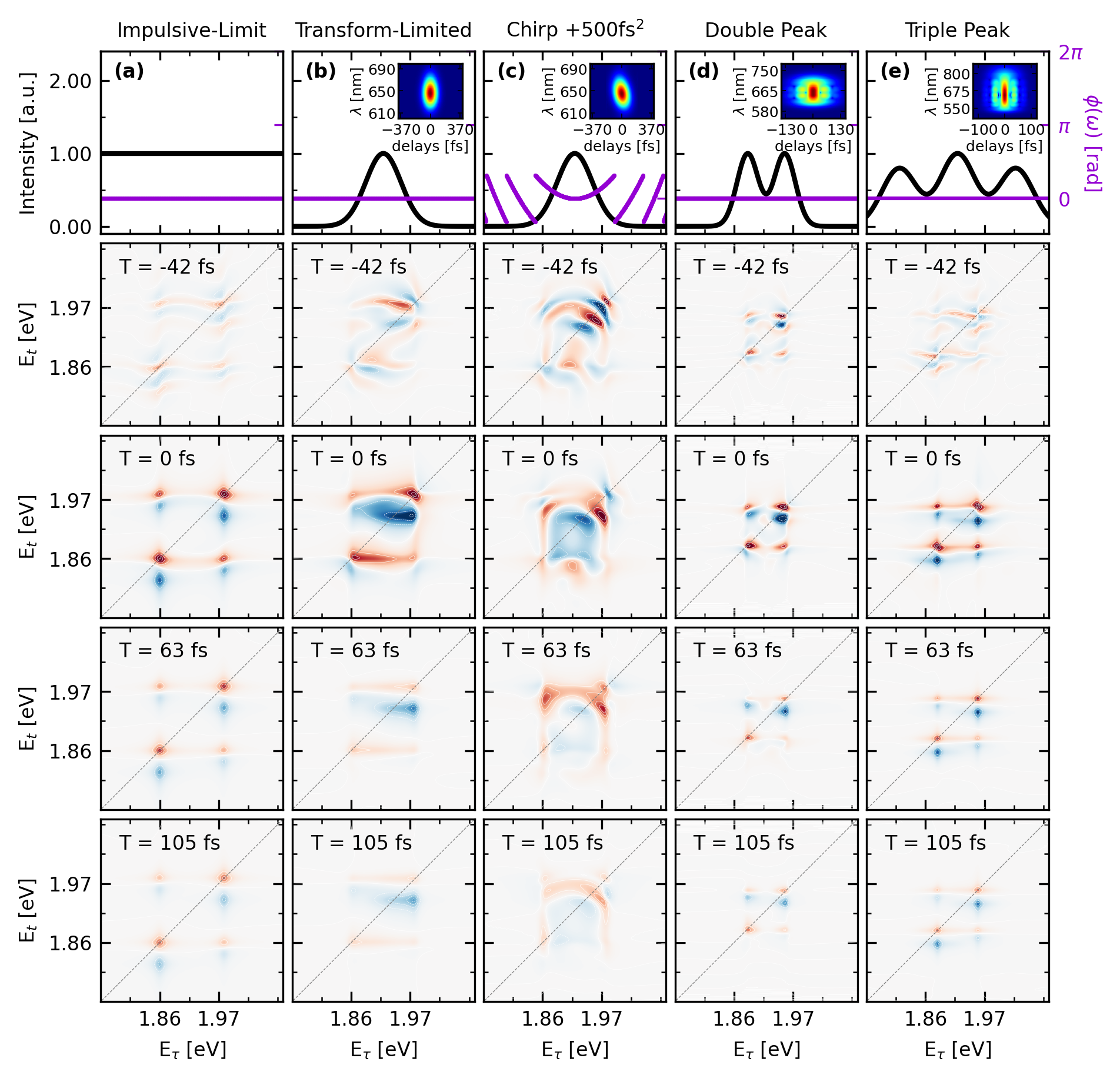}
    \vspace{-10pt}
    \caption{Influence of the pulse shape on the 2D spectra of the dimer system. Each column corresponds to a different excitation condition: (a) impulsive limit (IL), (b) Gaussian transform-limited pulse (TL), (c) chirped Gaussian pulse (CH), (d) double-peaked pulse with flat spectral phase, and (e) triple-peaked pulse with flat spectral phase. The first panel of each column shows the spectral amplitude and phase of the corresponding pulse (purple line), together with the associated FROG trace (inset). The panels below display the temporal evolution of the simulated 2D spectra. The same pump and probe pulse shapes are used in each simulation. The color scale is normalized column-wise for each column. The double-peak has $FWHM \approx 80 fs$ while the triple-peak has $FWHM \approx 36 fs$}
    \label{fig:DI}
\end{figure}

In this section, we present two studies analogous to that shown in Fig.~\ref{fig:AO} for the anharmonic oscillator, focusing instead on the two-exciton in Fig.~\ref{fig:EX} and dimer models in Fig.~\ref{fig:DI}. As can be seen, the analysis carried out for the anharmonic oscillator leads to conclusions that are essentially reproduced in these two cases. The behavior observed in the two-exciton and dimer models closely mirrors that reported for the anharmonic oscillator, indicating that the main trends and underlying mechanisms are largely independent of the specific model considered.

\newpage
\section{Region of interest (ROI) analysis}
\label{app:ROI}

\begin{figure}[H]
    \centering
    \includegraphics[width=3.37in,height=5in]{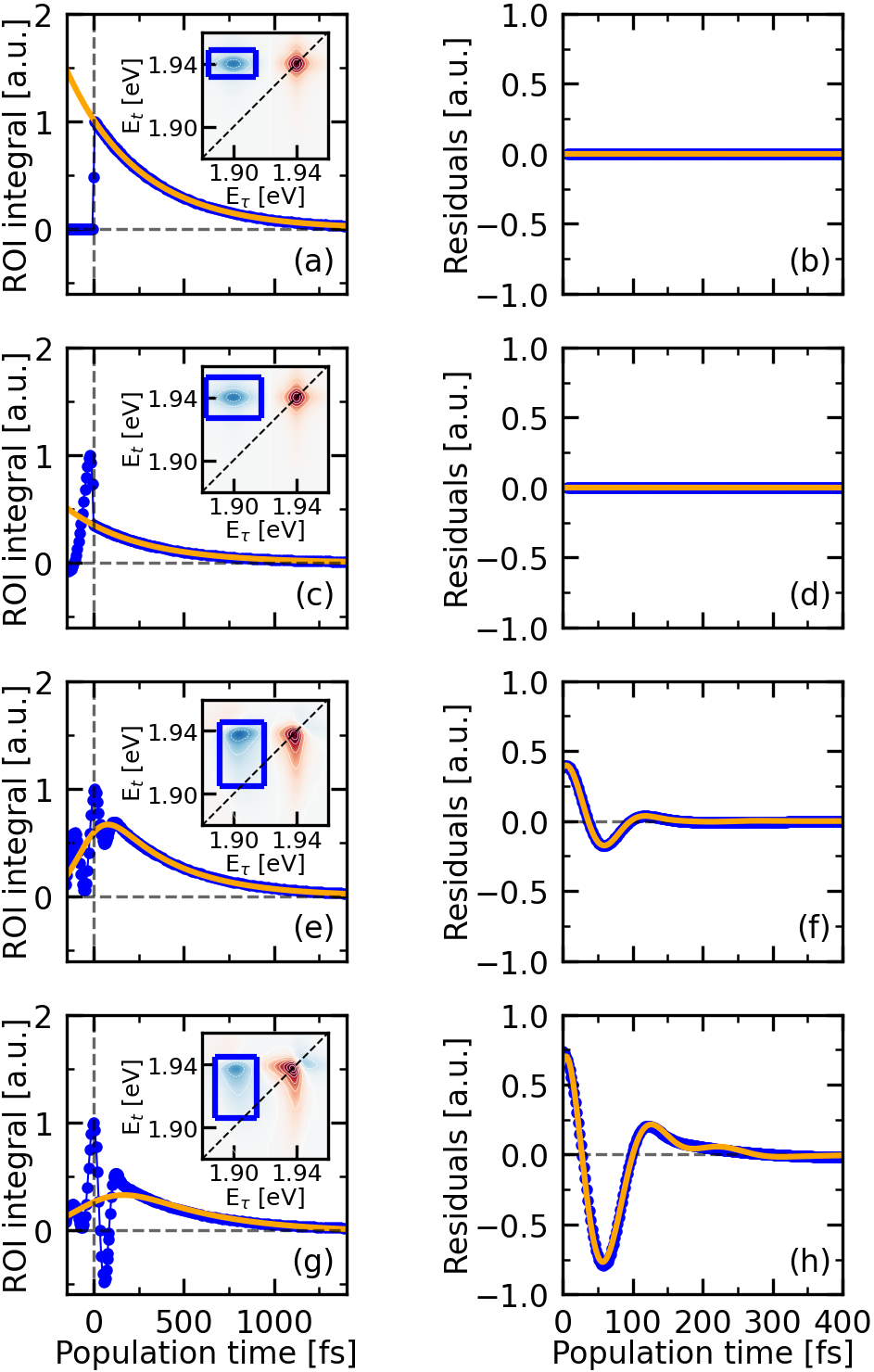}
    \caption{Time evolution of the integrated signal over a selected region of interest (ROI) corresponding to the off-diagonal (blue) peak of the anharmonic oscillator 2D spectrum. Panels (a, c, e, g) show the dynamics of the ROI signal for increasing levels of model complexity: impulsive limit without time-ordering effects (IL-noTO), impulsive limit with
    time-ordering (IL-TO), transform-limited pulses with finite duration (TL-TO), and chirped pulses with quadratic spectral phase (CHIRP). Panels (b, d) show the corresponding residuals obtained by subtracting a single-exponential fit from the long-time dynamics, while panels (f, h) report the corresponding residuals obtained after subtraction of an exponentially decaying function convoluted with a Gaussian response. The insets show snapshots of the 2D spectra at $T = 0$ fs.}
    \label{fig:ROI_AO_C}
\end{figure}

\begin{figure}[H]
    \centering
    \includegraphics[width=3.37in,height=5in]{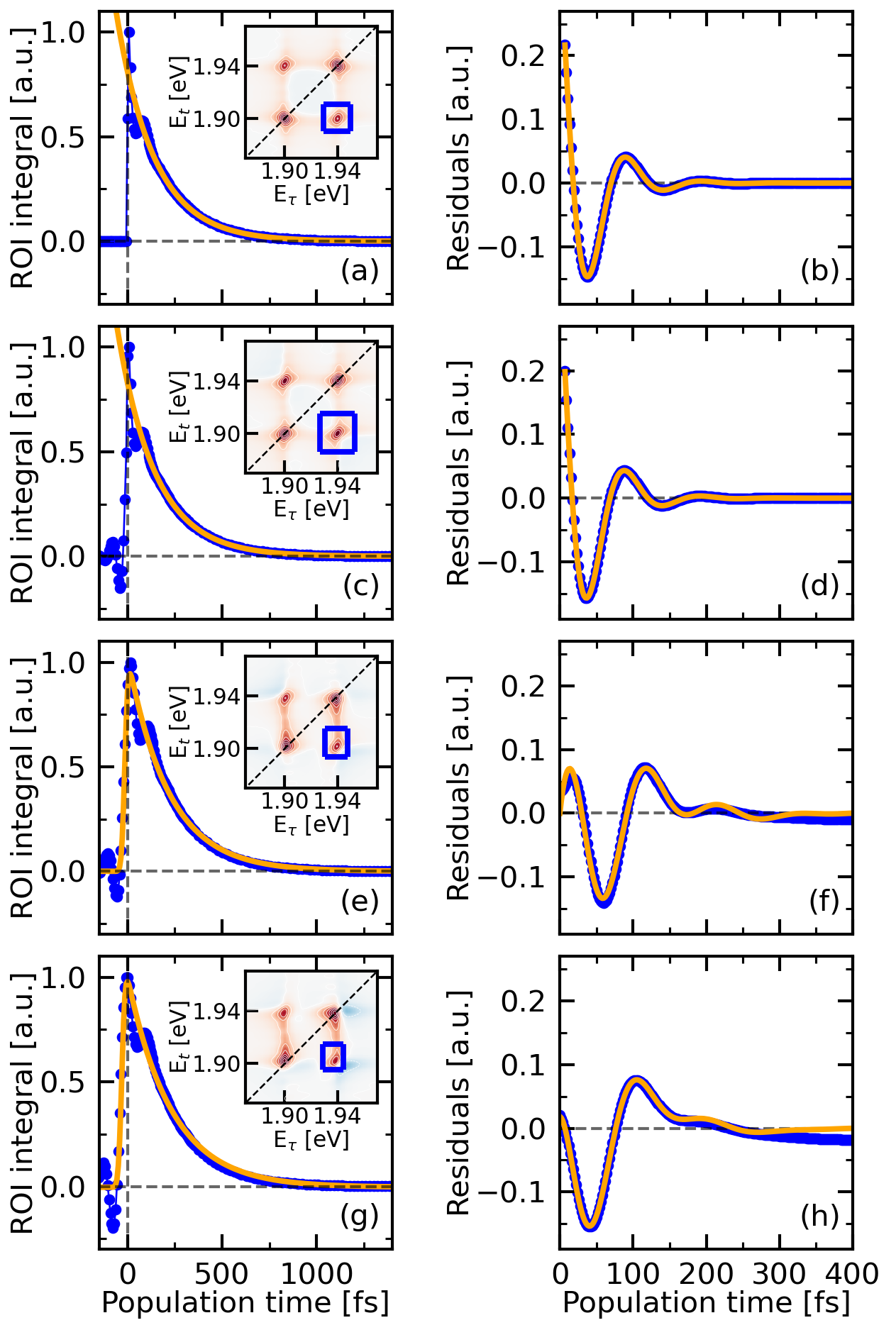}
    \caption{Time evolution of the integrated signal over a selected region of interest (ROI) corresponding to the cross-peak of the two-excitons system's 2D spectrum. Panels (a, c, e, g) show the dynamics of the ROI signal for increasing levels of model complexity: impulsive limit without time-ordering effects (IL-noTO), impulsive limit with
    time-ordering (IL-TO), transform-limited pulses with finite duration (TL-TO), and chirped pulses with quadratic spectral phase (CHIRP). Panels (b, d) show the corresponding residuals obtained by subtracting a single-exponential fit from the long-time dynamics, while panels (f, h) report the corresponding residuals obtained after subtraction of an exponentially decaying function convoluted with a Gaussian response.. The insets show snapshots of the 2D spectra at $T = 0$ fs.}
    \label{fig:ROI_EX_C}
\end{figure}

In this section, we present two studies analogous to those shown in Fig.~\ref{fig:ROI_AO} for the anharmonic oscillator and in Fig.~\ref{fig:ROI_EX} for the two-excitons system but discussing the respective cross-peaks features (in Fig.~\ref{fig:ROI_AO_C} and Fig.~\ref{fig:ROI_EX_C}, respectively).  As can be seen, the analysis carried out for the main peaks in both systems leads to conclusions that are essentially reproduced also in these two cases. 

\twocolumngrid 
\newpage

\onecolumngrid

\begin{table}[H]
	\centering
	\begin{tabular}{c c c}
		\toprule
		 & \textbf{Extracted lifetimes [fs]} & \textbf{Residual parameters} \\
		\midrule
		IL-noTO &  400.00 $\pm$ 0.09 $\cdot$ $10^{-6}$&  - \\
		IL-TO   &  400.00 $\pm$ 0.14 $\cdot$ $10^{-6}$&  - \\
		TL-TO   &  391.21 $\pm$ 0.22 & \makecell{$\tau = 33.9 fs$ \\ $\omega = 0.040 eV$} \\
		CHIRP   &  405.23 $\pm$ 0.14 &  \makecell{${\tau}_1 = 33.5 fs$,  ${\tau}_2 = 39.5 fs$\\ ${\omega}_1 = 0.015 eV$, ${\omega}_2 = 0.041 eV$} \\
		\bottomrule
	\end{tabular}
	\caption{The fitted parameters corresponding to Fig.~\ref{fig:ROI_AO} are summarized in this table for all investigated cases. The first column reports the exponential decay constants extracted from the fits of the ROI dynamics
    (compare to App.~\ref{app:pulse_flat}). The second column lists the parameters obtained from the analysis of the residuals in the transform-limited (TL) and chirped pulse (CHIRP) cases. In these configurations, the residuals are described using one or a sum of two damped harmonic oscillators, namely. From these models we retrieve the decay time ($\tau$) and frequencies ($\omega$) of the damped oscillations.}
	\label{tab:ROI_AO}
\end{table}

\begin{table}[H]
	\centering
	\begin{tabular}{c c c}
		\toprule
		 & \textbf{Extracted lifetimes [fs]} & \textbf{Residual parameters} \\
		\midrule
		IL-noTO &  400.00 $\pm$ 0.01 $\cdot$ $10^{-6}$&  - \\
		IL-TO   &  400.00 $\pm$ 0.56 $\cdot$ $10^{-6}$&  - \\
		TL-TO   &  401.33 $\pm$ 0.38 & \makecell{${\tau}_1$ = 73.1 fs,  ${\tau}_2$ = 44.2 fs\\ ${\omega}_1$ = 0.024 eV, ${\omega}_2$ = 0.041 eV} \\
		CHIRP   &  400.00 $\pm $ 0.19  &  \makecell{${\tau}_1$ = 84.5 fs,  ${\tau}_2$ = 52.9 fs\\ ${\omega}_1$ = 0.013 eV, ${\omega}_2 $= 0.035 eV} \\
		\bottomrule
	\end{tabular}
	\caption{The fitted parameters corresponding to Fig.~\ref{fig:ROI_AO_C} are summarized in this table for all investigated cases. The first column reports the exponential decay constants extracted from the fits of the ROI dynamics
    (compare to App.~\ref{app:pulse_flat}). The second column lists the parameters obtained from the analysis of the residuals in the transform-limited (TL) and chirped pulse (CHIRP) cases. In these configurations, the residuals are described a sum of two damped harmonic oscillators. From these models we retrieve the decay time ($\tau$) and frequencies ($\omega$) of the damped oscillations.}
	\label{tab:ROI_AO_C}
\end{table}

\begin{table}[H]
	\centering
	\begin{tabular}{c c c}
		\toprule
		 & \textbf{Extracted lifetimes [fs]} & \textbf{Residual parameters} \\
		\midrule
		IL-noTO &  200.1 $\pm$ 0.02 &  $\tau = 40.1fs$,  $\omega = 0.040 eV$ \\
		IL-TO   &  200.1 $\pm$ 0.01 &  $\tau = 40.1 fs$,  $\omega = 0.040 eV$ \\
		TL-TO   &  195.77 $\pm$ 0.34 & \makecell{${\tau}_1 = 76.0 fs$,  ${\tau}_2 = 64.6 fs$\\ ${\omega}_1 = 0.025 eV$, ${\omega}_2 = 0.042 eV$} \\
		CHIRP   &  192.9 $\pm$ 0.35 &  \makecell{${\tau}_1 = 51.5 fs$,  ${\tau}_2 = 63.5 fs$\\ ${\omega}_1 = 0.035 eV$, ${\omega}_2 = 0.048 eV$} \\
		\bottomrule
	\end{tabular}
	\caption{The fitted parameters corresponding to Fig.~\ref{fig:ROI_EX} are summarized in this table for all investigated cases. The first column reports the exponential decay constants extracted from the fits of the ROI dynamics
    (compare to App.~\ref{app:model_systems}). The second column lists the parameters obtained from the analysis of the residuals in the impulsive-limit without time-ordering effects (IL-noTO), impulsive-limit with time-ordering effects (IL-TO), transform-limited (TL) and chirped pulse (CHIRP) cases. In these configurations, the residuals are described using a sum of damped harmonic oscillators, characterized by their decay times ($\tau$) and frequencies ($\omega$).}
	\label{tab:ROI_EX}
\end{table}

\twocolumngrid

\onecolumngrid

\begin{table}[H]
	\centering
	\begin{tabular}{c c c}
		\toprule
		 & \textbf{Extracted lifetimes [fs]} & \textbf{Residual parameters} \\
		\midrule
		IL-noTO &  200.08 $\pm$ 0.10&  $\tau = 40.1 fs$,  $\omega = 0.040 eV$ \\
		IL-TO   &  200.09 $\pm$ 0.08&  $\tau = 40.1 fs$,  $\omega = 0.040 eV$ \\
		TL-TO   &  217 $\pm$ 6& \makecell{${\tau}_1 = 82.0 fs$,  ${\tau}_2 = 69.5 fs$\\ ${\omega}_1 = 0.014 eV$, ${\omega}_2 = 0.039 eV$} \\
		CHIRP   &  232 $\pm$ 10&  \makecell{${\tau}_1 = 76.5 fs$,  ${\tau}_2 = 55.5 fs$\\ ${\omega}_1 = 0.012 eV$, ${\omega}_2 = 0.038 eV$} \\
		\bottomrule
	\end{tabular}
	\caption{The fitted parameters corresponding to Fig.~\ref{fig:ROI_EX_C} are summarized in this table for all investigated cases. The first column reports the exponential decay constants extracted from the fits of the ROI dynamics
    (compare to App.~\ref{app:model_systems}). The second column lists the parameters obtained from the analysis of the residuals in the impulsive-limit without time-ordering effects (IL-noTO), impulsive-limit with time-ordering effects (IL-TO), transform-limited (TL) and chirped pulse (CHIRP) cases. In these configurations, the residuals are described using a sum of damped harmonic oscillators, characterized by their decay times ($\tau$) and frequencies ($\omega$).}
	\label{tab:ROI_EX_C}
\end{table}

\begin{table}[H]
	\centering
	\begin{tabular}{c c}
		\toprule
		   \textbf{Extracted lifetimes [fs]} & \textbf{Residual parameters} \\
		\midrule
		  202.42  $\pm$ 0.32& \makecell{${\tau}_1 = 38.3 fs$,  ${\tau}_2 = 60.7 fs$\\ ${\omega}_1 = 0.012 eV$, ${\omega}_2 = 0.039 eV$} \\
		\bottomrule
	\end{tabular}
	\caption{The fitted parameters corresponding to Fig.~\ref{fig:ROI_EX_A} are summarized in this table for all investigated cases. The first column reports the exponential decay constants extracted from the fits of the ROI dynamics (compare to App.~\ref{app:model_systems}). The second column lists the parameters obtained from the analysis of the residuals in the chirped pulse (CHIRP) case. In this configuration, the residuals are described using a sum of damped harmonic oscillators, characterized by their decay times ($\tau$) and frequencies ($\omega$).}
	\label{tab:ROI_EX_A}
\end{table}

\end{document}